\documentclass[aps,prd,twocolumn,showpacs,floatfix,preprintnumbers,amsfont,amsmath,amssymb,nofootinbib,superscriptaddress]{revtex4-1}

\usepackage{aas_macros}
\usepackage{graphicx}
\usepackage{dcolumn}
\usepackage{bm}
\usepackage[normalem]{ulem}
\usepackage[usenames,dvipsnames,svgnames,table]{xcolor}

\usepackage{xcolor}

\definecolor{colorLink}{rgb}{0.9,0,0} 
\definecolor{colorCite}{rgb}{0,0.7,0} 
\definecolor{colorURL} {rgb}{0,0,0.8} 
\usepackage[colorlinks=true,linktocpage=true,linkcolor=colorLink,citecolor=colorCite,urlcolor=colorURL]{hyperref}
\usepackage{bbold}
\usepackage{multirow}
\usepackage{longtable}
\usepackage{fontawesome}

\newcommand{\be}{\begin{equation}}
\newcommand{\ee}{\end{equation}}

\newcommand{\sk}[1]{}

\def\chieff{\chi_{\rm eff}}

\def\pastro{p_{\rm astro}}
\def\msun{{\rm M_{\odot}}}

\DeclareMathSymbol{\mhyphen}{\mathord}{AMSa}{"39}
\defcitealias{Ols22_ias_o3a}{O22}
\defcitealias{Wad23_Pipeline}{W24}

\newcommand{\neviasnew}{11}

\graphicspath{{./}{figs/}}

\begin{document}

\title{New black hole mergers in the LIGO--Virgo O3 data\\ from a gravitational wave search including higher-order harmonics}

 \author{Digvijay Wadekar}
  \email{jayw@ias.edu}
 \affiliation{\mbox{School of Natural Sciences, Institute for Advanced Study, 1 Einstein Drive, Princeton, NJ 08540, USA}}
 \affiliation{\mbox{Department of Physics and Astronomy, Johns Hopkins University,
3400 N. Charles Street, Baltimore, Maryland, 21218, USA}}
  \author{Javier Roulet}
\affiliation{TAPIR, Walter Burke Institute for Theoretical Physics, California Institute of Technology, Pasadena, CA 91125, USA}
 \author{Tejaswi Venumadhav}
\affiliation{\mbox{Department of Physics, University of California at Santa Barbara, Santa Barbara, CA 93106, USA}}
\affiliation{\mbox{International Centre for Theoretical Sciences, Tata Institute of Fundamental Research, Bangalore 560089, India}}
\author{Ajit Kumar Mehta}
\affiliation{\mbox{Department of Physics, University of California at Santa Barbara, Santa Barbara, CA 93106, USA}}
\author{Barak Zackay}
\affiliation{\mbox{Department of Particle Physics \& Astrophysics, Weizmann Institute of Science, Rehovot 76100, Israel}}
\author{Jonathan Mushkin}
\affiliation{\mbox{Department of Particle Physics \& Astrophysics, Weizmann Institute of Science, Rehovot 76100, Israel}}
\author{Seth Olsen}
\affiliation{\mbox{Department of Physics, Princeton University, Princeton, NJ 08540, USA}}
  \author{Matias Zaldarriaga}
\affiliation{\mbox{School of Natural Sciences, Institute for Advanced Study, 1 Einstein Drive, Princeton, NJ 08540, USA}}
 \date{\today}

\begin{abstract}
Nearly all of the previous gravitational wave (GW) searches in the LIGO--Virgo data included GW waveforms with only the dominant quadrupole harmonic, i.e., omitting higher-order harmonics which are predicted by general relativity. We improved the IAS pipeline by efficiently introducing higher harmonics in the GW templates using the techniques in \citet{Wad23_TemplateBanks,Wad23_Pipeline}. Using the \texttt{IAS-HM}\footnote{\label{foot}The $\tt{IAS-HM}$ pipeline is publicly available at \url{https://github.com/JayWadekar/gwIAS-HM}} pipeline on the public LIGO--Virgo data from the O3 run, we find \neviasnew\ new candidate BBH mergers with $0.52\leq p_\mathrm{astro}\leq 0.88$ (we use the detection threshold as the astrophysical probability, $p_\mathrm{astro}$, being over 0.5, following the approach of other pipelines). We broadly recover the high-significance events from earlier catalogs, except a few which were vetoed. We also find that including higher harmonics in our search raises the significance of a few previously reported marginal events (e.g., GW190711\_030756).

A few notable properties of our new candidate events are as follows. At $>95$\% credibility, 4 candidates have primary masses in the intermediate-mass black hole (IMBH) range (i.e., above $\sim$100 $M_\odot$). 5 candidates have median mass ratio $q \leq 0.5$. 5 candidates have median redshift $z \geq 0.8$. 3 candidates have non-zero $\chi_{\rm eff}$ at $>95\%$ credibility. While our new candidate events have modest false alarm rates ($\gtrsim 1.5 $/yr), a population inference study including these can better inform the parameter space of BHs corresponding to the pair instability mass gap, high redshifts and asymmetric mass ratios.
 \href{https://github.com/JayWadekar/GW_higher_harmonics_search}{\faGithub}

\end{abstract}
\maketitle

\section{Introduction}

Gravitational waves (GW) from roughly 100 binary black hole (BBH) mergers have been detected in the data from the first three observing runs of the LIGO--Virgo--Kagra collaboration (LVK), combining detections by the LVK \cite{O1catalog_LVC2016, gwtc1_o2catalog_LVC2018, lvc_o3a_gwtc2_catalog_2021, lvc_o3a_deep_gwtc2_1_update_2021, lvc_gwtc3_o3_ab_catalog_2021, LVK_O3_IMBH_search, Kum24} and by other independent pipelines \cite{ias_pipeline_o1_catalog_new_search_prd2019, ias_o2_pipeline_new_events_prd2020, Ols22_ias_o3a, NitzCatalog_1-OGC_o1_2018, NitzCatalog_2-OGC_o2_2020, nitz_o3a_3ogc_catalog_2021, nitz_4ogc_o3_ab_catalog_2021,Chi23,Meh23_ias_o3b, Kol24_Ares_ML_Search} who access the data made publicly available through the Gravitational Wave Open Science Center (GWOSC) \cite{GWOSC}. Nearly all the previous template-bank based searches have approximated GW waveforms to have only the dominant quadrupole mode, $(\ell, |m|) = (2,2)$, where the frequency of GWs is twice the orbital frequency.

\begin{figure}[!h]
    \centering
    \includegraphics[width=0.52\textwidth]{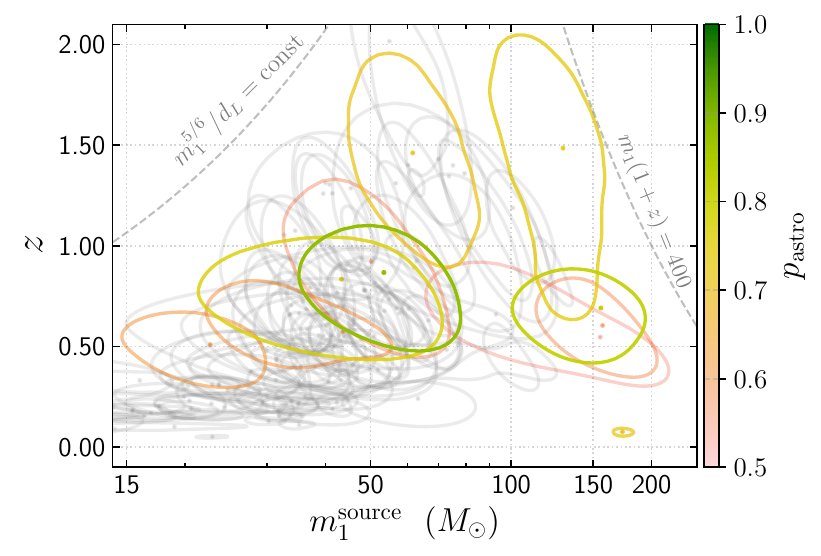}
    \includegraphics[width=0.52\textwidth]{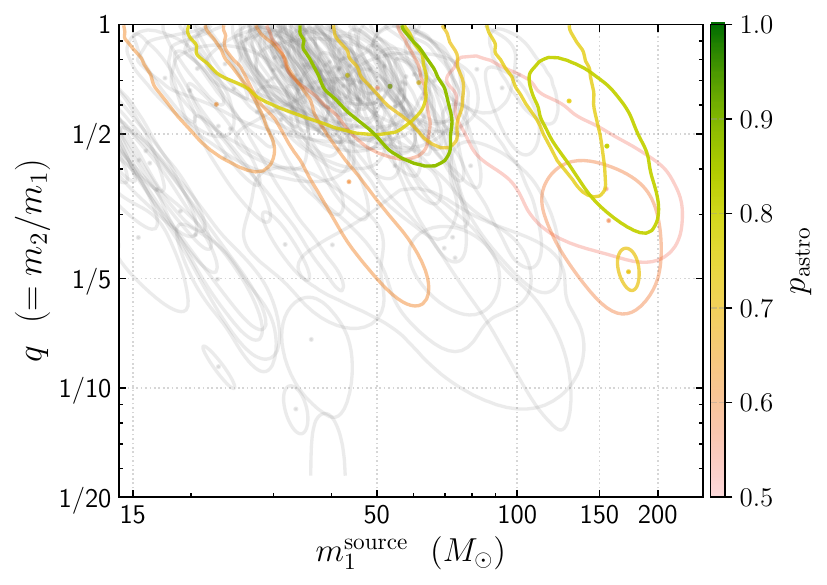}
    \caption{In colored contours, we show properties of the \neviasnew{} new candidate events found in our O3 search with $\pastro >0.5$ ($\pastro$ is the probability of events being of astrophysical origin, as opposed to being a noise transient). The color of the contours corresponds to $\pastro$ values of the events. All the previously reported events from O1--O3 runs (combining the GWTC-3 LVK catalog, the OGC-4 catalog and the previous IAS catalogs) are shown in transparent gray contours \cite{O1catalog_LVC2016, ias_pipeline_o1_catalog_new_search_prd2019, gwtc1_o2catalog_LVC2018, ias_o2_pipeline_new_events_prd2020, lvc_gwtc3_o3_ab_catalog_2021, lvc_o3a_deep_gwtc2_1_update_2021, Ols22_ias_o3a}. The source-frame primary mass and redshift (mass ratio) for events are shown in the top (bottom) panels. 
The posterior contours are obtained from parameter estimation runs using \texttt{cogwheel} \cite{Rou22_cogwheel} and  enclose $50\%$ of the probability and median values are represented by dots. The dashed lines in the upper panel give a rough indication of the limits of our search under the current detector sensitivity, see the text in Section~\ref{sec:PE} for further details. Interestingly, some of the new candidate events have support in the ranges corresponding to IMBH masses, pair-instability mass gap, high redshift, and asymmetric mass ratio.}
    \label{fig:m1_z}
\end{figure}

\begin{figure}
    \centering
\includegraphics[width=0.52\textwidth]{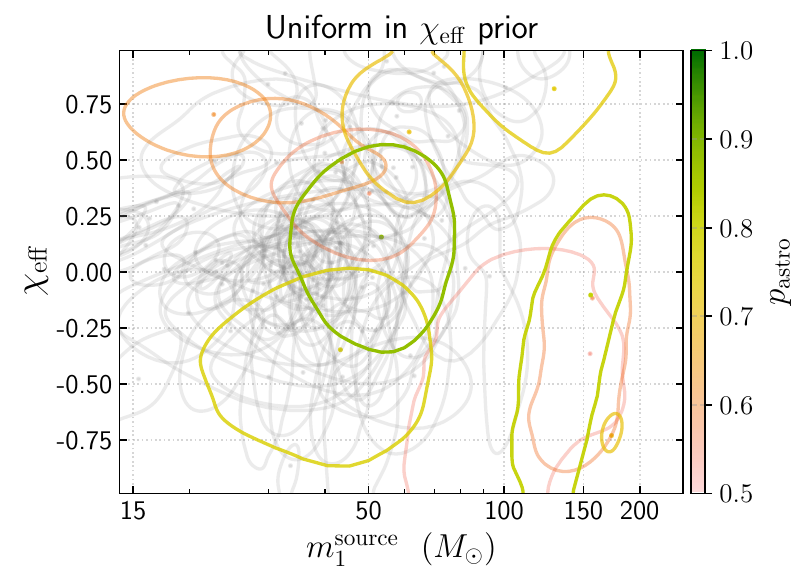}
\includegraphics[width=0.52\textwidth]{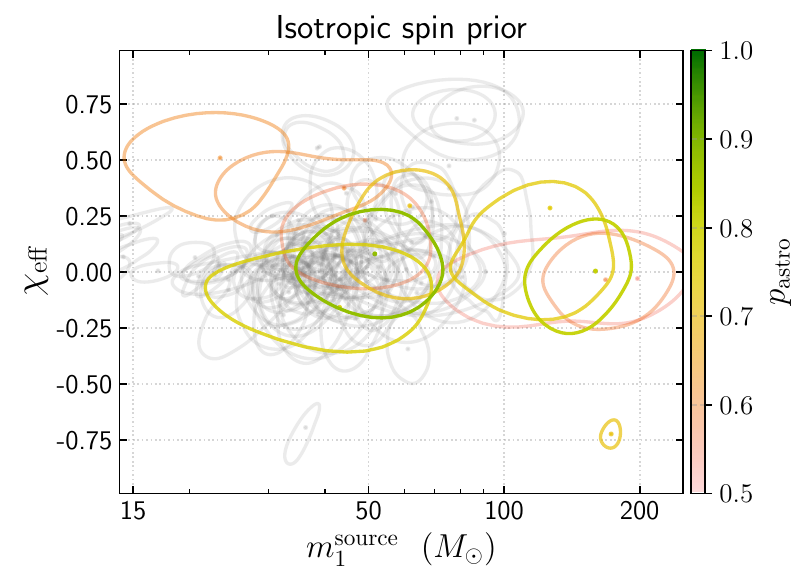}
    \caption{Same as Fig.~\ref{fig:m1_z}, but for the effective spin. \textbf{Top}: Uniform prior on $\chieff$ (which is the fiducial prior adopted throughout this paper). \textbf{Bottom:} Results with isotropic priors on individual spins (which favors $\chi_\mathrm{eff}$ closer to zero), which is also used in the GWTC-3 and 4-OGC catalogs. The relative significance for low SNR positive/negative $\chi_\mathrm{eff}$ values indeed decreases in the isotropic prior, but there are a few events which still show preference for positive/negative $\chi_\mathrm{eff}$. We show similar plots for $\chieff$--$q$ and $\chieff$--$z$ combinations in Fig.~\ref{fig:chieff_extra} in the Appendix.}
    \label{fig:mtot_chieff}
\end{figure}

The $(2,2)$ mode appears at the lowest order in post-Newtonian (PN) expansion, but there are a number of higher harmonics originating from the subsequent orders of the PN expansion (e.g., the $(3,3)$ mode has thrice the orbital frequency) \cite{Tho80, Pan11}.
Approximate formulae for the relative strength of frequency-domain amplitudes of two of the higher-order modes (HM) are given by \cite{Wad23_TemplateBanks}
\be\begin{split}
\bigg|\frac{h_{33}(3f)}{h_{22}(2f)} \bigg| \simeq& \frac{3\sqrt{3}}{4\sqrt{2}}\, \left[\frac{1-q}{1+q}\right]\, (2\pi M_\mathrm{tot} f)^{1/3}\, \sin \iota\\
\bigg|\frac{h_{44}(4f)}{h_{22}(2f)} \bigg| \simeq& \frac{2\sqrt{2}}{3}\, \left[1-\frac{3 q}{(1+q)^2}\right](2\pi M_\mathrm{tot} f)^{2/3}\, \sin^2 \iota
\end{split}
\label{eq:AmpFormulae}\ee
where $q$ is the mass ratio ($=m_2/m_1$), $f$ is the orbital frequency and $\iota$ is the binary inclination ($\iota=\pi/2$ corresponds to the edge-on configuration).
These higher harmonics have a number of important applications for characterizing properties of BBHs.
If one uses the $(2,2)$ mode waveform alone in inferring the BBH properties, we have the following important degeneracies between different binary parameters: luminosity distance and $\iota$; $q$ and effective spin ($\chi_\mathrm{eff}$); polarization and initial orbital phase ($\phi_\mathrm{inital}$). 
All three degeneracies are broken by HM \cite{Mil21, Var14}. 
This follows from the amplitudes of HM in Eq.~\eqref{eq:AmpFormulae} having a different dependence on $\iota$ and $q$ as compared to that of the $(2, 2)$ mode, and similarly the phases of HM have different dependence on $\phi_\mathrm{inital}$. 
It is worth mentioning that the degeneracy between distance and inclination is currently one of the most important limiting factors in inferring the Hubble parameter $H_0$ from GW systems and HM can be crucial in breaking this degeneracy. Similarly, events which have HM also have the best measured mass ratios (e.g., \cite{GW190412, GW190814}). HM become important in the strong gravity regime close to merger and are thus useful to test the multipolar structure of
gravity \cite{Kas18,Kas19,Dha19,Isl20,hm_observation_qn_ringdown_capano2021, Isl21_IMRI, Isl21_190412}, and to study properties of the merger remnant such as the recoil kick \cite{Cal18_HM_BH_kicks}.

Multiple waveform models including HM have been recently made available (e.g., \cite{Lon18, Gar20, Meh17,Ram24_SEOBv5PHM}), and they are currently being used ubiquitously in parameter estimation studies. However, \emph{HM are still not included in current search pipelines} (as of now, the only exception is ~\citet{Cha22}, which however only searched for binaries with nearly edge-on inclinations). The reason behind this is that for making template banks with the $(2, 2)$ mode, one only needs to sample over the masses and spins of black holes ($m_1,m_2, \chi_{1z},\chi_{2z}$). However, for constructing banks including HM, one needs to additionally sample over the inclination and initial phase ($\iota$, $\phi_\mathrm{initial}$). These additional degrees of freedom can cause: $(i)$ the size of the template banks and the cost of search to increase significantly (by a factor of $\sim$100 for the case of HM \cite{Har18, Cha22, Sch23_NF_TemplateBank}), $(ii)$ loss of sensitivity in the search as there is increased number of background triggers \cite{Cap14}. To tackle the first issue, we found more efficient ways of including HM in our template banks (which we presented in detail in our first companion paper \cite{Wad23_TemplateBanks}). Regarding the second issue, we devise new detection statistics which includes marginalization over the amplitudes and phases of higher modes in our second companion paper \cite{Wad23_Pipeline}. We present an overview of these new techniques in Section~\ref{sec:Improvements}.



We see from Eq.~\eqref{eq:AmpFormulae} that, at a given orbital frequency, the relative HM amplitudes increase as a power-law with $M_\mathrm{tot}$. 
Moreover, due to the shape of the current detector power spectral density (PSD), the signal-to-noise ratio (SNR) contribution of HM is additionally boosted at high masses and redshifts. 
In this regime, the lower-frequency $(2, 2)$ mode can cut off below the detector's sensitive band (the detector PSD sharply increases below $\sim100$\,Hz) and HM (being at higher frequencies) naturally have an increased contribution to the SNR. 
The background corresponding to the high-mass regime is however heavily dominated by short-duration noise transients (called instrumental ``glitches"). This is because the high-mass waveforms have a very short duration in band, which makes it easy for transients like blip glitches to mimic real signals. We therefore introduce additional methods in the IAS pipeline to downweight the background at high masses. Some of these methods are detailed in our second companion paper \citet{Wad23_Pipeline} (hereafter \citetalias{Wad23_Pipeline}) and we present an overview in Section~\ref{sec:Downweight_glitches}. 

Running our new search pipeline over the public LIGO--Virgo data from the third observing run (O3), we found new candidate events with properties shown in Figs.~\ref{fig:m1_z} and~\ref{fig:mtot_chieff}. We further discuss these new events and also show our search results for the events from the GWTC-3 and 4-OGC catalogs in Section~\ref{sec:Results}. We review the interesting astrophysical properties of the new candidate events in Section~\ref{sec:astro_properties}. We analyze the impact of our astrophysical prior and comment upon the gain/loss of sensitivity due to adding HM in Section~\ref{sec:Discussion}, and we conclude in Section~\ref{sec:Conclusions}.

\section{Pipeline improvements}
\label{sec:Improvements}

Here, we present an outline of the additions that we made to the previous $(2, 2)$-only version of the IAS pipeline \cite{Ols22_ias_o3a, ias_pipeline_o1_catalog_new_search_prd2019}.

\subsection{Adding HM in template banks}
\label{sec:HM_TemplateBanks}

In this sub-section, we summarize the methodology for constructing template banks for HM from our companion paper: \citet{Wad23_TemplateBanks}.
For each $(2, 2)$ mode template, we use a combination of post-Newtonian formulae and machine learning tools to model normalized aligned-spin templates corresponding to the $(3, 3)$ and $(4, 4)$ modes.
We then filter the data separately with the three templates and store the complex SNR timeseries $(\rho_{22}, \rho_{33}, \rho_{44})$, see figure~1 of \cite{Wad23_TemplateBanks}. As a result, our HM search matched-filtering cost is just $3\times$ that of a $(2, 2)$-only search (in contrast to being $\sim\! 100 \times$ as in previous HM search methods, where the different modes are combined and templates sample over inclination and initial orbital phase \cite{Cha22, Sch23_NF_TemplateBank, Har18}).

We divide our search space into 17 template banks, which are divided according to the shape of normalized $(2, 2)$ waveforms corresponding to the binary parameters (the shape of normalized waveforms is roughly related to the binary merger frequency). 
The parameter space corresponding to each bank is shown in the bottom panel of Fig.~\ref{fig:Template_Prior_Banks} in the Appendix. Some of the banks are further divided into subbanks based on the range of chirp masses they cover. 
Note that the division into banks and subbanks is only to facilitate covering the parameter space with templates of the form that we use, and the final ranking statistic we assign to each trigger is weighed against those of all other triggers (over the entire parameter space) to estimate its global significance. 
Other than adding HM, we also improved our $(2, 2)$ template banks by using a machine learning tool called random forest regressor (which helps us reduce the dimensionality of our template banks and enables us to search with 30$-$40\% fewer templates than before), see Fig.~5 of \cite{Wad23_TemplateBanks}.

\subsection{New statistics for combining SNR timeseries from different harmonics}
\label{sec:HM_statistics}

We present  the techniques for combining the $(\rho_{22}, \rho_{33}, \rho_{44})$ timeseries in our companion paper \citepalias{Wad23_Pipeline} and here we provide a brief overview.
Combining SNR timeseries of different modes incorrectly can lead to a large loss of sensitivity of the pipeline \cite{Cap14}. 
We use the Neyman--Pearson lemma to develop an optimal statistic to combine the SNR timeseries for different harmonics; our detection statistic is the ratio of the Bayesian evidence under the signal and the noise hypothesis \cite{neymanpearson}. Ideally, one would marginalize over all the intrinsic and extrinsic binary parameters to calculate the evidence under the signal hypothesis. 
We develop two different approximate versions of the statistic to be used in different parts of our pipeline as detailed in the sub-sections below. 

\subsubsection{Single-detector collection statistic}
\label{sec:single_det_statistic}

We first developed an approximate statistic to collect individual detector triggers.
Similar to other pipelines, we set a particular SNR threshold ($\rho_\mathrm{collect}$) to collect triggers from individual detectors. The primary reason behind this is to reduce the number of triggers passed to the next steps of the pipeline (coincidence analysis, veto and coherent score calculation) as these steps are computationally expensive (another reason is storage space constraints when saving a large number of triggers on disk). For collecting the triggers in a $(2, 2)$-only search, one can simply use $|\rho_{22}|^2>\rho^2_\mathrm{collect}$. For the HM case, if we naively use $|\rho_{22}|^2+|\rho_{33}|^2+|\rho_{44}|^2>\rho^2_\mathrm{collect}$, the number of collected triggers increases by a factor of over $100$ (using the same value of $\rho_\mathrm{collect}$ as the $(2, 2)$-only case \citepalias{Wad23_Pipeline}). To ameliorate this, we develop a new marginalized statistic to collect triggers ($\rho^2_\mathrm{single-det}>\rho^2_\mathrm{collect}$ in~\citetalias{Wad23_Pipeline}). We exploit the fact that the values of $|\rho_{33}/\rho_{22}|$ and $|\rho_{44}/\rho_{22}|$ are bounded for physical signals (as opposed to Gaussian noise triggers where the values $|\rho_{33}/\rho_{22}|$ and $|\rho_{44}/\rho_{22}|$ can be arbitrarily large), see Fig.~2 of \citepalias{Wad23_Pipeline}.
We therefore marginalize over the space of physically possible amplitude ratios of higher harmonics using a pre-computed prior samples for $|\rho_{ii}/\rho_{22}|$ (which are derived from our astrophysical prior). Note that the prior on $|\rho_{ii}/\rho_{22}|$ can be very different for templates in different regions of parameter space.
For example, the median of our prior for $|\rho_{33}/\rho_{22}|$ is 0.04 for Bank 0 (a low-mass bank), while it is 0.3 for Bank 14 (a high-mass bank). If we did not include such a prior, we would indeed expect to encounter a large loss of sensitivity on including HM.
To make the computation of $\rho^2_\mathrm{single-det}$ faster, we maximize (instead of marginalizing) over the phases of HM and also over all the remaining intrinsic and extrinsic binary parameters. 

\subsubsection{Coherent marginalized statistic}
Once we collect single detector triggers, we check if the same template(s) fired in other detectors at a similar time. To rank the resulting coincident triggers (for calculating their false-alarm rate (FAR) and subsequently $p_\mathrm{astro}$), we develop a more accurate statistic which also includes coherence between different detectors for ranking the events. For example, we exploit the fact that astrophysical signals have correlated phase and time differences between the detectors, while the noise triggers have random distributions \cite{CoherentScore}.
In the previous sub-section, we maximized over extrinsic binary parameters, but marginalizing over them is the more optimal strategy.
Our coherent ranking statistic thus includes marginalization over extrinsic parameters in a similar way to the IAS $(2, 2)$-only search methodology given in Appendix~D of \citet{Ols22_ias_o3a} (hereafter \citetalias{Ols22_ias_o3a}), except generalized to account for higher modes (see \cite{Rou23_CoherentScore,relative_binning_higher_modes} for details). In addition, we also include marginalization over the amplitude of HM (which effectively corresponds to marginalizing over mass ratios); the only parameters that we currently do not marginalize over are the intrinsic binary parameters corresponding to the 22 waveform, as we use only the best-fit 22 template in our calculations. Details regarding our coherent statistic are provided in our companion papers \cite{Wad23_Pipeline, Rou23_CoherentScore}.

\subsection{Removing/downweighting noise transients}
\label{sec:Downweight_glitches}

In addition to the changes involving HMs, we also improve the sensitivity of our search for high masses compared to previous IAS pipelines by developing new noise reduction tools. 

\subsubsection{Trigger Inference using Extended strain representation (TIER)}
\label{sec:TIER}

We developed a machine learning (ML) framework called \texttt{TIER} (Trigger Inference using Extended strain Representation) for improving the sensitivity of any generic pipeline \cite{Wad25_TIER_GW}. Typically, search pipelines only use a small region of strain data in the vicinity of a candidate signal to construct the detection statistic. However, extended strain data ($\sim 10$ s) in the candidate's vicinity can also carry valuable complementary information. We included this extended information into our detection statistic using ML classifiers (see Fig.~1 of \cite{Wad25_TIER_GW}). Using TIER, we found up to$\sim$ 20\% improvement in sensitive volume of our pipeline, with improvements being concentrated in regions of high masses and unequal mass ratios. We refer the reader to \cite{Wad25_TIER_GW} for further details regarding TIER.

\subsubsection{Band eraser tool for data preprocessing}

In our $(2, 2)$-only pipeline, in a data pre-processing step (before we matched-filter with templates), we first identify regions of noisy data segments (see section~C of \cite{ias_pipeline_o1_catalog_new_search_prd2019}), and remove the data in the flagged regions (i.e., make “holes”). We then fill the holes using an “inpainting filter” (see Fig.~6 of \cite{psd_drift}).
Note that the holes remove the entire data from a particular time chunk. Sometimes, however, the noise transients are only localized to particular frequency bands (particularly the low-frequency end $\lesssim$ 100 Hz) and do not affect astrophysical signals which dominantly have power in a different frequency band.

Therefore, before making holes in entire time-chunks, we try to smoothly remove bad time-frequency regions (or bands) in the spectrogram by developing a new tool called \emph{band eraser}. First, we divide the spectrogram into multiple bands (each with dimensions $64\,\text{s} \times 2\,\text{Hz}$). We further split each band into multiple chunks (of $2\,\text{s} \times 0.5\,\text{Hz}$) and measure the number of chunks which have power above a certain threshold. If this number of chunks in a given band is above a certain threshold, we remove the entire band from our analysis. We leave further discussion of the band eraser tool to~\citetalias{Wad23_Pipeline}.



\subsubsection{Improving signal consistency vetos using ML}
\label{sec:Veto_ML}

The $(2, 2)$-only IAS pipeline included a series of signal consistency checks in order to veto triggers that are likely to arise from noise transients, in order to improve the search sensitivity (see section~I of \cite{ias_pipeline_o1_catalog_new_search_prd2019}). 
On the coincident triggers, we perform similar veto procedures as the $(2, 2)$-only pipeline, except for the following change.
For a particular trigger, we first subtract our best-fit $(3, 3)$ and $(4, 4)$ waveforms from data and then run the same $(2, 2)$ signal consistency tests as earlier on the residual data. 
In the limit in which the higher harmonics are nearly orthogonal to the $(2, 2)$ waveform, the consistency tests can be left largely unchanged. We aim to modify these tests to also check the consistency of the $(3, 3)$ and $(4, 4)$ parts of the signals in the future.

One of the downsides of our veto tests is that if a loud signal displays effects which are not currently present in our templates (e.g. precession or eccentricity), the chances of a veto test mis-identifying it as a glitch are much higher (e.g. GW190521 was a high-mass merger which got vetoed in a previous IAS search \citetalias{Ols22_ias_o3a} as it shows a mild evidence for precession). To remedy this, we run our pipeline on precessing GW signals injected in the data. We then use a ML classifier to distinguish between properties of GW signals which got vetoed, as opposed to the properties of the noise transients which got vetoed. The ML classifier therefore identifies the precise combination of signal consistency tests which are more likely to fire when a GW signal is precessing. The ML classifier output probability is then used to set a criteria for overriding the vetos in a select few cases. This helps us recover loud precessing signals
and increases the volume-time sensitivity of our pipeline by $\sim$ 10\% across nearly all of the parameter space. We leave further discussion of the ML-based veto overriding procedure to a future paper~\cite{Wad25_TIER_GW}.


In previous quadrupole-only searches, we reduced the noise background by accounting for the non-stationarity of the PSD \cite{psd_drift}. 
In this procedure, we matched-filter a reference template from the bank with the data and register its SNR $\rho (t)$. 
We then estimate the variation in PSD as a function of time by averaging the power $\langle \rho^2 \rangle$ within rolling windows of length $\sim 15$\,s.
In the limit in which the fluctuations of the PSD over all frequencies are described by a single scalar, this procedure returns an unbiased estimate of the drift of the PSD.
In our HM search, we calculate the correction for the PSD drift in the same way as before by only using a reference $(2, 2)$ waveform from the template bank for calculating $\langle \rho^2 \rangle$. We currently do not include (3,3) or (4,4) waveforms in estimating PSD drift and leave this direction to future work.

\subsubsection{Ranking statistic}
\label{sec:noise_ranking_statistic}

Even after performing the above noise-removal steps, there are still some noise triggers which survive. Our ranking statistic is the final tool we can use to downweight such triggers. Beyond upweighting triggers which have good coherence among multiple detectors, an important job of the ranking statistic is to also downweight triggers based on the probability with which similar triggers occur in the background. We simulate 2000 instances of background runs using the method of timeslides (where we shift one of the detectors by unphysical amounts which are more than the light crossing time between the detectors).

In the $(2, 2)$-only case, for a trigger with SNR $\rho_{22}$ associated with a template $\alpha$, this probability under the noise hypothesis $\mathcal N$ is $P(|\rho_{22}|^2 \mid \alpha,\mathcal N)$. In the case of Gaussian noise, $|\rho_{22}|^2$ follows a $\chi_\mathrm{2\, d.o.f}^2$ (i.e., $P(|\rho_{22}|^2 \mid \alpha,\mathcal N) \propto \exp(-|\rho_{22}|^2/2)$). However, we empirically measure $P(|\rho_{22}|^2 \mid \alpha,\mathcal  N)$ by constructing a histogram of the $|\rho_{22}|^2$ values of background triggers associated with the template $\alpha$. 
In a similar way for the HM case, if the noise were Gaussian, $|\rho_{22}|^2, |\rho_{33}|^2, |\rho_{44}|^2$ separately follow $\chi_\mathrm{2\, d.o.f}^2$ distributions (note that we orthogonalize templates of different harmonics), and their sum thus follows a $\chi_\mathrm{6\, d.o.f}^2$ distribution with 6 d.o.f.
This motivated us to make the approximation $P(\rho_{22}, \rho_{33}, \rho_{44} \mid \alpha,\mathcal  N) \propto P(|\rho_{22}|^2 + |\rho_{33}|^2 + |\rho_{44}|^2 \mid \alpha,\mathcal  N)$ and again measure this probability distribution empirically by constructing a 1D histogram of $(|\rho_{22}|^2 + |\rho_{33}|^2 + |\rho_{44}|^2)$ values of background triggers associated with the template $\alpha$ (see~\citetalias{Wad23_Pipeline}). 

Note that in the previous IAS studies, $P(|\rho_{22}|^2 \mid \alpha,\mathcal  N)$ is assumed to be the same for all templates in a subbank. However, especially for high masses, this term can have different values even for different templates within a subbank, see e.g., \cite{LIGO_O1, CoherentScore, Wad23_Pipeline}. In order to optimize our analysis, we follow the technique in~\citetalias{Wad23_Pipeline}, where we first cluster the templates within each sub-bank based on their sensitivity to glitches (as quantified empirically by the relative fraction of loud and faint triggers belonging to each template). We then separately construct the histograms of background triggers for each template group to estimate the noise probabilities of triggers associated to templates within that group.



\bgroup
\def\arraystretch{1.5}
\begin{table*}
    \centering
\caption{Properties of new candidate events found in our search with $p_{\rm astro} > 0.5$. We also report the inverse false-alarm rate (IFAR) and $\rho_{\rm{H,L}}$ denotes the incoherent SNR of the triggers in the Hanford and Livingston detectors ($\rho_{\rm{H,L}}$ values here include contributions from all harmonics, their separate contributions are shown separately in Table~\ref{tab:subthreshold_LVC}). The parameter estimation (PE) results are obtained using the \texttt{IMRPhenomXODE} model. The errorbars correspond to $90\%$ confidence range. Note that the PE results here incorporate precession and use Virgo data when available, unlike our search. We also perform PE separately with $(2, 2)$-only aligned-spin waveforms and report the change in evidence when aligned-spin HM and precession effects are iteratively included in PE.}
    \begin{tabular}{|c|c|c|ccccc|cc|cc|cc|c|}
        \hline
        Sr. & New candidate & \multirow{2}{*}{Bank} &\multicolumn{5}{c|}{PE (90\% credibility)} & \multicolumn{2}{c|}{$\Delta \log_{10} $(evidence)}  & \multirow{2}{*}{$\rho_{\rm H}^2$}  & \multirow{2}{*}{$\rho_{\rm L}^2$} &\multicolumn{2}{c|}{IFAR (yr)} & \multirow{2}{*}{$p_\mathrm{astro}$} \\
        \cline{4-10} \cline{13-14}
         No. & event& & $m_1^{\textrm{src}} {\scriptstyle( \rm \msun)}$ & $q$ & $\chi_{\rm eff}$ & $z$ & $\rho^2_\mathrm{network}$ & \scriptsize{ HM} & \scriptsize{HM+Prec.} & & &  \scriptsize{per bank} & \scriptsize{overall}  &
         \\ 
         \hline 

         \rowcolor{gray!30} 1 & GW200104\_184028 & \texttt{8,1} & $55_{-15}^{+29}$\!&$0.7_{-0.4}^{+0.3}$\!&$0.1_{-0.8}^{+0.6}$\!&$0.8_{-0.4}^{+0.5}$\!&$72.2$ &$-0.0$ &$0.16$ &$46.5$ & $30.1$ & $5.3$&$ 0.67$ & $0.88$ \\
2 & GW190425\_133124 & \texttt{14,0} & $150_{-50}^{+60}$\!&$0.5_{-0.3}^{+0.4}$\!&$-0.2_{-0.7}^{+0.7}$\!&$0.7_{-0.3}^{+0.3}$\!&$87.0$ &$0.26$ &$0.31$ &$32.4$ & $44.3$ & $14.5$&$ 0.34$ & $0.81$ \\
\rowcolor{gray!30} 3 & GW200319\_022728 & \texttt{7,1} & $43_{-16}^{+23}$\!&$0.7_{-0.4}^{+0.3}$\!&$-0.4_{-0.5}^{+0.6}$\!&$0.8_{-0.4}^{+0.5}$\!&$67.4$ &$0.0$ &$-0.04$ &$29.0$ & $43.3$ & $2.3$&$ 0.23$ & $0.76$ \\
4 & GW190605\_025957 & \texttt{14,0} & $130_{-40}^{+80}$\!&$0.6_{-0.4}^{+0.4}$\!&$0.7_{-1.1}^{+0.3}$\!&$1.3_{-0.7}^{+0.8}$\!&$110.0$ &$-0.01$ &$-0.08$ &$30.5$ & $45.1$ & $10.8$&$ 0.20$ & $0.74$ \\
\rowcolor{gray!30} 5 & GW200114\_020818 & \texttt{8,1} & $175_{-15}^{+11}$\!&$0.21_{-0.08}^{+0.03}$\!&$-0.72_{-0.15}^{+0.12}$\!&$0.09_{-0.07}^{+0.03}$\!&$220.0$ &$5.22$ &$5.24$ &$30.3$ & $149$ & $1.7$&$ 0.16$ & $0.71$ \\
6 & GW190524\_134109 & \texttt{9,0} & $64_{-18}^{+31}$\!&$0.7_{-0.4}^{+0.3}$\!&$0.6_{-0.8}^{+0.3}$\!&$1.4_{-0.7}^{+0.7}$\!&$67.6$ &$-0.33$ &$-0.34$ &$22.4$ & $40.2$ & $1.4$&$ 0.16$ & $0.71$ \\
\rowcolor{gray!30} 7 & GW191213\_204950 & \texttt{1,2} & $23_{-7}^{+18}$\!&$0.6_{-0.4}^{+0.4}$\!&$0.7_{-0.3}^{+0.2}$\!&$0.5_{-0.2}^{+0.2}$\!&$75.4$ &$0.0$ &$-0.01$ &$27.2$ & $55.2$ & $2.3$&$ 0.09$ & $0.61$ \\
8 & GW190530\_030659 & \texttt{3,2} & $37_{-15}^{+26}$\!&$0.5_{-0.3}^{+0.4}$\!&$0.5_{-0.3}^{+0.3}$\!&$0.6_{-0.3}^{+0.3}$\!&$75.8$ &$-0.12$ &$-0.08$ &$34.7$ & $39.4$ & $1.7$&$ 0.09$ & $0.60$ \\
\rowcolor{gray!30} 9 & GW200210\_035448 & \texttt{9,0} & $150_{-50}^{+70}$\!&$0.31_{-0.16}^{+0.29}$\!&$-0.3_{-0.6}^{+0.6}$\!&$0.6_{-0.3}^{+0.4}$\!&$86.2$ &$2.15$ &$2.24$ &$23.7$ & $44.0$ & $0.72$&$ 0.08$ & $0.57$ \\
10 & GW190908\_161547 & \texttt{7,1} & $50_{-20}^{+30}$\!&$0.7_{-0.5}^{+0.3}$\!&$0.3_{-0.4}^{+0.4}$\!&$0.9_{-0.5}^{+0.6}$\!&$67.0$ &$-0.11$ &$-0.13$ &$48.2$ & $33.1$ & $0.81$&$ 0.08$ & $0.56$ \\
\rowcolor{gray!30} 11 & GW190710\_150742 & \texttt{10,0} & $130_{-60}^{+100}$\!&$0.4_{-0.2}^{+0.5}$\!&$-0.3_{-0.6}^{+0.8}$\!&$0.6_{-0.3}^{+0.5}$\!&$78.6$ &$0.41$ &$0.25$ &$23.2$ & $45.5$ & $0.59$&$ 0.06$ & $0.52$ \\
         
         \hline 
    \end{tabular}
    \label{tab:NewEvents}
\end{table*}

\begin{figure*}
    \includegraphics[width=1\textwidth]{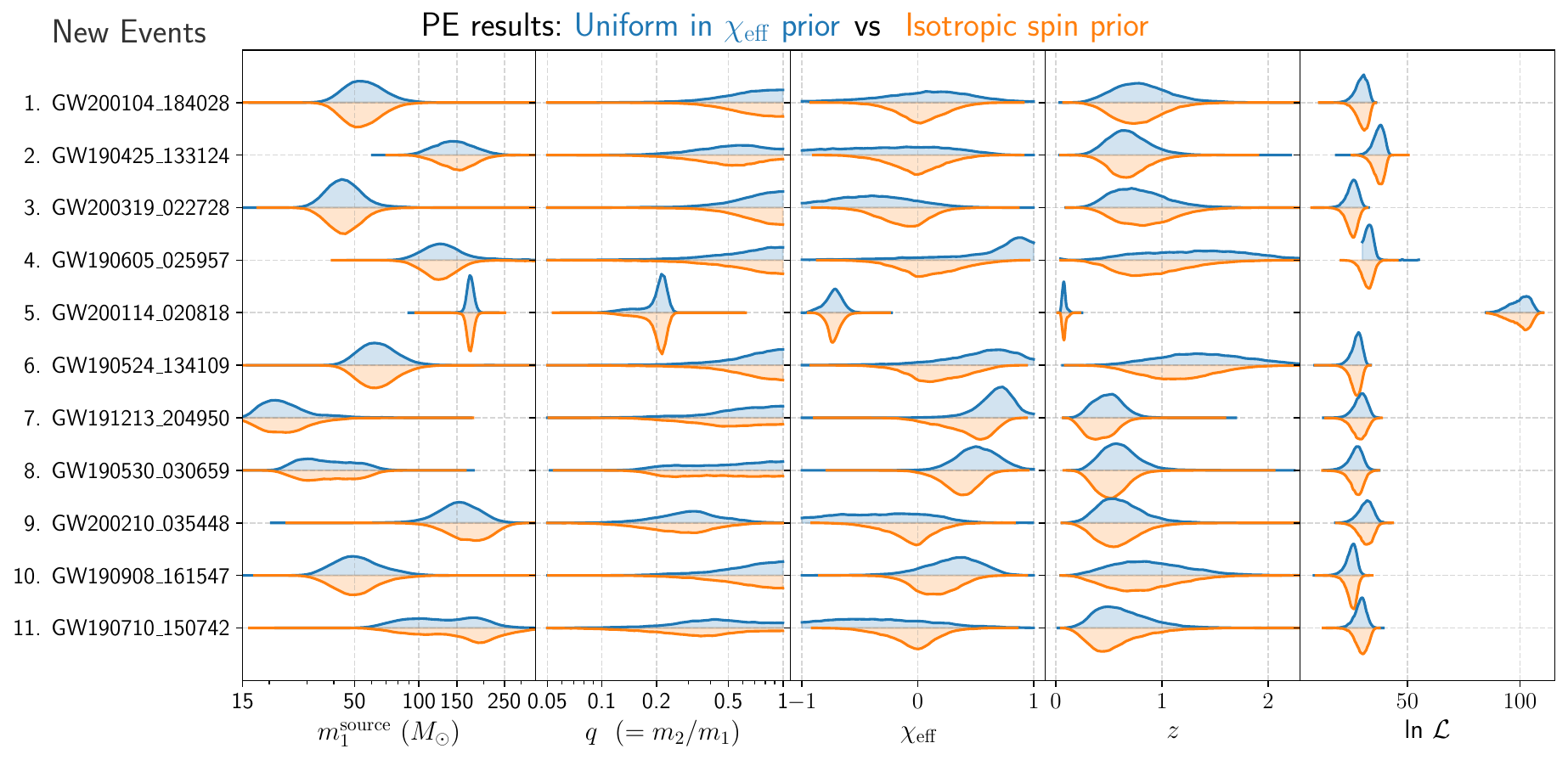}
    \caption{Parameter estimation results for the new events under two different priors on BH spins. The largest difference between the two cases is when the log-likelihood is relatively low but the posterior exhibits support for non-zero $\chi_\mathrm{eff}$ under the uniform-in-$\chieff$ prior (blue). More detailed corner plots from different runs are provided in Figs.~\ref{fig:PE_Events_1} and~\ref{fig:PE_Events_2}, and comparison with the PE runs using \texttt{IMRPhenomXPHM} are in Fig.~\ref{fig:XODE_XPHM} in the Appendix.}
    \label{fig:PE_1D_spin}
\end{figure*}
\twocolumngrid
\egroup

\subsection{Astrophysical prior}
\label{sec:astrophysical_prior}
In this section, we discuss the astrophysical prior used in our search. This prior is used to assign weights to different templates according to the probability of the template $\alpha$ firing under the signal hypothesis $\mathcal S$: $P(\alpha\mid \mathcal S$), and also to downweight triggers with unphysical $|\rho_{33}/\rho_{22}|$ and $|\rho_{44}/\rho_{22}|$ ratios as discussed earlier in Section~\ref{sec:HM_statistics}.

We only probe binary black hole (BBH) mergers in our study and leave the search for systems containing neutron stars for future work. 
The range of our search was restricted to the following space:
\be\begin{split}
3 \,\msun < m^\mathrm{det}_2 &< m^\mathrm{det}_1 < 400\, \msun \\
1/18 &< q < 1\\
|\chi_1|, |\chi_2| &<0.99
\end{split}\label{eq:prior}\ee
where $m^\mathrm{det}=m^\mathrm{source} (1+z)$ refer to detector-frame (redshifted) masses of the binary.
We chose the mass ratio cut for BBHs to restrict to the calibration regime of the IMRPhenomXAS approximant to numerical relativity simulations \cite{Pratten:2020fqn}.

We picked a simple prior which broadly covers the parameter space (picking a broad prior can help with reweighting the results to a different prior in the future \cite{ias_popO2_Roulet_2020}).
We use a uniform prior on the mass ratio $q$ for $1/18 < q < 1$; the effective spin parameter $\chi_\mathrm{eff}$ is flat within $-0.95<\chi_\mathrm{eff}<0.95$ (with $|\chi_1|, |\chi_2| <0.99$); the total mass $M_\mathrm{tot}$ is distributed in a power-law fashion, $P(M_\mathrm{tot})\propto M_\mathrm{tot}^{-2}$. 
Note that we use a prior on the detector-frame masses, and we do not assume a separate prior over redshift $z$.

We drew $\sim 10^{6}$ samples according to this prior (each sample corresponding to the physical parameter set $[m_1, m_2, \chi_{1,z}, \chi_{2,z}]$), and assigned a weight to each sample based on its observable volume. We improve upon the astrophysical prior used in \citetalias{Ols22_ias_o3a} by incorporating the fact that the observable changes across the parameter space. For example, for inspiral dominated waveforms, the observable volume roughly scales as $M_\mathrm{tot}^{2.2}$ \cite{Fis17}.
As a proxy for the observable distance, we simply calculated the SNR for the sample at a fixed distance by simulating its waveform using \texttt{IMRPhenomXHM} under a reference PSD (weight$_\mathrm{ sample}$ $\propto$ SNR$^3$).

Using these samples, we then utilized normalizing flows to predict the effective astrophysical prior density at the location of our templates. We also use normalizing flows to generate physical $|\rho_{33}/\rho_{22}|$ and $|\rho_{44}/\rho_{22}|$ samples corresponding to each template (which are required by our marginalization integral). We leave further discussion of our use of normalizing flows to an upcoming paper \cite{Wad25_NormFlow_HM_search}. 
Fig.~\ref{fig:Template_Prior_Banks} in the Appendix shows the cumulative probabilities for our banks, each of which is obtained by adding the weights of all the samples falling within the corresponding bank.

\section{Results}
\label{sec:Results}

\subsection{New candidate events}
Using our new search pipeline on the Hanford--Livingston data from the third observing run (O3), we find \neviasnew{} candidate events with $p_\mathrm{astro}>0.5$ which have not been previously reported in the literature.
We list the properties of these new candidates in Table~\ref{tab:NewEvents}. In all tables in this paper, $\rho^2_{\rm H}$, $\rho^2_{\rm L}$ columns correspond to incoherent SNR obtained from $|\rho_{22}|^2+|\rho_{33}|^2+|\rho_{44}|^2$ in the particular detector (note that we have orthogonalized the templates of different harmonics, so that they can be added in quadrature).
The incoherent $\rho_{22}, \rho_{33}, \rho_{44}$ values may not correspond to those possible for a physical trigger. Thus, in some cases we see that $\rho^2_{\rm H} + \rho^2_{\rm L}$ can even exceed the best-fit parameter estimation (PE) values $\rho^2_\mathrm{network}$. Further discussion of the PE runs are given below in Section~\ref{sec:PE}, and we report 90\% CL values of some of the binary parameters in Table~\ref{tab:NewEvents}. We show the SNR contribution of individual modes separately in Table~\ref{tab:subthreshold_LVC} in the Appendix.

We also mention the best bank\_id, subbank\_id for each event; the bank\_id roughly increases with mass (see Fig.~\ref{fig:Template_Prior_Banks}). Note that the high-mass banks in our case have much fewer templates than the low-mass banks (see Table~I of \cite{Wad23_TemplateBanks}), which leads to considerable heterogeneity in the number of background triggers in the banks. 
Hence, the $\rho^2$ values needed for events to be above threshold can be different for different banks.

Other than running the pipeline on coincident triggers, we also artificially simulate background using 2000 timeslides (where we shift one of the detectors by unphysical amounts which are more than the light crossing time between the detectors) and run the pipeline on these in exactly the same way.
This allows us to measure the inverse false alarm rate (IFAR) of our coincident triggers. We report this value within each bank and the overall IFAR after combining the background lists from all of our 17 banks. 
Additionally, for each trigger, the statistic $\pastro$ is its probability of being of astrophysical origin; computing it involves estimating the astrophysical rate of events in addition to their IFARs. 
We estimate the $\pastro$ values in a manner that is similar to that outlined in Appendix~B of \citetalias{Ols22_ias_o3a}, from the distribution of foreground and background triggers according to
 \be
 p_\mathrm{astro} (\rho_\mathrm{rank}^2) = \dfrac{\dfrac{d N}{d\rho^2} (\rho_\mathrm{rank}^2\mid \mathcal S)}{\dfrac{d N}{d\rho^2} (\rho_\mathrm{rank}^2\mid \mathcal S)+ \dfrac{d N}{d\rho^2} (\rho_\mathrm{rank}^2 \mid \mathcal N)}
 \ee
where $\rho_\mathrm{rank}^2$ in the above formula corresponds to our final ranking statistic value for the candidate events (see Eq.~16 of Ref.~\citetalias{Wad23_Pipeline} for the full expression), and $d N/d(\rho_\mathrm{rank}^2)$ corresponds to number density values from a 1D histogram of the ranking statistic (which we separately calculate for the signal $\mathcal{S}$ and background $\mathcal{N}$ cases). Note that as the templates in our banks have astrophysical probabilities associated with them (see Section~\ref{sec:astrophysical_prior}), and we use these values in calculating $\rho_\mathrm{rank}^2$. Hence, our IFAR and $\pastro$ values are also dependent on our assumption of astrophysical distributions of binary parameters. We expect this dependence to be more significant for the `overall' IFAR in comparison to the `per bank' IFAR values quoted in Table~\ref{tab:NewEvents}. We therefore separately provide the `per bank' values in our table, so one could reweight them in a rough way for different astrophysical distributions than the ones assumed here.

\bgroup
\begin{table*}
\def\arraystretch{1.5}
    \centering
\caption{Same as Table~\ref{tab:NewEvents} but for candidate events with $p_\mathrm{astro}>0.5$ in our search which overlap with those first reported in the IAS $(2, 2)$-only searches: \cite{Ols22_ias_o3a, Meh23_ias_o3b}.}
    \begin{tabular}{|c|c|ccccc|cc|cc|cc|c|}
        \hline
        \multirow{2}{*}{Event} & \multirow{2}{*}{Bank} &\multicolumn{5}{c|}{PE 
 (90\% credibility)} & \multicolumn{2}{c|}{$\Delta \log_{10}$(evidence)}  & \multirow{2}{*}{$\rho_{\rm H}^2$}  & \multirow{2}{*}{$\rho_{\rm L}^2$} &\multicolumn{2}{c|}{IFAR (yr)} & \multirow{2}{*}{$p_\mathrm{astro}$} \\
    \cline{3-9} \cline{12-13}
        & & $m_1^{\textrm{src}} {\scriptstyle( \rm \msun)}$ & $q$ & $\chi_{\rm eff}$ & $z$ & $\rho^2_\mathrm{network}$ & \scriptsize{HM} & \scriptsize{HM+Prec.} & & &  \scriptsize{per bank} & \scriptsize{overall}  &
        \\ 
        \hline 
\rowcolor{gray!30} GW190711\_030756 & \texttt{6,2} & $60_{-20}^{+54}$\!&$0.4_{-0.2}^{+0.3}$\!&$0.1_{-0.6}^{+0.5}$\!&$0.41_{-0.19}^{+0.34}$\!&$102.8$\!&$1.14$\!&$0.9$\!&$30.9$ & $68.5$ & $581$&$10.96$ & $0.99$ \\
GW200109\_195634 & \texttt{9,0} & $70_{-30}^{+40}$\!&$0.7_{-0.4}^{+0.3}$\!&$0.5_{-1.0}^{+0.4}$\!&$1.0_{-0.5}^{+0.7}$\!&$80.8$\!&$-0.07$\!&$-0.02$\!&$40.0$ & $37.4$ & $9.3$&$1.14$ & $0.92$ \\
\rowcolor{gray!30} GW190818\_232544 & \texttt{8,1} & $70_{-30}^{+20}$\!&$0.6_{-0.3}^{+0.4}$\!&$0.7_{-0.3}^{+0.3}$\!&$1.0_{-0.5}^{+0.8}$\!&$77.2$\!&$-0.01$\!&$0.01$\!&$45.9$ & $35.5$ & $6.2$&$0.81$ & $0.90$ \\
GW190707\_083226 & \texttt{7,1} & $52_{-13}^{+24}$\!&$0.6_{-0.3}^{+0.4}$\!&$-0.1_{-0.6}^{+0.5}$\!&$0.6_{-0.3}^{+0.5}$\!&$79.0$\!&$0.19$\!&$0.15$\!&$52.4$ & $32.2$ & $6.5$&$0.55$ & $0.87$ \\
\rowcolor{gray!30} GW191117\_023843 & \texttt{10,0} & $70_{-20}^{+50}$\!&$0.7_{-0.4}^{+0.3}$\!&$-0.4_{-0.5}^{+0.8}$\!&$0.8_{-0.5}^{+0.6}$\!&$78.8$\!&$-0.16$\!&$-0.11$\!&$42.6$ & $29.1$ & $0.52$&$0.06$ & $0.52$ \\
         \hline 
    \end{tabular}
    \label{tab:22_IAS_events_significant}

\def\arraystretch{1}
\LTcapwidth=\textwidth
\begin{longtable*}{|c|c|cc|c|ccc|ccc|}
\caption{Hanford--Livingston coincident events already reported in the GWTC-3 catalog \cite{lvc_gwtc3_o3_ab_catalog_2021} as detected by our pipeline and the 4-OGC catalog \cite{nitz_4ogc_o3_ab_catalog_2021}. The
inverse false alarm rate (IFAR) values in the GWTC-3 column are taken from the \href{https://gwosc.org/eventapi/json/GWTC-3-confident/}{$\tt{GWOSC}$} catalog, which
corresponds to whichever LVK pipeline achieved the highest astrophysical probability for that event in the GWTC-3
analysis. Note that we report our IFAR values after combining all our template banks (in contrast to some of the previous IAS searches where IFARs per bank were reported \cite{Ols22_ias_o3a, ias_o2_pipeline_new_events_prd2020, ias_pipeline_o1_catalog_new_search_prd2019}).
At the bottom of the table, we show the comparison with the 4-OGC events not in GWTC-3 catalog. We broadly recover the confident LVK detections, except some events which were vetoed (see section~\ref{sec:LVC} for a discussion).
}\\
    \hline
    \multirow{2}{*}{GWTC-3 events} & Bank & \multirow{2}{*}{$\rho_{\rm H}^2$} & \multirow{2}{*}{$\rho_{\rm L}^2$} & \multirow{2}{*}{${}^{\rm GWTC-3}$}  & \multicolumn{3}{c|}{$p_{\rm astro}$} & \multicolumn{3}{c|}{IFAR (yr)} \\
     \cline{6-11}
     & /comment & & & $\rho_{\rm Network}^2$ & \, IAS \, $|$ & $|$ GWTC-3 $|$ & $|$ 4-OGC & \,
     IAS \, $|$ & $|$ GWTC-3 $|$ & $|$ 4-OGC \\

     \hline
     \rowcolor{gray!30} GW190403\_051519 & \texttt{BBH\_6} & $28.1$ & $33.2$ & $57.8$ & $0.070$ & $0.61$ & $--$ & $0.004$ & $0.13$ & $--$ \\
GW190408\_181802 & \texttt{BBH\_2} & $106.6$ & $108.4$ & $213.2$ & $1.00$ & $1.0$ & $1.0$ & $> 1000$ & $> 1000$ & $> 1000$ \\
\rowcolor{gray!30} GW190412\_053044 & \texttt{BBH\_1} & $80.8$ & $267.8$ & $392.0$ & $1.00$ & $1.0$ & $1.0$ & $> 1000$ & $> 1000$ & $> 1000$ \\
GW190413\_052954 & \texttt{BBH\_5} & $29.8$ & $53.9$ & $81.0$ & $0.99$ & $0.93$ & $1.0$ & $12.6$ & $1.2$ & $1.4$ \\
\rowcolor{gray!30} GW190413\_134308 & \texttt{BBH\_7} & $24.1$ & $67.3$ & $112.4$ & $1.00$ & $0.99$ & $1.0$ & $64.5$ & $5.6$ & $6.4$ \\
GW190421\_213856 & \texttt{BBH\_5} & $72.4$ & $44.6$ & $114.5$ & $1.00$ & $1.00$ & $1.0$ & $> 1000$ & $357$ & $> 1000$ \\
\rowcolor{gray!30} GW190426\_190642 & Veto (\texttt{BBH\_12}) & 20.88 & 58.34 & $75.7$ & $ -- $ & $0.75$ & $--$ & $ -- $ & $0.24$ & $--$ \\
GW190503\_185404 & \texttt{BBH\_5} & $87.8$ & $52.2$ & $148.8$ & $1.00$ & $1.0$ & $1.0$ & $> 1000$ & $> 1000$ & $> 1000$ \\
\rowcolor{gray!30} GW190512\_180714 & \texttt{BBH\_1} & $42.4$ & $120.0$ & $161.3$ & $1.00$ & $1.0$ & $1.0$ & $> 1000$ & $> 1000$ & $> 1000$ \\
GW190513\_205428 & \texttt{BBH\_3} & $91.0$ & $63.2$ & $156.2$ & $1.00$ & $1.00$ & $1.0$ & $> 1000$ & $> 1000$ & $> 1000$ \\
\rowcolor{gray!30} GW190514\_065416 & \texttt{BBH\_6} & $41.8$ & $31.7$ & $64.0$ & $0.95$ & $0.76$ & $0.82$ & $2.1$ & $0.36$ & $0.19$ \\
GW190517\_055101 & \texttt{BBH\_3} & $54.4$ & $60.3$ & $116.6$ & $1.00$ & $1.0$ & $1.0$ & $> 1000$ & $> 1000$ & $66.1$ \\
\rowcolor{gray!30} GW190519\_153544 & \texttt{BBH\_7} & $91.0$ & $130.1$ & $252.8$ & $1.00$ & $1.0$ & $1.0$ & $> 1000$ & $> 1000$ & $> 1000$ \\
GW190521\_030229 & \texttt{BBH\_11} & $64.2$ & $115.1$ & $204.5$ & $1.00$ & $1.00$ & $1.0$ & $> 1000$ & $769$ & $805$ \\
\rowcolor{gray!30} GW190521\_074359 & \texttt{BBH\_5} & $152.8$ & $443.6$ & $670.8$ & $1.00$ & $1.0$ & $1.0$ & $> 1000$ & $> 1000$ & $> 1000$ \\
GW190527\_092055 & \texttt{BBH\_5} & $26.9$ & $51.1$ & $64.0$ & $0.95$ & $0.85$ & $0.94$ & $1.9$ & $4.3$ & $0.37$ \\
\rowcolor{gray!30} GW190602\_175927 & \texttt{BBH\_8} & $46.5$ & $112.8$ & $174.2$ & $1.00$ & $1.0$ & $1.0$ & $> 1000$ & $> 1000$ & $391$ \\
GW190701\_203306 & Veto (\texttt{BBH\_7}) & 33.32 & 53.2 & $125.4$ & $ -- $ & $1.00$ & $1.0$ & $ -- $ & $175$ & $0.13$ \\
\rowcolor{gray!30} GW190706\_222641 & \texttt{BBH\_8} & $91.5$ & $72.4$ & $179.6$ & $1.00$ & $1.0$ & $1.0$ & $> 1000$ & $> 1000$ & $> 1000$ \\
GW190707\_093326 & \texttt{BBH\_0} & $66.1$ & $111.9$ & $171.6$ & $1.00$ & $1.0$ & $1.0$ & $> 1000$ & $> 1000$ & $> 1000$ \\
\rowcolor{gray!30} GW190719\_215514 & \texttt{BBH\_5} & $33.4$ & $39.1$ & $62.4$ & $0.90$ & $0.92$ & $0.92$ & $0.77$ & $1.6$ & $0.25$ \\
GW190720\_000836 & \texttt{BBH\_0} & $48.1$ & $62.2$ & $118.8$ & $1.00$ & $1.0$ & $1.0$ & $> 1000$ & $> 1000$ & $559$ \\
\rowcolor{gray!30} GW190725\_174728 & \texttt{BBH\_0} & $28.6$ & $62.2$ & $82.8$ & $0.87$ & $0.96$ & $0.96$ & $0.56$ & $2.2$ & $0.41$ \\
GW190727\_060333 & \texttt{BBH\_5} & $74.8$ & $64.9$ & $136.9$ & $1.00$ & $1.0$ & $1.0$ & $> 1000$ & $> 1000$ & $> 1000$ \\
\rowcolor{gray!30} GW190728\_064510 & \texttt{BBH\_0} & $58.4$ & $117.5$ & $171.6$ & $1.00$ & $1.0$ & $1.0$ & $> 1000$ & $> 1000$ & $> 1000$ \\
GW190731\_140936 & \texttt{BBH\_6} & $29.5$ & $43.9$ & $77.4$ & $0.92$ & $0.83$ & $0.92$ & $1.1$ & $3.0$ & $0.43$ \\
\rowcolor{gray!30} GW190803\_022701 & \texttt{BBH\_5} & $33.2$ & $46.7$ & $86.5$ & $0.99$ & $0.97$ & $1.0$ & $12.9$ & $13.7$ & $2.4$ \\
GW190805\_211137 & \texttt{BBH\_6} & $23.6$ & $47.6$ & $65.6$ & $0.92$ & $0.95$ & $--$ & $1.1$ & $1.6$ & $--$ \\
\rowcolor{gray!30} GW190828\_063405 & \texttt{BBH\_3} & $119.1$ & $142.9$ & $272.2$ & $1.00$ & $1.0$ & $1.0$ & $> 1000$ & $> 1000$ & $> 1000$ \\
GW190828\_065509 & \texttt{BBH\_1} & $60.7$ & $58.8$ & $104.0$ & $1.00$ & $1.00$ & $1.0$ & $> 1000$ & $> 1000$ & $> 1000$ \\
\rowcolor{gray!30} GW190909\_114149\footnote{GW190909\_114149 is a marginal candidate in GWTC-3 which has $p_\mathrm{astro}>0.5$ from our pipeline.} & \texttt{BBH\_5} & $35.5$ & $35.2$ & $63.7$ & $0.72$ & $0.52$ & $--$ & $0.17$ & $0.45$ & $--$ \\
GW190915\_235702 & \texttt{BBH\_3} & $97.9$ & $74.9$ & $171.6$ & $1.00$ & $1.0$ & $1.0$ & $> 1000$ & $> 1000$ & $> 1000$ \\
\rowcolor{gray!30} GW190916\_200658 & \texttt{BBH\_6} & $35.7$ & $38.7$ & $65.6$ & $0.94$ & $0.66$ & $0.90$ & $1.8$ & $0.21$ & $0.22$ \\
GW190917\_114630 & \texttt{BBH\_0} & $26.6$ & $55.1$ & $68.9$ & $0.12$ & $0.77$ & $--$ & $0.007$ & $1.5$ & $--$ \\
\rowcolor{gray!30} GW190924\_021846 & \texttt{BBH\_0} & $37.6$ & $104.7$ & $144.0$ & $1.00$ & $1.0$ & $1.0$ & $290$ & $> 1000$ & $> 1000$ \\
GW190926\_050336 & \texttt{BBH\_6} & $46.8$ & $27.5$ & $65.6$ & $0.94$ & $0.54$ & $0.92$ & $1.7$ & $0.91$ & $0.27$ \\
\rowcolor{gray!30} GW190929\_012149 & \texttt{BBH\_8} & $43.6$ & $58.1$ & $94.1$ & $1.00$ & $0.87$ & $0.99$ & $194$ & $6.2$ & $3.1$ \\
GW190930\_133541 & \texttt{BBH\_0} & $44.7$ & $54.0$ & $94.1$ & $0.97$ & $1.00$ & $1.0$ & $3.4$ & $83.3$ & $295$ \\     
\hline
\rowcolor{gray!30} GW191103\_012549 & \texttt{BBH\_0} & $43.8$ & $50.9$ & $79.2$ & $0.98$ & $0.94$ & $--$ & $6.5$ & $2.2$ & $--$ \\
GW191105\_143521 & \texttt{BBH\_0} & $34.7$ & $63.1$ & $94.1$ & $0.99$ & $0.99$ & $1.0$ & $26.4$ & $83.3$ & $316$ \\
\rowcolor{gray!30} GW191109\_010717 & \texttt{BBH\_7} & $83.4$ & $166.0$ & $299.3$ & $1.00$ & $0.99$ & $1.0$ & $> 1000$ & $> 1000$ & $> 1000$ 
\label{tab:lvc_events}
\end{longtable*}
\end{table*}

\begin{table*}
\def\arraystretch{1}
\begin{longtable*}{|c|c|cc|c|ccc|ccc|}
 \hline
\multirow{2}{*}{GWTC-3 events} & Bank & \multirow{2}{*}{$\rho_{\rm H}^2$} & \multirow{2}{*}{$\rho_{\rm L}^2$} & \multirow{2}{*}{${}^{\rm GWTC-3}$}  & \multicolumn{3}{c|}{$p_{\rm astro}$} & \multicolumn{3}{c|}{IFAR (yr)} \\
     \cline{6-11}
     & /comment & & & $\rho_{\rm Network}^2$ & \, IAS \, $|$ & $|$ GWTC-3 $|$ & $|$ 4-OGC & \,
     IAS \, $|$ & $|$ GWTC-3 $|$ & $|$ 4-OGC \\
\hline
GW191113\_071753 & \texttt{BBH\_1} & $40.4$ & $32.1$ & $62.4$ & $0.002$ & $0.68$ & $--$ & $< 0.001$ & $0.038$ & $--$ \\
\rowcolor{gray!30} GW191126\_115259 & \texttt{BBH\_0} & $36.5$ & $52.5$ & $68.9$ & $0.81$ & $0.70$ & $1.0$ & $0.33$ & $0.31$ & $4.9$ \\
GW191127\_050227 & \texttt{BBH\_7} & $52.2$ & $39.3$ & $84.6$ & $0.99$ & $0.74$ & $0.99$ & $16.5$ & $4.0$ & $0.15$ \\
\rowcolor{gray!30} GW191129\_134029 & \texttt{BBH\_0} & $79.7$ & $102.0$ & $171.6$ & $1.00$ & $0.99$ & $1.0$ & $> 1000$ & $> 1000$ & $> 1000$ \\
GW191204\_110529 & \texttt{BBH\_2} & $36.6$ & $71.0$ & $79.2$ & $0.94$ & $0.74$ & $0.99$ & $2.2$ & $0.30$ & $1.6$ \\
\rowcolor{gray!30} GW191204\_171526 & \texttt{BBH\_0} & $97.7$ & $190.8$ & $302.8$ & $1.00$ & $0.99$ & $1.0$ & $> 1000$ & $> 1000$ & $> 1000$ \\
GW191215\_223052 & \texttt{BBH\_2} & $53.7$ & $69.4$ & $125.4$ & $1.00$ & $0.99$ & $1.0$ & $> 1000$ & $> 1000$ & $869$ \\
\rowcolor{gray!30} GW191222\_033537 & \texttt{BBH\_6} & $83.7$ & $68.5$ & $156.2$ & $1.00$ & $0.99$ & $1.0$ & $> 1000$ & $> 1000$ & $> 1000$ \\
GW191230\_180458 & \texttt{BBH\_7} & $61.5$ & $56.4$ & $108.2$ & $1.00$ & $0.96$ & $1.0$ & $> 1000$ & $20.0$ & $497$ \\
\rowcolor{gray!30} GW200128\_022011 & \texttt{BBH\_6} & $56.1$ & $55.1$ & $112.4$ & $1.00$ & $0.99$ & $1.0$ & $> 1000$ & $233$ & $307$ \\
GW200129\_065458 & \texttt{BBH\_3} & $213.6$ & $347.5$ & $718.2$ & $1.00$ & $0.99$ & $1.0$ & $> 1000$ & $> 1000$ & $> 1000$ \\
\rowcolor{gray!30} GW200202\_154313 & \texttt{BBH\_1} & $28.3$ & $100.7$ & $116.6$ & $0.17$ & $0.99$ & $1.0$ & $0.010$ & $> 1000$ & $6.1$ \\
GW200208\_130117 & \texttt{BBH\_5} & $40.3$ & $61.3$ & $116.6$ & $1.00$ & $0.99$ & $1.0$ & $528$ & $> 1000$ & $917$ \\
\rowcolor{gray!30} GW200208\_222617 & \texttt{BBH\_6} & $35.2$ & $36.0$ & $54.8$ & $0.059$ & $0.70$ & $--$ & $0.003$ & $0.21$ & $--$ \\
GW200209\_085452 & \texttt{BBH\_5} & $59.9$ & $36.8$ & $92.2$ & $0.99$ & $0.97$ & $0.99$ & $16.5$ & $21.7$ & $1.1$ \\
\rowcolor{gray!30} GW200210\_092254 & Veto (\texttt{BBH\_1}) & 29.3 & 45.4 & $70.6$ & $ -- $ & $0.54$ & $--$ & $ -- $ & $0.83$ & $--$ \\
GW200216\_220804 & \texttt{BBH\_7} & $47.7$ & $42.3$ & $65.6$ & $0.95$ & $0.77$ & $0.78$ & $2.8$ & $2.9$ & $0.093$ \\
\rowcolor{gray!30} GW200219\_094415 & \texttt{BBH\_5} & $39.2$ & $77.0$ & $114.5$ & $1.00$ & $0.99$ & $1.0$ & $528$ & $> 1000$ & $22.9$ \\
GW200220\_061928 & \texttt{BBH\_11} & $22.6$ & $47.4$ & $51.8$ & $0.64$ & $0.62$ & $--$ & $0.11$ & $0.15$ & $--$ \\
\rowcolor{gray!30} GW200220\_124850 & \texttt{BBH\_6} & $38.4$ & $32.2$ & $72.2$ & $0.91$ & $0.83$ & $--$ & $0.97$ & $0.033$ & $--$ \\
GW200224\_222234 & \texttt{BBH\_5} & $162.3$ & $163.8$ & $400.0$ & $1.00$ & $0.99$ & $1.0$ & $> 1000$ & $> 1000$ & $> 1000$ \\
\rowcolor{gray!30} GW200225\_060421 & \texttt{BBH\_1} & $102.9$ & $59.1$ & $156.2$ & $1.00$ & $0.99$ & $1.0$ & $> 1000$ & $> 1000$ & $> 1000$ \\
GW200306\_093714 & \texttt{BBH\_2} & $39.4$ & $37.4$ & $60.8$ & $0.74$ & $0.81$ & $0.51$ & $0.19$ & $0.042$ & $0.018$ \\
\rowcolor{gray!30} GW200308\_173609 & Below $\rho^2_\mathrm{collect}$ & $ -- $ & $ -- $ & $22.1$ & $ -- $ & $0.86$ & $--$ & $ -- $ & $0.42$ & $--$ \\
GW200311\_115853 & \texttt{BBH\_3} & $152.3$ & $114.0$ & $316.8$ & $1.00$ & $0.99$ & $1.0$ & $> 1000$ & $> 1000$ & $817$ \\
\rowcolor{gray!30} GW200316\_215756 & \texttt{BBH\_1} & $36.0$ & $74.4$ & $106.1$ & $1.00$ & $0.99$ & $1.0$ & $132$ & $> 1000$ & $22.4$ \\
GW200322\_091133 & \texttt{BBH\_1} & $46.3$ & $30.7$ & $20.2$ & $0.041$ & $0.62$ & $--$ & $0.002$ & $0.007$ & $--$ \\
\hline \\[1pt]
\hline
    \multirow{2}{*}{4-OGC events} & Bank & \multirow{2}{*}{$\rho_{\rm H}^2$} & \multirow{2}{*}{$\rho_{\rm L}^2$} & \multirow{2}{*}{${}^{\rm 4-OGC}$}  & \multicolumn{3}{c|}{$p_{\rm astro}$} & \multicolumn{3}{c|}{IFAR (yr)} \\
     \cline{6-11}
     & /comment & & & $\rho_{\rm Network}^2$ & \, IAS \, $|$ & $|$ GWTC-3 $|$ & $|$ 4-OGC & \,
     IAS \, $|$ & $|$ GWTC-3 $|$ & $|$ 4-OGC\\
\hline
\rowcolor{gray!30} GW191224\_043228 & \texttt{BBH\_0} & $36.6$ & $50.0$ & 73.0 & $0.82$ & $--$ & $0.87$ & $0.37$ & $--$ & $0.13$ \\
GW200106\_134123 & \texttt{BBH\_6} & $37.9$ & $37.2$ & 55.0 & $0.78$ & $--$ & $0.69$ & $0.25$ & $--$ & $0.059$ \\
\rowcolor{gray!30} GW200129\_114245 & \texttt{BBH\_7} & $25.6$ & $45.7$ & 63.0 & $0.95$ & $--$ & $0.53$ & $2.3$ & $--$ & $0.037$ \\
GW200210\_005122 & \texttt{BBH\_1} & 32.4 & 42.1 & 70.0& $<0.001$ & $--$ & $0.74$ & $<0.001$ & $--$ & $0.042$ \\
\rowcolor{gray!30} GW200214\_223306 & \texttt{BBH\_7} & $33.6$ & $30.9$ & 55.0 & $0.79$ & $--$ & $0.72$ & $0.28$ & $--$ & $0.079$ \\
GW200305\_084739 & Below $\rho^2_\mathrm{collect}$ & $--$ & $--$ & 59.0& $--$ & $--$ & $0.59$ & $--$ & $--$ & $0.019$ \\
\rowcolor{gray!30} GW200318\_191337 & \texttt{BBH\_7} & $28.6$ & $43.8$ & 63.0 & $0.96$ & $--$ & $0.97$ & $3.9$ & $--$ & $0.50$ \\
\hline
\end{longtable*}
\end{table*}
\egroup

\subsubsection{Parameter estimation runs}
\label{sec:PE}
We perform parameter estimation (PE) runs using the $\tt{cogwheel}$ package\footnote{\url{https://github.com/jroulet/cogwheel}} \cite{Rou22_cogwheel} with the $\tt{NAUTILUS}$ sampler \cite{Lange2023}. We use priors that are uniform in detector-frame constituent masses and comoving volume-time ($VT$) throughout this paper. 
Our fiducial prior for the BH spins is flat in effective spin $\chieff$, but we also compare results with those obtained using an isotropic prior on individual spins (which is used for PE in the GWTC-3 and 4-OGC catalogs). 
Other extrinsic parameters have the standard geometric priors used in GWTC-3 \cite{lvc_gwtc3_o3_ab_catalog_2021}. 
We compute likelihoods using the relative binning/heterodyning method described in Ref.~\cite{relative_binning, Rou23_CoherentScore} as implemented in \texttt{cogwheel}.
The relation between redshift and luminosity distance is based on the Planck15 cosmology \cite{cosmology_planck2015}.
Our fiducial parameter estimation runs are based on the \texttt{IMRPhenomXODE} waveform approximant \cite{Yu23_XODE} and include the effect of both HM and precession. 
Table~\ref{tab:NewEvents} reports marginalized 1D posterior values for a relevant subset of parameters.

\begin{figure}
    \includegraphics[width=\linewidth]{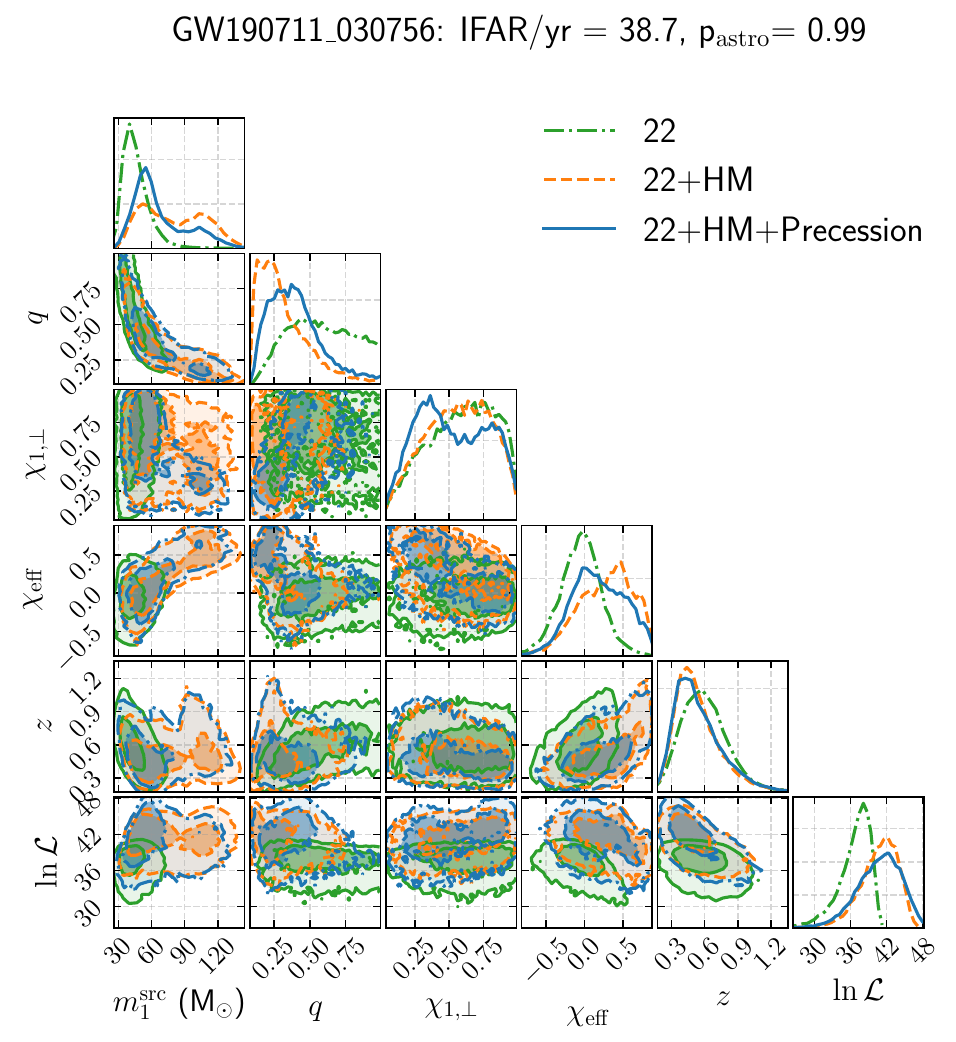}
    \caption{GW190711\_030756 is among the most significant candidates recovered in our search. 
    It was first reported in our previous IAS $(2, 2)$-only search \citetalias{Ols22_ias_o3a}, however, its IFAR in our HM search is comparatively much larger (i.e., 11\,yr instead of $\lesssim 2$\,yr). 
    Our search also recovers an additional 10 points in the total SNR$^2$ compared to the $(2, 2)$-only search. 
    Indeed, we also see a significant evidence for higher modes in the PE results. 
    The equal mass ratio ($q$) solution is ruled out at $>95\%$ CL. Its total mass ($=90_{-30}^{+50}\, \msun$) implies that the remnant lies within the upper mass gap. Note that subsequent to our work, this event was independently confirmed by other pipelines \cite{Kum24, Kol24_Ares_ML_Search, Mis24_CWB}.}
    \label{fig:GW190711}
    \end{figure}

\begin{figure}
\includegraphics[width=\linewidth]{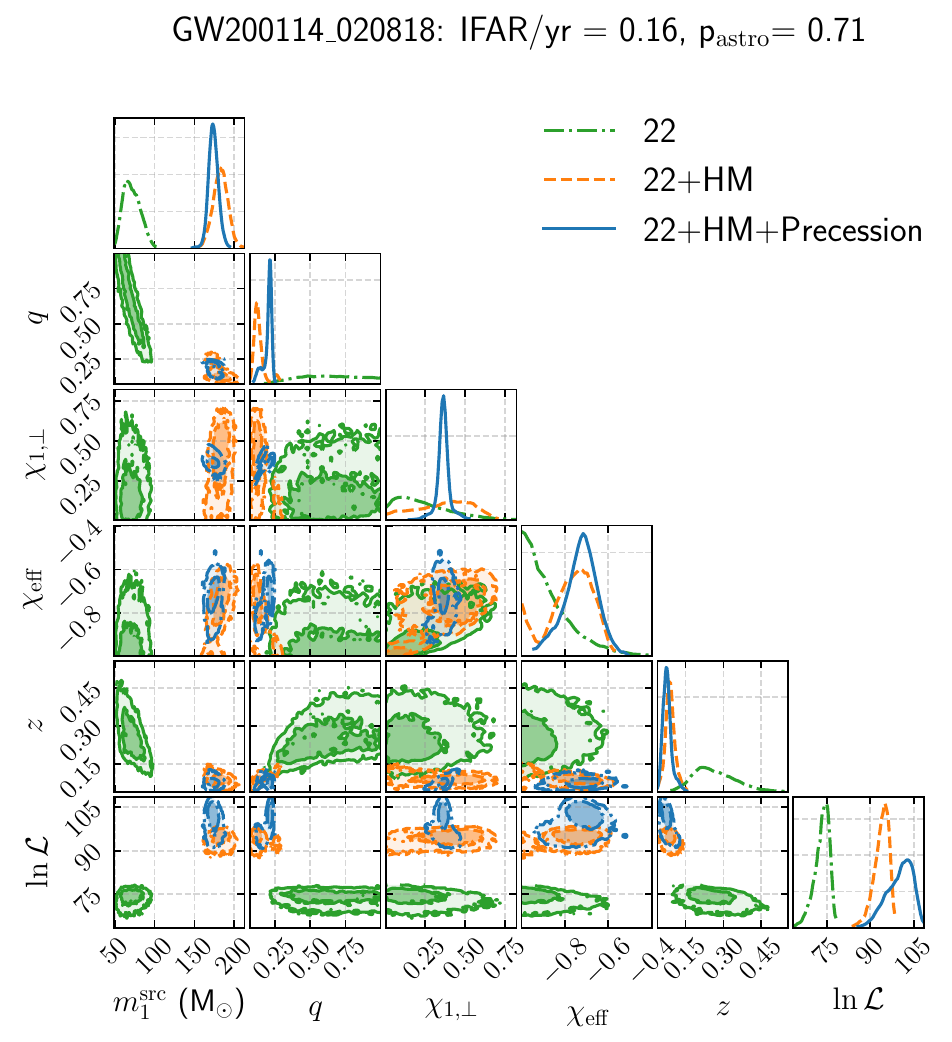}
\includegraphics[width=\linewidth]{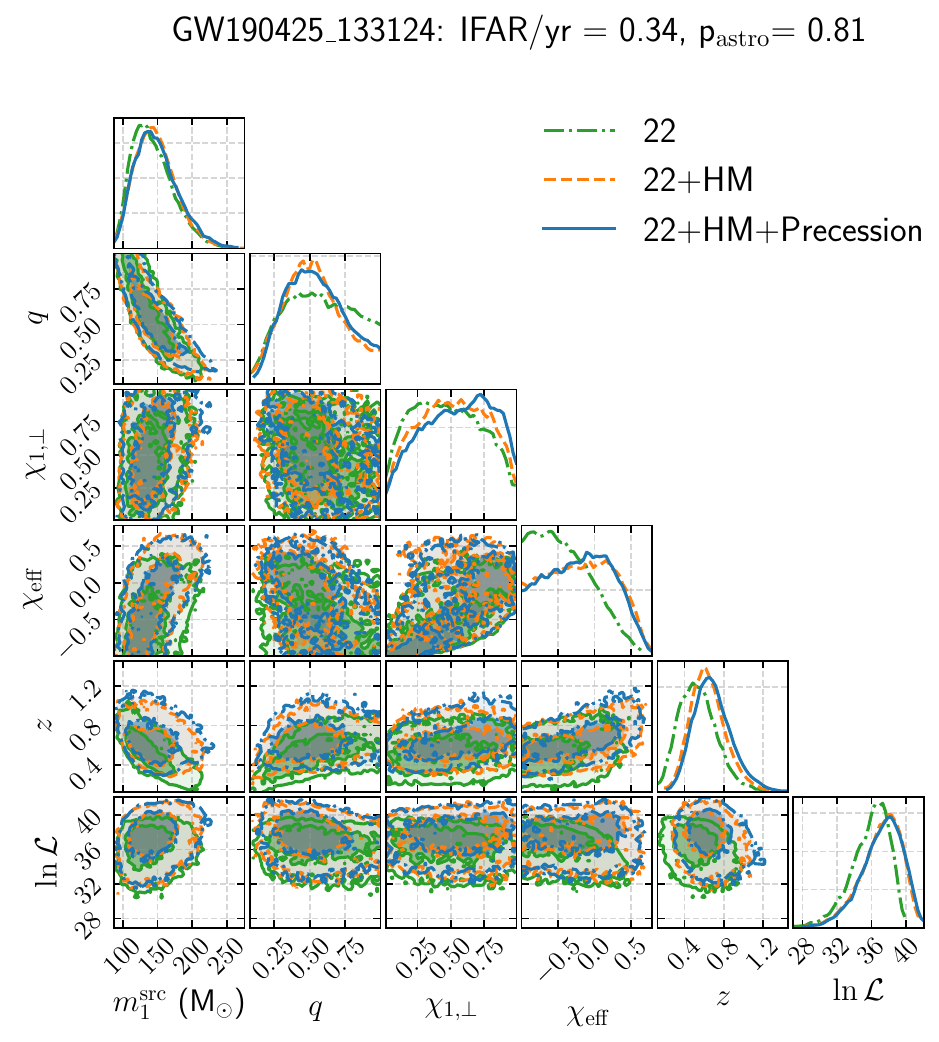}
\caption{Posteriors of two of the most interesting new candidate events from our search. We perform the parameter estimation runs in the three labelled cases and find significant support for HM in both the cases (which could be one of the reasons our HM search recovered these candidate). The primary masses of both candidates are confidently in the IMBH regime. Additionally, GW200114\_020818 has asymmetric mass ratio and has strongly anti-aligned primary spin.}
\label{fig:GW200114}
\end{figure}

Other than the fiducial runs, we also perform two separate PE runs without precession, i.e., restricting to binaries with aligned spins: one run with a $(2, 2)$-only waveform approximant and the other with an approximant that includes HM as well. 
This helps us calculate the relative evidence (Bayes factor) due to including the effects of HM and HM + precession separately for the new candidate events; we show the logarithm of the Bayes factors in Table~\ref{tab:NewEvents}.
Two of the new candidate events have Bayes factors larger than unity for HM, and the detection of these candidates likely benefited the most from our search containing HM. 

We show a few select 2D posteriors of our new candidate events in colored contours in Figs.~\ref{fig:m1_z}, \ref{fig:mtot_chieff} and \ref{fig:chieff_extra}.
We also overplot the contours from events in GWTC-3, 4-OGC and the IAS $(2, 2)$-only O3 catalogs in light gray. 
Our search only covers events with Hanford--Livingston coincidence triggers, but we include all binary BH events that are part of the LVK catalogs through O1--O3 in these figures. 

We add dashed lines to mark the approximate limits of sensitivity in the top panel of Fig.~\ref{fig:m1_z}.
The dashed line in the top-right of the panel marks the region where the binary starts moving out of band (we roughly take this to happen when the detector frame mass equals $400\,\msun$, which is the limit of our template bank). 
The dashed line in the top-left marks the contour of constant detector-frame luminosity for an inspiral-dominated equal-mass binary; it roughly delineates the parameter limits past which binaries are too faint to be observable. 
It is also worth mentioning that the size of the contours increases for higher masses as the number of cycles in band starts to decrease. 
In the inspiral phase, the chirp mass is well-measured, but this no longer continues to be the case for high-mass events where the SNR is dominated by the merger and ringdown phase and the total mass of the binary starts to be measured better.

We show the marginalized 1D posteriors for our new candidate events under the aforementioned two different spin priors in Fig.~\ref{fig:PE_1D_spin}.
Figs.~\ref{fig:PE_Events_1} and~\ref{fig:PE_Events_2} in the Appendix show full corner plots from different PE runs. 
We additionally highlight the posteriors for two most interesting new candidate events (GW200114\_020818, GW190425\_133124) in Fig.~\ref{fig:GW200114}.
GW200114\_020818 has the largest evidence for HM ($\sim 10^5$). We show three PE runs with and without HM and precessing effects.
Note that for the non-precessing (green and orange) cases, the distribution of the in-plane spin component $\chi_{\perp}=\sqrt{\chi^2_{x}+\chi^2_{y}}$ is same as that of the prior conditioned on the other parameters (we set contribution of in-plane spin components to be zero inside the likelihood for these cases).
Note also that the total SNR$^2$ ($\sim$180) recovered by our non-precessing Hanford+Livingston search is significantly smaller than $\rho^2_\mathrm{network}\sim 220$ obtained from a PE run which includes both the effects of orbital precession and data from the Virgo detector. Similarly, SNR$^2\sim$77 and $\rho^2_\mathrm{network}\sim 87$ for GW190425\_133124. 
Therefore, we predict that the significance of these event could be improved in future searches that include the effects of precession and/or Virgo data.

We also compare our results against PE runs using the \texttt{IMRPhenomXPHM} approximant \cite{Pra21_XPHM} (using the default MSA \cite{Cha17_MSA} prescription for evolving the spins) in Fig.~\ref{fig:XODE_XPHM} in the Appendix; we find good agreement in most of the cases. It would be interesting to compare the PE for these events with NR surrogate models (e.g.~\cite{Isl23_NRsurrogate}) in the future. We will discuss interesting astrophysical properties of our new candidate events in Section~\ref{sec:astro_properties} below.

\subsection{Comparison to the IAS (2, 2)-only catalogs}
In Table~\ref{tab:22_IAS_events_significant}, we report the candidate events with $p_\mathrm{astro}>0.5$ from our pipeline which have already been reported in the previous $(2, 2)$-only IAS searches in the O3 data \cite{Ols22_ias_o3a, Meh23_ias_o3b}. The full comparison catalog including lower $\pastro$ candidates is given in Table~\ref{tab:22_IAS_AllEvents}. Overall, we recover 5 events with $\pastro \geq 0.5$.

We caution against directly comparing the sensitivities of the $(2, 2)$-only and $(2, 2)$ + HM pipelines from Table~\ref{tab:22_IAS_AllEvents} as there are multiple differences: ($i$) including HM in the search, ($ii$) using a different astrophysical prior (see Section~\ref{sec:astrophysical_prior}) ($iii$) additional glitch mitigation at high masses (see Section~\ref{sec:Downweight_glitches}).
It is worth mentioning though that we recover the events which were reported in previous IAS catalogs with  $\pastro\gtrsim0.6$, except GW190704\_104834 (this event was in BBH-0 in \citetalias{Ols22_ias_o3a} and we checked that the difference in the astrophysical priors between the pipelines was large for that bank). We collect a larger number of triggers in the HM search as compared to a $(2, 2)$-only search, so we raised the collection threshold to mitigate the increased memory requirements (see Section~\ref{sec:single_det_statistic}). 
This caused us to miss low-SNR events that were reported by $(2, 2)$-only pipelines with lower collection thresholds in the previous literature. 

 
One notable event in Table~\ref{tab:22_IAS_events_significant} is GW190711\_030756. The IFAR of this event in our HM search is $\sim$11\,yr as opposed to $\sim 2$\,yr in the $(2, 2)$-only search \citetalias{Ols22_ias_o3a} (after accounting for the trials factor in \citetalias{Ols22_ias_o3a} due to the number of banks searched). We show a corner plot for this event in Fig.~\ref{fig:GW190711}. Indeed, GW190711\_030756 also shows a significant evidence for higher modes and our search recovers an additional 10 points in SNR$^2$ compared to the $(2, 2)$-only search. See Fig.~3 and section III A.2 of~\citetalias{Ols22_ias_o3a} for a detailed discussion of the properties of this event. 
  
\subsection{Comparison to the LVK and OGC $(2, 2)$-only catalogs}
\label{sec:LVC}
 
In Table~\ref{tab:lvc_events}, we compare our pipeline's results for the O3 Hanford--Livingston coincident events published in the GWTC-3 catalog\footnote{\url{https://gwosc.org/GWTC-3/}} \cite{lvc_gwtc3_o3_ab_catalog_2021}.
The GWTC-3 catalog combines results from four pipelines: three that search using templates which only contain the $(2, 2)$ mode: GstLAL, Multi-Band Template Analysis (MBTA) and PyCBC \cite{gstlal, PYCBCPipeline, mbta_o3a_pastro_andres2022}; and one that searches for transient signals with minimal assumptions about sources, cWB \cite{Klimenko:2016}.
We also include the comparison with the 4-OGC catalog taken from the summary data in the repository\footnote{\url{https://github.com/gwastro/4-ogc/blob/master/search/4OGC_top.txt}} listed in \cite{nitz_4ogc_o3_ab_catalog_2021}.
Note that the quoted GWTC-3 squared SNR, $\rho_{\rm Network}^2$, is from LVK's PE runs for the events. 
The search SNR$^2$ is generally lower, and varies between the different pipelines that the LVK uses.

We broadly recover the confident LVK detections, except some vetoed/missed events which are discussed in more detail in the subsections below.
Overall, three GWTC-3 events in Table~\ref{tab:NewEvents} were vetoed in our pipeline.  GW190701\_203306 was vetoed due to excess power present after subtracting the best-fit waveform in Livingston (note that the data around this event was treated for scattered light glitches \cite{lvc_o3a_gwtc2_catalog_2021}). 
GW200308\_173609 has GWTC-3 $\rho^2_\mathrm{Network}=22.1$ and it fell below our SNR collection threshold. Six GWTC-3 events were detected with $p_\mathrm{astro}<0.5$ in our pipeline (GW190403\_051519, GW190917\_114630, GW191113\_071753, GW200202\_154313, GW200208\_222617, GW200322\_091133). 
Note that all the GWTC-3 events we missed (except GW200202\_154313) were confidently detected by only one of the LVK pipelines. We find GW200202\_154313 gets penalized in the multi-detector coherence module of our pipeline possibly due to having a large SNR difference in the two detectors.


At the end of Table~\ref{tab:lvc_events}, we include comparison with the list of 7 events in 4-OGC which are not present in GWTC-3 catalog (from which we recover 5 with $p_\mathrm{astro}>0.5$). GW190909\_114149 is a marginal candidate in GWTC-3 which has $p_\mathrm{astro}>0.5$ from our pipeline. We still include it in Table~\ref{tab:lvc_events} as it was first a significant candidate in GWTC-2 \cite{lvc_o3a_gwtc2_catalog_2021}, but later updated to be in their marginal list \cite{lvc_o3a_deep_gwtc2_1_update_2021}. 
 We also checked the overlap of our new candidate event list with the sub-threshold candidate list (i.e., which have $\pastro<0.5$ but FAR $<2$ day$^{-1}$) released by the LVK collaboration: GWTC-2.1 and GWTC-3 \cite{lvc_o3a_deep_gwtc2_1_update_2021, lvc_gwtc3_o3_ab_catalog_2021}; we show the corresponding comparison in Table~\ref{tab:subthreshold_LVC} in the Appendix. In the same table, we also show the overlap of our new candidate list with sub-threshold candidates from the IAS $(2, 2)$-only pipeline \cite{Ols22_ias_o3a, Meh23_ias_o3b}.



As we discussed earlier in Section~\ref{sec:HM_statistics}, we set a particular SNR threshold ($\rho_\mathrm{collect}$) to collect triggers from individual detectors. 
We do this in order to regulate the number of single-detector triggers that are passed to the more costly coincidence step of the pipeline (see section~\ref{sec:single_det_statistic} for a detailed discussion). 
Due to this, we missed low-SNR events that were reported by $(2, 2)$-only pipelines in the previous literature. 
We label the events which fell below this threshold as ``Below $\rho^2_\mathrm{collect}$" in the table.
 
We do not include the following events in Table~\ref{tab:lvc_events} because they were detected in Livingston--Virgo or Hanford--Virgo coincidence, or single detector search, all of which we did not run: 
GW190620\_030421, 
GW190630\_185205, 
GW190708\_232457,
GW190814,
GW190910\_112807, and 
GW190925\_232845. 
We also only searched for BBH mergers in this paper, and did not recover the NSBH candidates GW191219\_163120 and GW200115\_042309 (in addition, the posterior probability for GW200210\_092254 peaks in the NSBH region \cite{lvc_gwtc3_o3_ab_catalog_2021}, which could lead to it being vetoed in our search). 
We hope to perform coincident searches which would also include Virgo (and KAGRA) in the future. 
We also intend to perform a separate search for NSBH mergers including our HM pipeline in a future paper. 


\section{Interesting astrophysical properties of the new candidate events}
\label{sec:astro_properties}

In this section, we discuss interesting astrophysical properties of the \neviasnew\, new candidate events, whose parameters we have reported in Table~\ref{tab:NewEvents} and in Figs.~\ref{fig:m1_z},~\ref{fig:mtot_chieff} and~\ref{fig:PE_1D_spin}.


\subsection{Masses in the IMBH range and pair instability mass gap}
BHs roughly within the mass range $10^2$--$10^5\, \msun$ are designated as intermediate-mass black holes (IMBHs). There are a number of channels which can produce BHs in this mass regime, see e.g., \cite{imbh_from_globular_clusters2002, Fra24_UMG}.
Several IMBH candidates have been suggested by electromagnetic observations, but these lack conclusive confirmation \cite{Gre20}. 
Gravitational waves have the potential to definitively detect objects in this regime, e.g., \cite{GW190521, Fra18_IMBH_LISA}. 
In our search, at $>95$\% credibility, 4 candidate events have primary black hole masses in the IMBH range (\# 2, 4, 5, 9 in Table~\ref{tab:NewEvents}).

Another interesting regime in BH masses is the pair instability mass gap. This is the regime in which stars undergo pair-instability supernova (where they are completely disrupted and no remnant is left behind). Otherwise, the stars undergo pulsational pair-instability supernova
and iteratively lose their mass before collapsing to a black hole. These processes are expected to produce a gap roughly between $65$--$135\, \msun$ in the distribution of BH mass. Many of our new candidate events have support for masses in the gap. Note however that the exact boundaries of such a gap are not well determined, as they depend on multiple uncertain factors such as the rate of carbon to oxygen burning in the core ${}^{12}\rm{C}(\alpha, \gamma){}^{16} O$, angular momentum transport, metallicity, wind loss, etc. \cite{Marchant:2020haw, Farmer:2019jed, VanSon20_UMG_Pollution, Mehta:2021fgz, Hen23_UMG, Gol23_UMG, Fra24_UMG}. 

Recent analyses have not found any conclusive evidence for the upper mass gap in the observed BH mass distribution, e.g., \cite{LVKO3bpopulation, ias_o3a_population_analysis_prd2021roulet}. This either challenges the current theoretical models of stellar evolution, or the gap might be filled by black holes formed via alternative mechanisms such as hierarchical mergers \cite{hierarchical_7merger_scenario2020b, hierarchical_from_dynamical_in_any_star_cluster2020b, hierarchical_mergerFamily_dynamical_mass_dist_matters2021, hierarchical_rate_sensitive_to_natal_spins_Fragione2021kocsis}, stellar collisions \cite{UMG_Kre20_StellarCollisions}, or from the gaseous disk channel \cite{hierarchical_mergers_agn_kocsis2019,agn_bbh_population_chieff_q_simulation_mckernan_ford2019, bbh_spin_evolution_agn_Tagawa_2020a, bbh_evolution_agn_merger_timescale_ishibashi2020a, migration_traps_spins_rates_mckernan_ford2020a, 
mass_gap_agn_bbh_mergers2021, agn_accretion_disk_merger_population2020a, Tag20_AGN, Sam22_AGN}. 
 Our new candidate events could be helpful in estimating the abundance of BH (or rate of BH mergers) in/beyond the pair-instability mass regime. This can also shed light on BH progenitor channels. Other potential explanation for high-mass merger detections could be that they are lower-mass mergers but are lensed by a foreground galaxy \cite{Dai16_Lensing}.

\subsection{High redshifts}
We find two new candidates with redshift $z>0.5$ at $>95$\% CL (\# 4, 6 in Table~\ref{tab:NewEvents}). 5 of our new candidates have median redshift $z \geq 0.8$. Finding high-redshift BHs can be important to answer questions such as: How does the merger rate of BHs evolve with redshift \cite{Fis18_merger_redshift}? Does the merger rate change in a different manner compared to the the star formation rate (which peaks at $z\sim 2$ \cite{Mad14_SFR})? Furthermore, different BH formation channels have different predictions for the merger rate at high redshift \cite{Map22_MergerRate, Van22_MergerRate, Boe24_DelayTime_COMPASS, Boe24_RedshiftEvolution_COMPASS}. For example, the chemically homogeneous evolution channel or the stable mass transfer channel produces BBHs of relatively higher masses and lower redshifts than the common envelope evolution channel \cite{Man16_CHE, Van22_MergerRate}. In the dynamical channel, massive BHs can have relatively longer delay times compared to star formation \cite{Map19_MergerRate}. The redshifts of binaries can be also be correlated with other properties of BHs, e.g., spins \cite{Bis22, Kus16_MergerTime_150914, chieff_isolated_field_binary_matias2018}. We aim to test some of these models in a future population inference study.

\subsection{Preference for non-zero spins}

Measuring spins of the black holes in a binary is one of the best ways to distinguish between isolated and dynamical formation channels.
In the isolated channel, accretion can lead to spins of BHs which are aligned in the direction of the binary angular momentum (leading to positive values of the effective spin parameter $\chi_\mathrm{eff}$). Dynamical channels, on the other hand, predict the astrophysical $\chi_\mathrm{eff}$ distribution to be symmetric around 0. This is because the spins BH captured into a binary are not expected to be preferentially aligned along any particular direction.
From the PE runs under our fiducial prior (flat in $\chi_\mathrm{eff}$), two new candidates have positive effective spin $(\chi_\mathrm{eff} > 0)$ at $>95\%$ CL (GW191213\_204950, GW190530\_030659). On the other hand, GW200114\_020818 has $\chi_\mathrm{eff} < -0.6$ at $>95\%$ CL. 
Even under the isotropic spin prior, some of the new candidate events still have mild preference for $\chi_\mathrm{eff}$ being non-zero.

\subsection{Asymmetric mass ratio systems}

5 candidates have median mass ratio $q \leq 0.5$. The $q$ posteriors are broad in most cases, but there are two candidates with a strong preference for asymmetric mass-ratios (GW200114\_020818 and GW200210\_035448). 
Finding sources with asymmetric mass ratios can shed light on the formation channels \cite{GW190814}. 
The globular cluster (dynamical) channel predicts that most merging BBHs have nearly equal masses \cite{Rod16_Dynamical}. 
This is because higher mass BHs in a globular cluster sink to their cores due to dynamical friction and vice versa for lighter BHs. 
Although there is a caveat here that if clusters can retain merger products, they can produce ``second-generation" mergers whose mass ratios can deviate significantly from unity \cite{Rodriquez_2019_2g}. 
In the isolated binary channel, in certain cases, an initially asymmetric mass ratio system can evolve toward a more symmetric configuration due to mass accretion episodes \cite{Iva13_common_envelope, Pos14_mass_transfer}; note that the reverse scenario has been proposed as well (see e.g.,~\cite{Ger13_MassRatioReversal, Broekgaarden_2022_reversal}). 
Another interesting property of asymmetric mass ratio and high total mass systems is that subdominant ringdown
modes have a higher chance of being observable in such systems, see e.g. \cite{hm_observation_qn_ringdown_capano2021}.

\section{Discussion}
\label{sec:Discussion}

\subsection{Injection campaigns to quantify gain in sensitivity due to including HM}
\label{sec:HM_sensitivity}

We quantify the improvement in the sensitive volume-time ($VT$) of the search pipeline in our companion paper: \citet{Meh25_VT_HM}. We use the injection catalog released by the LVK collaboration on Zenodo~\cite{zenodoLVK}. We also compare the results of our \texttt{IAS-HM} pipeline with its (2,2)-only limit. We find significant improvement in $VT$ for systems with asymmetric mass ratios and larger total masses, with gains in sensitivity even exceeding 100\% at certain high masses (see Fig.~2 of \cite{Meh25_VT_HM}). We also do not find a loss of sensitivity in the \texttt{IAS-HM} pipeline for near equal mass-ratio mergers. We also find that the $VT$ of the IAS pipeline
is either larger than or comparable to that of the individual LVK pipelines (see Fig.~1 of \cite{Meh25_VT_HM}).

It is quite encouraging to see that the parameter space where we get a large $VT$ improvement ($m_1\gtrsim 100 M_\odot$) over the LVK and the (2,2)-only IAS pipelines is similar to the parameter space where we detect most of the new candidate events, thus adding more credibility to our newly reported detections. We refer the reader to \cite{Meh25_VT_HM} for further details on the injection campaigns.

Previous studies have noted that the larger variety in waveform morphologies (brought by the inclusion of higher modes) can lead to a loss of sensitivity in the search, as the number of background triggers is increased, see e.g.,~\cite{Cap14, Cha22}.
This can be understood from the fact that, with HMs, the shape of the waveform also depends on the inclination and orbital phase, and these additional degrees of freedom increase the trials factor when finding the best-matching template. As discussed in section~\ref{sec:HM_statistics} and \citetalias{Wad23_Pipeline}, we do not naively add additional degrees of freedom to the templates throughout the parameter space.
Instead, our new detection statistic (see Section~\ref{sec:HM_statistics} and~\citetalias{Wad23_Pipeline}) includes HM while simultaneously marginalizing the likelihood over the extrinsic parameters and mass ratio.
This marginalization naturally downweights fine-tuned solutions, approaching the optimal statistic and preventing an undue increase of the trials factor (e.g., the edge-on solutions get naturally downweighted when we marginalize over distance, as their observable volume is smaller).

The downside of using a prior distribution on $|\rho_{ii}/\rho_{22}|$ is that it is partly dependent on our choice of astrophysical prior, which is currently not precisely known. Especially, the mass ratio distribution is not very well constrained and affects our prior for $|\rho_{ii}/\rho_{22}|$. Using a different distribution for $q$ than the true astrophysical distribution can indeed lead to some loss of sensitivity. 
We chose a broad prior (flat in $q$) in our search, with the motivation that such a prior can be easily reweighted later in a hierarchical population analysis.

Even though including HM can cause some loss in sensitivity for near-threshold face-on equal mass systems, note that detecting even a few events with asymmetric-mass ratios can inform our population estimates much more than detecting a slightly larger number of equal-mass systems.
It is worth drawing an analogy here to the case of spin distribution of templates used in $(2, 2)$-only searches.
Neglecting templates with negative $\chieff$ indeed improves the search sensitivity to near-threshold $\chieff \gtrsim 0$ systems, but finding even a few $\chieff<0$ systems can greatly help us distinguish between progenitor channels of BHs. We thus argue the goal of a search should not solely be to maximize the total number of detections, but to holistically cover most of the interesting regions of the binary parameter space.
A similar argument also holds for when including spin-precession or eccentricity effects in the search templates. Detecting even a few events with significantly misaligned spins can shed more light on the formation channels of binaries as compared to detecting a larger number of near-threshold aligned/zero spin mergers.
Some of the challenges for including these effects are similar to the ones for HM: significant increase in the template bank size and loss of sensitivity due to collecting additional background \cite{McI23, Sch24_Precession_Pipeline, Sch24_Precession_O3_Search, Dhu22, Dhu23_eccentricity_search, Nit20_eccentricity_search}. We will develop methods similar to the ones implemented in our search to help tackle those searches in future work \cite{Zho25_Prec_TemplateBank,Isl25_Eccentricity_TemplateBank}.

\subsection{Dependence on the choice of astrophysical prior}
\label{sec:astro_prior_dependence}
 Our estimate of $p_\mathrm{astro}$ and our overall IFAR depend on the astrophysical prior used in the search, and that dependence is strongest in the case of near-threshold detections like some of the new candidates in our catalog list in Table~\ref{tab:NewEvents}.

To facilitate further analysis and comparisons between different astrophysical priors, we make our posterior samples, the selection function of our pipeline and other relevant information public at \url{https://github.com/JayWadekar/GW_higher_harmonics_search}. These can be used to reweight our results with other choices of priors (see e.g., \cite{reweighting_importance_sampling_thrane2019}). The biggest difference in our astrophysical prior choice compared to other catalogs \cite{lvc_gwtc3_o3_ab_catalog_2021, nitz_4ogc_o3_ab_catalog_2021} is for the BH spins. We use a prior that is flat in the effective spin parameter $\chieff$. Oue motivation lies in the fact that $\chieff$ is often the best-measured spin parameter, and hence we want to give equal weights to all physically possible values of that parameter \cite{flat_chieff_prior_o1event1216_formation_channels2019}. On the other hand, using an isotropic spin prior will significantly down-weight large (positive or negative) effective spins and can make it harder to reweight the results to a different prior. 
We also provide PE samples for posteriors sampled under the isotropic spin prior in the above repository. 

Another difference in our priors is for the case of total mass and mass ratios. 
We used a search prior uniform in $q$, whereas, Ref.~\cite{LVKO3bpopulation} modeled the distribution as a power law $q^{\beta}$ and found $\beta = 1.1^{+1.7}_{-1.3}$. We used a simple power law prior for the detector frame total masses $P(M^\mathrm{det}_\mathrm{tot})\propto M_\mathrm{tot}^{-2}$ and did not include any separate prior over redshift. Ref.~\cite{LVKO3bpopulation} modeled the source primary mass distribution as a power law $m_1^{\alpha}$ with $\alpha=3.5^{+0.6}_{-0.56}$ supplemented by a Gaussian peak at $34^{+2.6}_{-4.0} M_\odot$. Ref.~\cite{LVKO3bpopulation} modeled the merger rate as $(1+z)^\kappa$ and found $\kappa=2.9^{+1.7}_{-1.8}$. The motivation for our prior choices was again to make it easy to reweight our results (both $\pastro$ and overall IFAR values) later in a hierarchical population analysis (especially for the high mass and low mass ratio parameter space), see e.g., \cite{ias_popO2_Roulet_2020}.

A full population analysis not only requires the PE samples but also the $\pastro$ of each event \cite{ias_popO2_Roulet_2020, Gae19_Pop_Subthreshold}.  For this reason, we also provide a file, \texttt{IAS\_HM\_O3\_triggers.json}, in the repository that contains the full list of triggers (including the sub-threshold ones). We also include the subthreshold events ($0.2<p_\mathrm{astro}<0.5$) in Table~\ref{tab:Subthreshold} in the Appendix for reference.

\subsection{Population study}

To probe interesting population inference questions in section~\ref{sec:astro_properties}, one needs to perform a hierarchical bayesian analysis and account for selection function of our search.
We leave this analysis to an upcoming work \cite{Meh25_Population_HM}.
The new candidate events reported here have moderate IFARs (the best event has ${\rm IFAR }\sim 0.67$\,yr), thus a methodology to combine confident and marginal events in population inference will be required \cite{ias_popO2_Roulet_2020}. Our sensitivity is higher than other searches especially for high mass and/or asymmetric mass ratio events due to addition of HM. Having higher sensitivity can improve over some regions of the population hyperparameter space even if we have null/marginal detections.

We also find that in our new search, the significance of a few previously reported marginal events is increased significantly. Especially for the case of GW190711\_030756, the increase in sensitivity is due to addition of HM. This will also be important in population level studies. Currently the LVK collaboration uses events with IFAR $>$1 year in their population studies. Due to the improved significance of our pipeline, more events can now be used in the IFAR $>$1 analysis

\section{Conclusions}
\label{sec:Conclusions}

Nearly all the previous gravitational-wave searches only include the quadrupole mode of the full GW waveform predicted by general relativity.
In our companion papers \citet{Wad23_TemplateBanks,Wad23_Pipeline}, we develop new efficient methods to introduce higher-order modes (HM) in our search pipeline (section~\ref{sec:Improvements}). This helped us perform the first search over the full binary black hole parameter space with including HM in our templates. Compared to our previous pipeline, we introduced additional improvements to remove/downweight noise transients (especially at low frequencies), which further improved our sensitivity to high mass and high redshift systems (section~\ref{sec:Downweight_glitches}).

We searched over public data from the third LIGO--Virgo observing run and found \neviasnew\, new candidate events with $\pastro>0.5$ (see Table~\ref{tab:NewEvents} and section~\ref{sec:Results}).
Some of the new candidate events have several interesting astrophysical properties (section~\ref{sec:astro_properties}) such as: ($i$) masses in the IMBH range and pair instability mass gap, ($ii$) high redshifts, ($iii$) non-zero effective spins, ($iv$) asymmetric mass ratios.

By simply summing over the complements of $p_\mathrm{astro}$ values of our new candidate events from Table~\ref{tab:NewEvents}, we expect that roughly 3 of the \neviasnew{} new events are noise transients instead of being astrophysical signals.
However, we note that estimates of $\pastro$ and source parameters depend on the choice of prior, and we discuss this dependence in section~\ref{sec:Discussion}\footnote{See \url{https://github.com/JayWadekar/GW_higher_harmonics_search} for obtaining our posterior samples and the selection function of our pipeline.}.

We also compared our pipeline's results for the events published in previous IAS catalogs (Tables~\ref{tab:22_IAS_events_significant} \&~\ref{tab:22_IAS_AllEvents}), and in the GWTC-3 and 4-OGC catalogs (Table~\ref{tab:lvc_events}). We broadly recovered the high-significance events, except some events which were either vetoed by our pipeline, which we discuss in detail in section~\ref{sec:LVC}.


Our methodology for efficiently including realistic effects like HM in GW templates will continue to be useful for high-precision searches in upcoming and future gravitational wave experiments \cite{Ng22_ET_HM, Fai23, einsteinTelescope_Maggiore:2019uih, Eva23, decigo_spaceDetector_kawamura2020, LISA, KAGRA:2020tym}. One particular example in which HM will be especially useful is when searching for intermediate to extreme mass-ratio inspirals \cite{Aru07_LISA, Aru07_LISA_2, Por08_LISA, Ber06_LISA, Bai20, Pit23_LISA, LISA_astrophysics, Gai08_LISA_harmonics}; we report results from such a search in a separate follow-up paper \cite{Che25_IMRI_search}.

\emph{Note added--} While this paper was on arXiv, a confident high-mass detection (GW231123) was reported by the LVK collaboration \cite{GW231123} in the O4a data. Incorporating this detection into the search astrophysical prior could make our high-mass detections ($m^\mathrm{source}_1\gtrsim 100\,M_\odot$) more significant.





\acknowledgments

We thank Horng Sheng Chia, Ajith Parameswaran, Will Farr, Vishal Baibhav, Katerina Chatziioannou, Vicky Kalogera, Sylvia Biscoveanu, Amanda Farah, Daniel Holz, Salvatore Vitale, Tom Edwards, Zoheyr Doctor, Suvodip Mukherjee and Mukesh Singh for helpful discussions. 
DW gratefully acknowledges support from the Friends of the Institute for Advanced Study Membership and the Keck foundation.
JR acknowledges support from the Sherman Fairchild Foundation.
TV acknowledges support from NSF grants 2012086 and 2309360, the Alfred P. Sloan Foundation through grant number FG-2023-20470, the BSF through award number 2022136, and the Hellman Family Faculty Fellowship.
MZ is supported by NSF 2209991 and NSF-BSF 2207583. 
BZ is supported by the Israel Science Foundation, NSF-BSF and by a research grant from the Willner Family Leadership Institute for the Weizmann Institute of Science. 
This research was also supported in part by the National Science Foundation under Grant No. NSF PHY-1748958. We also thank ICTS-TIFR for their hospitality during the completion of a part of this work. 

This research has made use of data, software and/or web tools obtained from the Gravitational Wave Open Science Center (\url{https://www.gw-openscience.org/}), a service of LIGO Laboratory, the LIGO Scientific Collaboration and the Virgo Collaboration. LIGO Laboratory and Advanced LIGO are funded by the United States National Science Foundation (NSF) as well as the Science and Technology Facilities Council (STFC) of the United Kingdom, the Max-Planck-Society (MPS), and the State of Niedersachsen/Germany for support of the construction of Advanced LIGO and construction and operation of the GEO600 detector. Additional support for Advanced LIGO was provided by the Australian Research Council. Virgo is funded, through the European Gravitational Observatory (EGO), by the French Centre National de Recherche Scientifique (CNRS), the Italian Istituto Nazionale di Fisica Nucleare (INFN) and the Dutch Nikhef, with contributions by institutions from Belgium, Germany, Greece, Hungary, Ireland, Japan, Monaco, Poland, Portugal, Spain.

\appendix

\section{Changes since previous arXiv version}

The previous version of our paper on the arXiv ([v1] 11 Dec 2023) included the list of events from an older version of our pipeline. Since that point, we have made a number of improvements to the pipeline. Furthermore, unlike the older version, the newer version of our \texttt{IAS-HM} pipeline has gone through an extensive injection recovery analysis to test its senstitivity (see section~\ref{sec:HM_sensitivity}). 
We list the details of other improvements below.

We implemented a few extra noise reduction algorithms like TIER (see section~\ref{sec:TIER}). We see an overall improvement in sensitivity due to it \cite{Wad25_TIER_GW}. We also however see both positive and negative fluctuations in IFAR and $\pastro$ values of the near-threshold new candidate events. This is expected, as some of the near-threshold candidate events were probabilistically likely to be glitches and the \texttt{TIER} algorithm is able to downweight those.


We fixed an issue with relative binning for template optimization in our bank BBH-1. This had earlier led to some of the LVK mergers being spuriously vetoed for this bank. We have now fixed the issue.

In the previous version of our catalog, some of the fairly high SNR events in GWTC-3 with imprints of precession were vetoed by our pipeline. We added a ML-based veto overriding procedure (see section~\ref{sec:Veto_ML}) to override the veto for these events and we also find an improvement in the pipeline sensitivity due to this procedure.

In section~\ref{sec:astrophysical_prior}, we discussed using an astrophysical prior for the templates. We improved the astrophysical prior applied to the templates using normalizing flows \cite{Wad25_NormFlow_HM_search}, superseding the earlier kernel density estimation based method given in Appendix~C of \citetalias{Ols22_ias_o3a}. We also improved our coherent marginalized statistic using normalizing flows (further details will be provided in a separate paper \cite{Wad25_NormFlow_HM_search}).

\section{Additional plots and tables}
\label{apx:Additional_plots}

\begin{figure}[!ht]
    \centering
\includegraphics[width=0.5\textwidth]{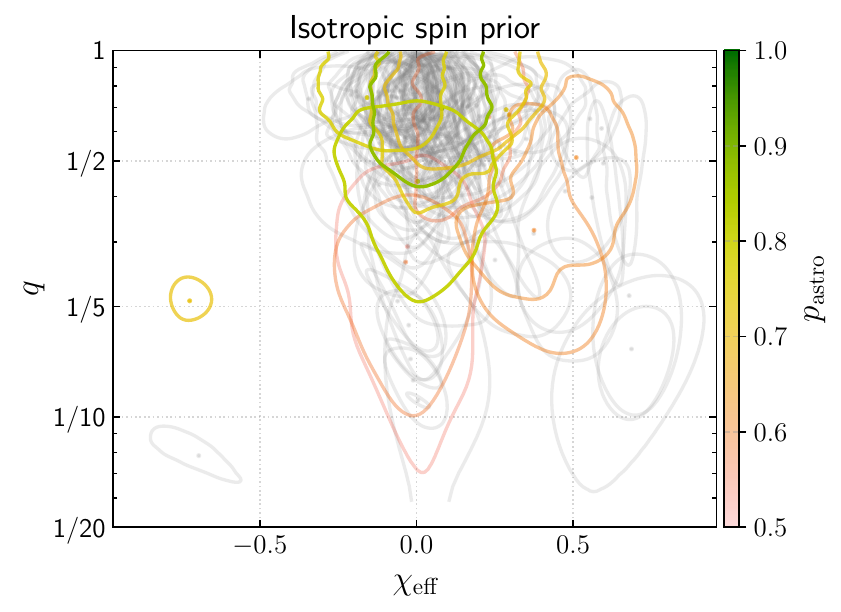}
\includegraphics[width=0.5\textwidth]{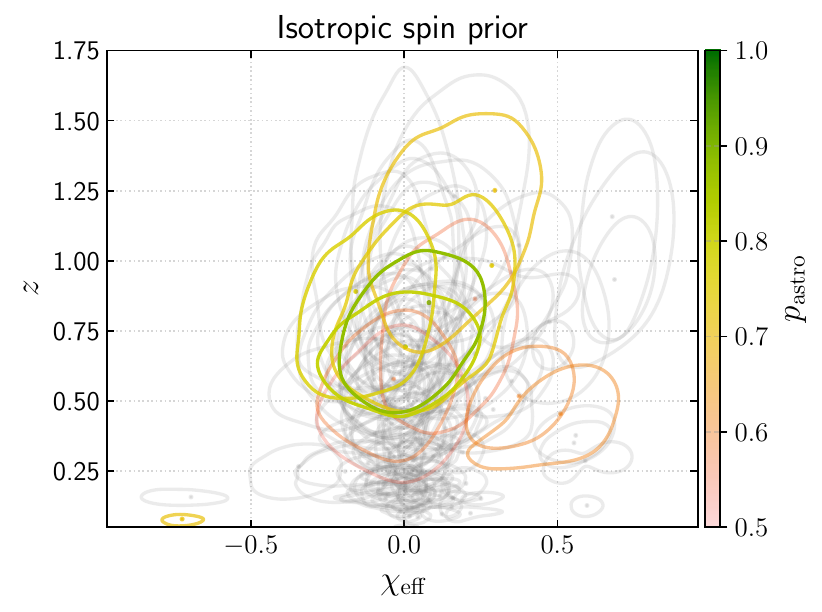}
    \caption{Same as Fig.~\ref{fig:m1_z}, but showing the 2D parameter estimation contours for $\chieff$ alongside mass ratio (redshift) in the top (bottom) panel.}
    \label{fig:chieff_extra}
\end{figure}

\begin{figure}
\centering
\includegraphics[width=0.45\textwidth,keepaspectratio=true]{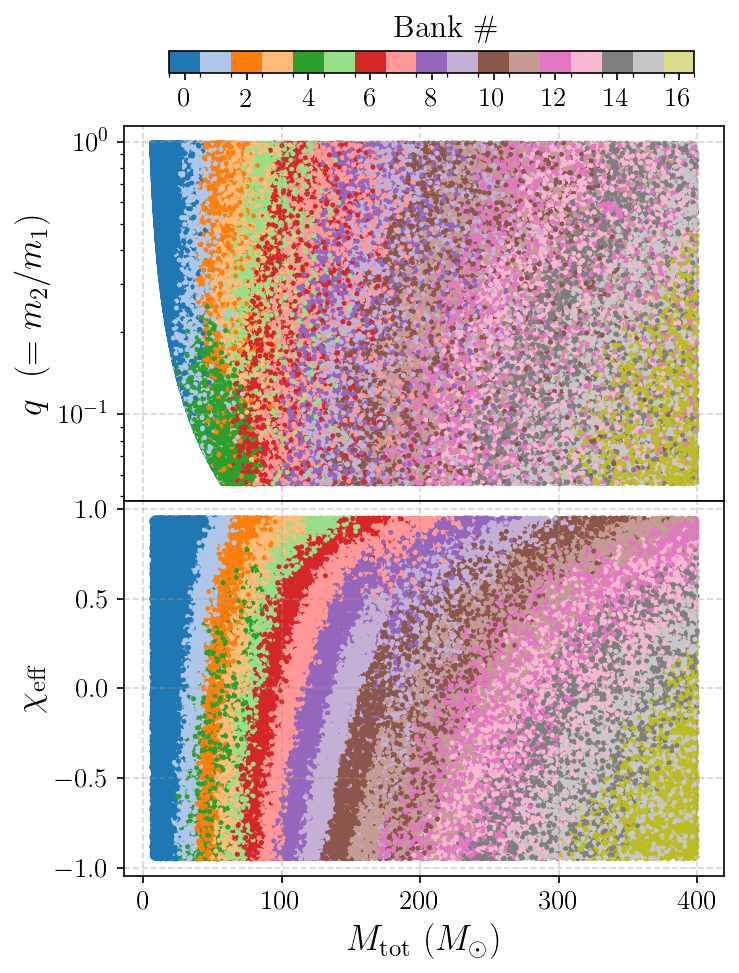}
\includegraphics[scale=0.7,keepaspectratio=true]{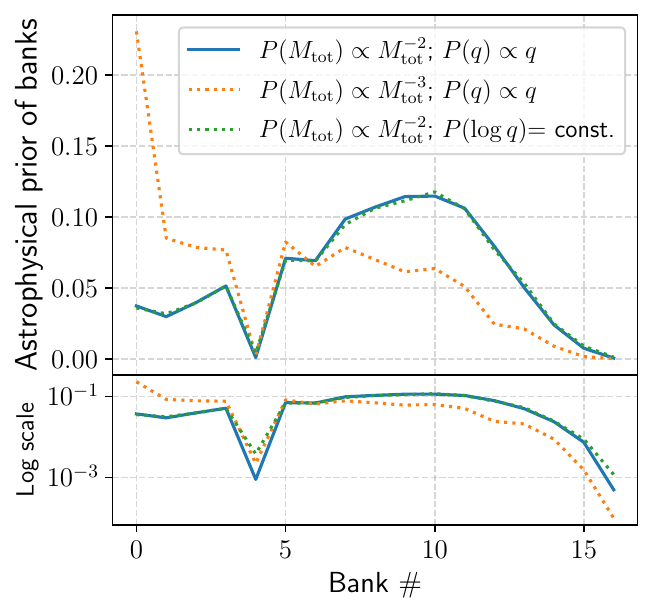}
\caption{
\textbf{Top:} Physical parameters corresponding to each bank (this panel is taken from our companion paper \cite{Wad23_TemplateBanks} and is shown here for reference).
\textbf{Bottom:}
Prior probability corresponding to the number of events expected in individual banks. Solid line corresponds to our fiducial astrophysical prior in Section~\ref{sec:astrophysical_prior}. For reference, we compare the probabilities for a different mass and mass ratio distribution in dotted lines. Note that bank 4 shows dips because it primarily includes low-$q$ systems which have a smaller observable distance due to their lower chirp masses.
The decrement seen for the high-mass banks is primarily due to the binary merger starting to fall out of band at lower frequencies.}
\label{fig:Template_Prior_Banks}
\end{figure}

\begin{figure*}
    \centering
    \includegraphics[width=0.99\textwidth]{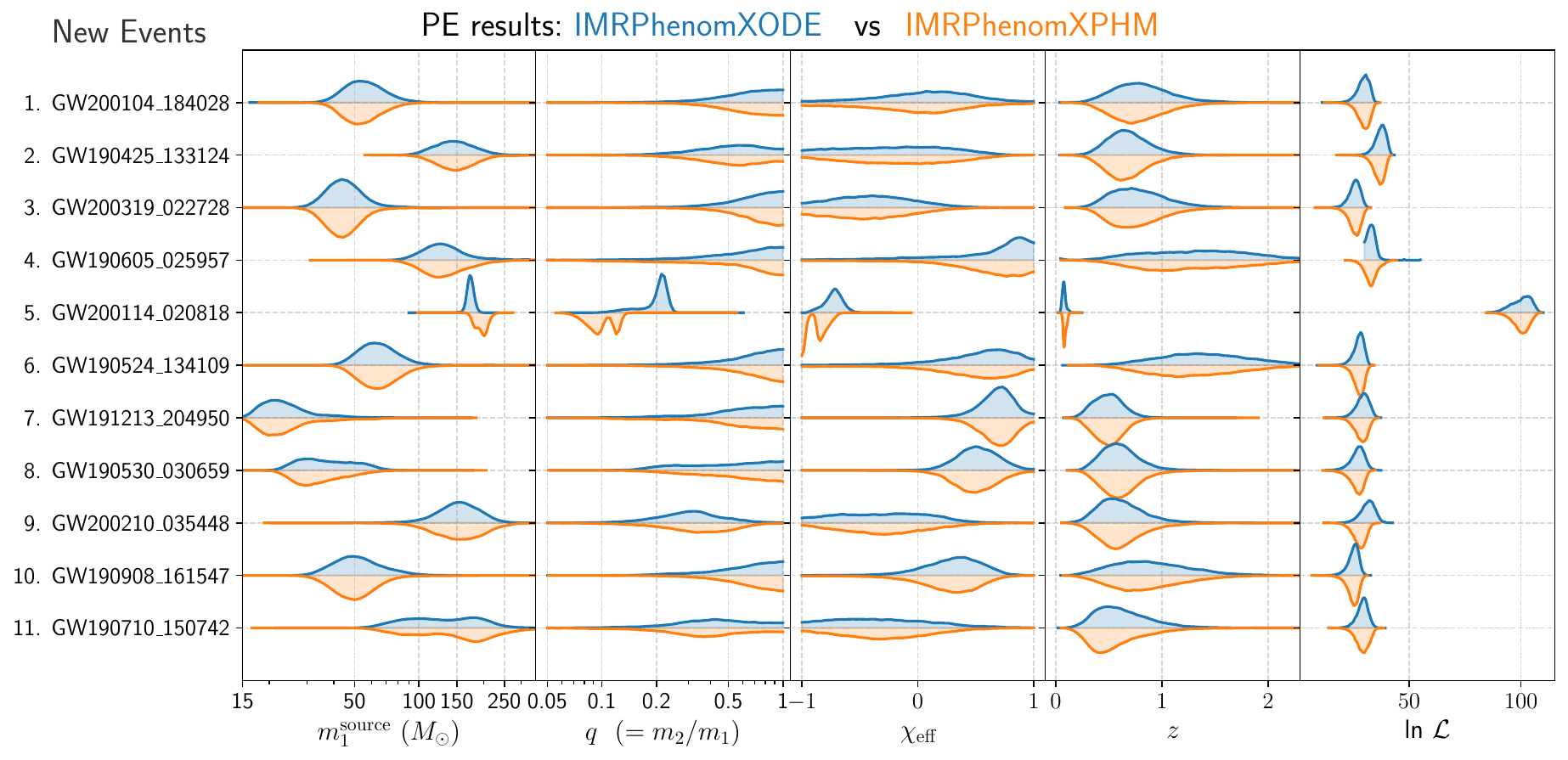}
    \caption{Similar to Fig.~\ref{fig:PE_1D_spin} but comparing our fiducial PE results in the main paper (using IMRPhenomXODE \cite{Yu23_XODE}) to the PE results from the widely-used IMRPhenomXPHM approximant. The results from the two approximants are broadly consistent. Only GW200114\_020818 shows differences, especially at negative $\chieff$ regime.}
    \label{fig:XODE_XPHM}
\end{figure*}

 \bgroup
\def\arraystretch{1.5}
\begin{table*}
    \centering
\caption{We show the SNR$^2$ contribution from the different harmonics for the new candidate events in our search (we sum over the Hanford and Livingston contributions in this table). We also cross-check our list with the sub-threshold candidates given in GWTC-3 \cite{lvc_gwtc3_o3_ab_catalog_2021} and the IAS $(2, 2)$-only catalogs \cite{Ols22_ias_o3a, Meh23_ias_o3b}; we report properties of the sub-threshold candidates which have GPS times differing by less than a second. GW200104\_184028 was reported by gstlal\_allsky pipeline.  GW191213\_204950, GW190530\_030659, GW190908\_161547 were reported by the pycbc\_highmass pipeline.}
    \begin{tabular}{|c|c|ccc|ccc|ccc|}
        \hline
        Sr. & New candidate &\multicolumn{3}{c|} {This search} & \multicolumn{3}{c|}{GWTC-3 sub-threshold} & \multicolumn{3}{c|}{IAS 22-only sub-threshold}  \\
        \cline{3-11} 
         No. & event& $\rho_{22}^2$ & $\rho_{33}^2$ & $\rho_{44}^2$ & $\rho_{22}^2$ & IFAR & $\pastro$ &
         $\rho_{22}^2$ & IFAR & $\pastro$ \\ 
         \hline 

         \rowcolor{gray!30} 1 & GW200104\_184028 &66.0 & $9.4$ & $1.2$ & $76.5$ & $0.011$ & $0.013$ & $67.4$ & $0.095$ & $0.32$ \\
         2 & GW190425\_133124 &67.6 & $9.1$ & $0.020$ & $--$ & $--$ & $--$ & $73.7$ & $0.011$ & $0.067$ \\
         \rowcolor{gray!30} 3 & GW200319\_022728 &61.1 & $0.80$ & $10.5$ & $--$ & $--$ & $--$ & $61.6$ & $0.036$ & $0.20$ \\
         4 & GW190605\_025957 &63.2 & $7.2$ & $5.1$ & $--$ & $--$ & $--$ & $--$ & $--$ & $--$ \\
         \rowcolor{gray!30} 5 & GW200114\_020818 &157 & $14.4$ & $8.0$ & $--$ & $--$ & $--$ & $--$ & $--$ & $--$ \\
         6 & GW190524\_134109 &56.8 & $3.7$ & $2.1$ & $--$ & $--$ & $--$ & $53.8$ & $0.015$ & $0.27$ \\
         \rowcolor{gray!30} 7 & GW191213\_204950 &66.2 & $8.0$ & $8.1$ & $62.0$ & $0.063$ & $0.39$ & $61.8$ & $0.015$ & $0.12$ \\
         8 & GW190530\_030659 &63.2 & $2.4$ & $8.5$ & $62.5$ & $0.016$ & $0.15$ & $61.6$ & $0.011$ & $0.21$ \\
         \rowcolor{gray!30} 9 & GW200210\_035448 &57.4 & $7.8$ & $2.5$ & $--$ & $--$ & $--$ & $--$ & $--$ & $--$ \\
         10 & GW190908\_161547 &58.2 & $3.8$ & $19.4$ & $56.7$ & $0.003$ & $0.023$ & $57.4$ & $0.030$ & $0.38$ \\
         \rowcolor{gray!30} 11 & GW190710\_150742 &59.2 & $7.9$ & $1.5$ & $--$ & $--$ & $--$ & $--$ & $--$ & $--$ \\
         \hline 
    \end{tabular}
    \label{tab:subthreshold_LVC}
\end{table*}

\bgroup
\def\arraystretch{1.3}
\begin{table*}
    \centering
    \caption{Same as Table~\ref{tab:22_IAS_events_significant} but for candidates found with $p_\mathrm{astro}>10^{-3}$ which overlap with the  events reported for the first time in IAS $(2, 2)$-only searches: \cite{Ols22_ias_o3a, Meh23_ias_o3b}. As we raised our $\rho^2$ bar for collecting triggers in the HM search, the following events from $(2, 2)$-only IAS searches were outside the domain of our search: GW190920\_113516, GW190718\_160159, GW200225\_075134.}
    \begin{tabular}{|c|c|cccc|cc|cc|}
        \hline
         \multirow{2}{*}{(O3a) Name} & \multirow{2}{*}{Bank} &\multicolumn{4}{c|}{Best fit template} 
         & \multirow{2}{*}{$\rho_{\rm H}^2$}  & \multirow{2}{*}{$\rho_{\rm L}^2$} & \multirow{2}{*}{IFAR (yr)}  & \multirow{2}{*}{$p_{\rm astro}$}\\
         \cline{3-6}
         & & $m_1^{\textrm{det}} {\scriptstyle( \rm \msun)}$ & $m_2^{\textrm{det}} {\scriptstyle(\rm \msun)}$ & $\chi_{\rm eff}$ & 
         $\mathcal{M}_c$ & & & & \\
         \hline 

         \rowcolor{gray!30} GW190711\_030756 & \texttt{[6, 2]} & $73.3$ & $37.4$ & $-0.088$ & $45.1$ & $30.9$ & $68.5$ & $11.0$ & $0.99$ \\
         GW190818\_232544 & \texttt{[8, 1]} & $139$ & $58.8$ & $0.65$ & $77.2$ & $45.9$ & $35.5$ & $0.81$ & $0.90$ \\
         \rowcolor{gray!30} GW190707\_083226 & \texttt{[7, 1]} & $90.9$ & $41.0$ & $-0.18$ & $52.4$ & $52.4$ & $32.2$ & $0.55$ & $0.87$ \\
         GW190814\_192009 & \texttt{[10, 0]} & $175$ & $99.2$ & $0.60$ & $114$ & $32.1$ & $33.5$ & $0.019$ & $0.27$ \\
         \rowcolor{gray!30} GW190704\_104834 & \texttt{[0, 0]} & $6.8$ & $4.1$ & $0.18$ & $4.6$ & $56.3$ & $39.1$ & $0.017$ & $0.25$ \\
         GW190906\_054335 & \texttt{[6, 1]} & $69.6$ & $15.5$ & $-0.15$ & $27.2$ & $25.3$ & $42.3$ & $0.003$ & $0.060$ \\
         \rowcolor{gray!30} GW190821\_124821 & \texttt{[0, 1]} & $13.1$ & $3.4$ & $-0.10$ & $5.6$ & $32.9$ & $54.3$ & $0.002$ & $0.029$ \\
         GW190910\_012619 & \texttt{[4, 0]} & $39.7$ & $3.5$ & $-0.82$ & $9.1$ & $46.1$ & $32.7$ & $< 0.001$ & $0.003$ \\
         \hline
        \hline
         (O3b) Name & Bank & $m_1^{\textrm{det}} {\scriptstyle( \rm \msun)}$ & $m_2^{\textrm{det}} {\scriptstyle(\rm \msun)}$ & $\chi_{\rm eff}$ & 
         $\mathcal{M}_c$
         & $\rho_{\rm H}^2$ & $\rho_{\rm L}^2$ & IFAR (yr) & $p_{\rm astro}$\\
         \hline 

         \rowcolor{gray!30} GW200109\_195634 & \texttt{[9, 0]} & $170$ & $76.4$ & $0.64$ & $97.8$ & $40.0$ & $37.4$ & $1.1$ & $0.92$ \\
         GW191117\_023843 & \texttt{[10, 0]} & $130$ & $30.8$ & $-0.92$ & $52.6$ & $42.6$ & $29.1$ & $0.058$ & $0.52$ \\
         \rowcolor{gray!30} GW200210\_100022 & \texttt{[7, 1]} & $176$ & $35.6$ & $0.88$ & $65.0$ & $29.0$ & $41.8$ & $0.036$ & $0.42$ \\
         GW191228\_085854 & \texttt{[0, 2]} & $14.5$ & $5.1$ & $-0.51$ & $7.3$ & $32.0$ & $51.2$ & $0.018$ & $0.27$ \\
         \rowcolor{gray!30} GW200316\_235947 & \texttt{[0, 0]} & $6.9$ & $4.6$ & $-0.14$ & $4.9$ & $46.1$ & $37.3$ & $0.005$ & $0.10$ \\

         \hline
    \end{tabular}
    
    \label{tab:22_IAS_AllEvents}
\end{table*}
\egroup

\bgroup
\def\arraystretch{1.1}
\begin{table*}
    \centering
    \caption{Same as Table~\ref{tab:NewEvents} but for candidates found with $0.2<p_\mathrm{astro}<0.5$ and  IFAR $>$ 1 within the respective bank for the respective observing run (O3a or O3b). The significance of these events depends on the assumed astrophysical prior.}
    \begin{tabular}{|c|c|cccc|cc|cc|c|}
        \hline
         \multirow{2}{*}{Candidate} & \multirow{2}{*}{Bank} &\multicolumn{4}{c|}{Best fit template} 
         & \multirow{2}{*}{$\rho_{\rm H}^2$}  & \multirow{2}{*}{$\rho_{\rm L}^2$} & \multicolumn{2}{c|}{IFAR (yr)}   & \multirow{2}{*}{$p_\mathrm{astro}$}\\
         \cline{3-6} \cline{9-10}
         & & $m_1^{\textrm{det}} {\scriptstyle( \rm \msun)}$ & $m_2^{\textrm{det}} {\scriptstyle(\rm \msun)}$ & $\chi_{\rm eff}$ & 
         $\mathcal{M}_c$ & & & per bank & overall & \\
         \hline 
         \rowcolor{gray!30} GW200104\_085935 & \texttt{[10, 0]} & $129$ & $36.9$ & $-0.39$ & $57.9$ & $27.1$ & $39.6$ & $0.45$ & $0.049$ & $0.48$ \\
GW190521\_075726 & \texttt{[14, 0]} & $241$ & $92.9$ & $-0.079$ & $127$ & $22.4$ & $47.6$ & $2.9$ & $0.052$ & $0.48$ \\
\rowcolor{gray!30} GW190529\_023648 & \texttt{[9, 0]} & $118$ & $32.4$ & $-0.55$ & $51.7$ & $24.7$ & $39.4$ & $0.48$ & $0.050$ & $0.47$ \\
GW190512\_001054 & \texttt{[6, 2]} & $66.6$ & $59.7$ & $0.11$ & $54.9$ & $22.5$ & $46.1$ & $0.70$ & $0.045$ & $0.45$ \\
\rowcolor{gray!30} GW191205\_033000 & \texttt{[7, 1]} & $79.6$ & $47.9$ & $-0.25$ & $53.4$ & $23.6$ & $54.2$ & $0.42$ & $0.038$ & $0.42$ \\
GW200210\_100022 & \texttt{[7, 1]} & $176$ & $35.6$ & $0.88$ & $65.0$ & $29.0$ & $41.8$ & $0.41$ & $0.036$ & $0.42$ \\
\rowcolor{gray!30} GW200217\_211137 & \texttt{[9, 0]} & $152$ & $91.0$ & $0.46$ & $102$ & $24.1$ & $38.7$ & $0.38$ & $0.036$ & $0.42$ \\
GW191228\_195439 & \texttt{[10, 0]} & $173$ & $138$ & $0.77$ & $135$ & $24.1$ & $44.9$ & $0.34$ & $0.036$ & $0.41$ \\
\rowcolor{gray!30} GW191127\_114537 & \texttt{[12, 0]} & $173$ & $52.6$ & $-0.82$ & $80.4$ & $23.1$ & $46.1$ & $0.38$ & $0.031$ & $0.38$ \\
GW190514\_065259 & \texttt{[3, 2]} & $68.7$ & $22.5$ & $0.72$ & $33.2$ & $27.0$ & $44.2$ & $0.52$ & $0.031$ & $0.37$ \\
\rowcolor{gray!30} GW191208\_080334 & \texttt{[5, 2]} & $52.3$ & $35.3$ & $0.29$ & $37.3$ & $44.8$ & $24.7$ & $0.30$ & $0.027$ & $0.35$ \\
GW191228\_062343 & \texttt{[11, 0]} & $313$ & $58.1$ & $0.81$ & $110$ & $26.9$ & $48.1$ & $0.28$ & $0.027$ & $0.35$ \\
\rowcolor{gray!30} GW190404\_121150 & \texttt{[9, 0]} & $172$ & $86.6$ & $0.92$ & $105$ & $42.4$ & $24.4$ & $0.29$ & $0.028$ & $0.35$ \\
GW200202\_160037 & \texttt{[5, 2]} & $47.8$ & $43.6$ & $-0.16$ & $39.7$ & $34.9$ & $31.6$ & $0.28$ & $0.025$ & $0.34$ \\
\rowcolor{gray!30} GW190810\_180039 & \texttt{[5, 2]} & $71.8$ & $37.5$ & $0.18$ & $44.7$ & $34.7$ & $35.1$ & $0.36$ & $0.027$ & $0.34$ \\
GW190605\_053717 & \texttt{[1, 2]} & $26.9$ & $25.7$ & $0.86$ & $22.9$ & $35.6$ & $40.5$ & $0.46$ & $0.026$ & $0.33$ \\
\rowcolor{gray!30} GW200129\_132219 & \texttt{[5, 0]} & $53.4$ & $8.8$ & $-0.66$ & $17.5$ & $61.2$ & $25.4$ & $0.27$ & $0.024$ & $0.33$ \\
GW200325\_090608 & \texttt{[10, 0]} & $175$ & $99.2$ & $0.60$ & $114$ & $25.8$ & $36.7$ & $0.23$ & $0.024$ & $0.33$ \\
\rowcolor{gray!30} GW200312\_093903 & \texttt{[11, 0]} & $145$ & $88.2$ & $-0.75$ & $97.8$ & $23.0$ & $40.0$ & $0.24$ & $0.023$ & $0.32$ \\
GW200224\_043639 & \texttt{[12, 0]} & $184$ & $20.0$ & $-0.31$ & $47.6$ & $28.7$ & $30.7$ & $0.26$ & $0.022$ & $0.31$ \\
\rowcolor{gray!30} GW200216\_200717 & \texttt{[7, 1]} & $96.6$ & $58.4$ & $0.41$ & $65.0$ & $32.6$ & $35.7$ & $0.25$ & $0.021$ & $0.31$ \\
GW190617\_041509 & \texttt{[8, 1]} & $80.3$ & $54.1$ & $-0.87$ & $57.1$ & $25.6$ & $35.3$ & $0.30$ & $0.023$ & $0.30$ \\
\rowcolor{gray!30} GW191230\_072040 & \texttt{[5, 2]} & $102$ & $37.1$ & $0.85$ & $52.4$ & $26.9$ & $49.9$ & $0.22$ & $0.020$ & $0.29$ \\
GW200114\_131740 & \texttt{[0, 2]} & $13.6$ & $5.8$ & $-0.17$ & $7.6$ & $29.4$ & $47.7$ & $0.31$ & $0.020$ & $0.29$ \\
\rowcolor{gray!30} GW191105\_105458 & \texttt{[6, 2]} & $55.4$ & $42.4$ & $-0.79$ & $42.1$ & $43.4$ & $26.1$ & $0.29$ & $0.018$ & $0.28$ \\
GW191228\_085854 & \texttt{[0, 2]} & $14.5$ & $5.1$ & $-0.51$ & $7.3$ & $32.0$ & $51.2$ & $0.27$ & $0.018$ & $0.27$ \\
\rowcolor{gray!30} GW190411\_075756 & \texttt{[9, 0]} & $118$ & $32.4$ & $-0.55$ & $51.7$ & $28.7$ & $30.4$ & $0.21$ & $0.019$ & $0.27$ \\
GW191224\_075833 & \texttt{[10, 0]} & $152$ & $49.0$ & $-0.005$ & $72.8$ & $24.5$ & $80.7$ & $0.17$ & $0.018$ & $0.27$ \\
\rowcolor{gray!30} GW200126\_225208 & \texttt{[6, 2]} & $88.2$ & $68.1$ & $0.92$ & $67.4$ & $41.9$ & $26.9$ & $0.26$ & $0.016$ & $0.26$ \\
GW200301\_211019 & \texttt{[2, 1]} & $29.2$ & $17.2$ & $-0.22$ & $19.4$ & $36.6$ & $36.0$ & $0.33$ & $0.016$ & $0.25$ \\
\rowcolor{gray!30} GW200128\_115507 & \texttt{[8, 1]} & $95.8$ & $58.2$ & $-0.20$ & $64.6$ & $33.4$ & $34.0$ & $0.24$ & $0.016$ & $0.25$ \\
GW190604\_103812 & \texttt{[11, 0]} & $254$ & $115$ & $0.88$ & $146$ & $27.0$ & $36.2$ & $0.15$ & $0.016$ & $0.24$ \\
\rowcolor{gray!30} GW190628\_185055 & \texttt{[1, 2]} & $35.0$ & $19.0$ & $0.91$ & $22.2$ & $44.8$ & $28.9$ & $0.24$ & $0.016$ & $0.24$ \\
GW190806\_033721 & \texttt{[9, 0]} & $170$ & $76.4$ & $0.64$ & $97.8$ & $27.7$ & $30.3$ & $0.18$ & $0.016$ & $0.23$ \\
\rowcolor{gray!30} GW190705\_164632 & \texttt{[7, 1]} & $118$ & $29.7$ & $0.30$ & $49.3$ & $30.2$ & $38.1$ & $0.19$ & $0.016$ & $0.23$ \\
GW191222\_085357 & \texttt{[8, 1]} & $79.3$ & $64.1$ & $-0.49$ & $62.0$ & $30.3$ & $32.6$ & $0.21$ & $0.014$ & $0.23$ \\
\rowcolor{gray!30} GW190610\_081435 & \texttt{[1, 2]} & $25.1$ & $18.9$ & $0.45$ & $18.9$ & $35.4$ & $40.5$ & $0.23$ & $0.015$ & $0.23$ \\
GW191102\_232120 & \texttt{[2, 2]} & $41.1$ & $18.4$ & $0.068$ & $23.5$ & $30.1$ & $39.7$ & $0.28$ & $0.014$ & $0.23$ \\
\rowcolor{gray!30} GW200303\_083614 & \texttt{[0, 1]} & $13.4$ & $3.9$ & $0.47$ & $6.1$ & $34.3$ & $50.4$ & $0.19$ & $0.013$ & $0.22$ \\
GW200222\_005735 & \texttt{[0, 4]} & $26.1$ & $16.9$ & $0.94$ & $18.2$ & $32.8$ & $42.7$ & $0.18$ & $0.013$ & $0.22$ \\
\rowcolor{gray!30} GW191115\_141720 & \texttt{[9, 0]} & $172$ & $86.6$ & $0.92$ & $105$ & $40.3$ & $33.7$ & $0.15$ & $0.013$ & $0.21$ \\
GW200308\_231646 & \texttt{[8, 1]} & $85.0$ & $77.5$ & $0.21$ & $70.6$ & $30.2$ & $37.2$ & $0.20$ & $0.013$ & $0.21$ \\
\rowcolor{gray!30} GW190813\_040521 & \texttt{[5, 2]} & $62.4$ & $57.3$ & $0.64$ & $52.1$ & $32.1$ & $43.2$ & $0.18$ & $0.014$ & $0.21$ \\
GW190826\_212543 & \texttt{[8, 1]} & $139$ & $58.8$ & $0.65$ & $77.2$ & $43.4$ & $22.7$ & $0.19$ & $0.013$ & $0.20$ \\
\rowcolor{gray!30} GW190422\_175505 & \texttt{[8, 1]} & $134$ & $64.4$ & $0.41$ & $79.6$ & $30.3$ & $38.7$ & $0.19$ & $0.013$ & $0.20$ \\

\hline
\end{tabular}

%
        
    \label{tab:Subthreshold}
\end{table*}
\egroup


\begin{figure*}
    \centering
    \includegraphics[width=1.02\linewidth]{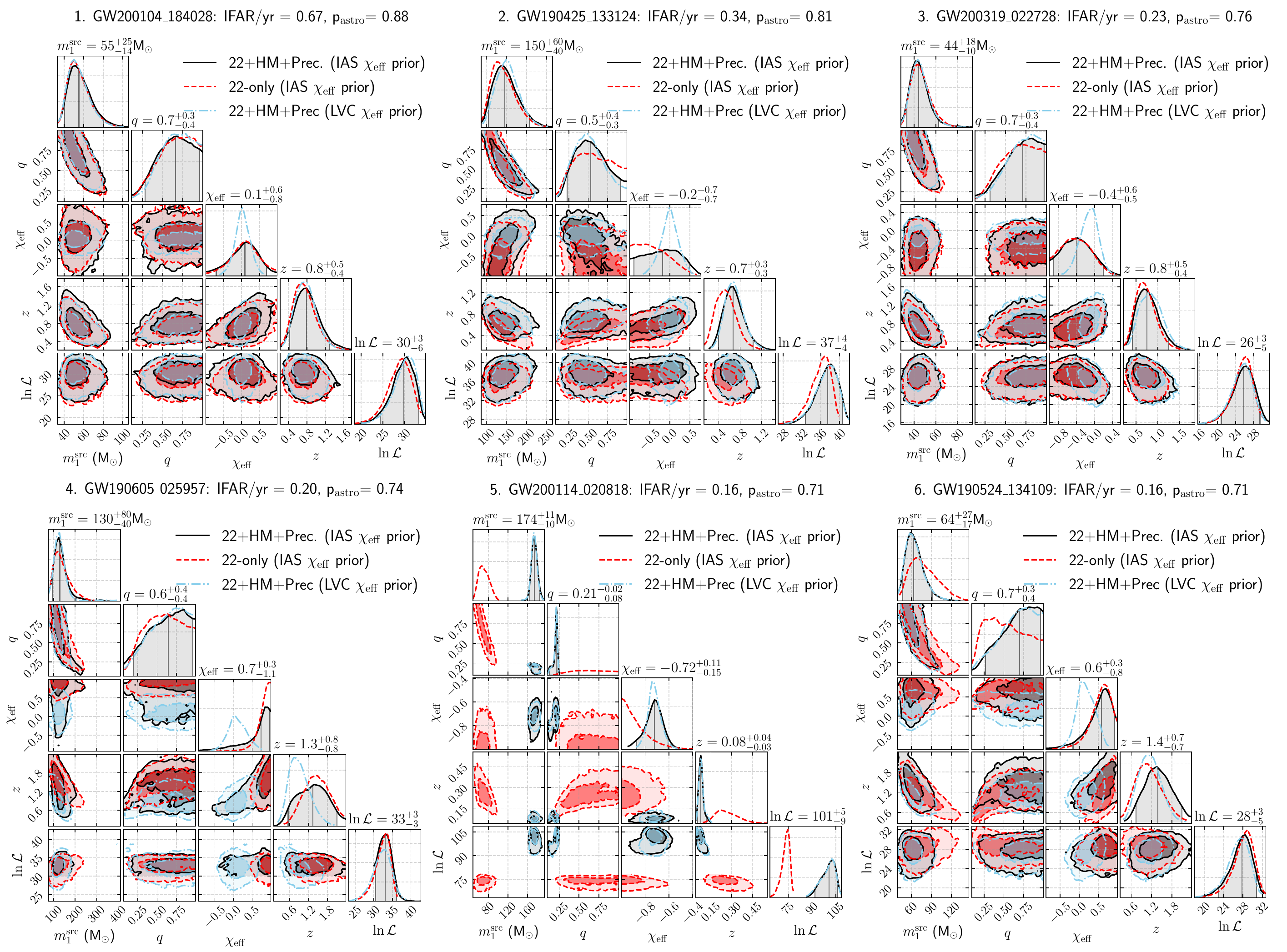}
    \caption{Corner plots for the first six of the new candidate events reported in the Table~\ref{tab:NewEvents} (ranked by $\pastro$ values). The marginalized values shown on the top of each column in the panels correspond to 90\% confidence intervals for our fiducial PE (shown in black) for the flat-in-$\chi_\mathrm{eff}$ prior (for the isotropic in component spin prior, we show the results in cyan). Our fiducial PE includes waveforms with HM and precession, but we also show in red the PE results with $(2, 2)$-only aligned-spin waveforms. We discuss the interesting astrophysical properties of these candidate events in Section~\ref{sec:astro_properties}.}
    \label{fig:PE_Events_1}
    \end{figure*}
    
    \begin{figure*}
        \centering
        \includegraphics[width=.95\linewidth]{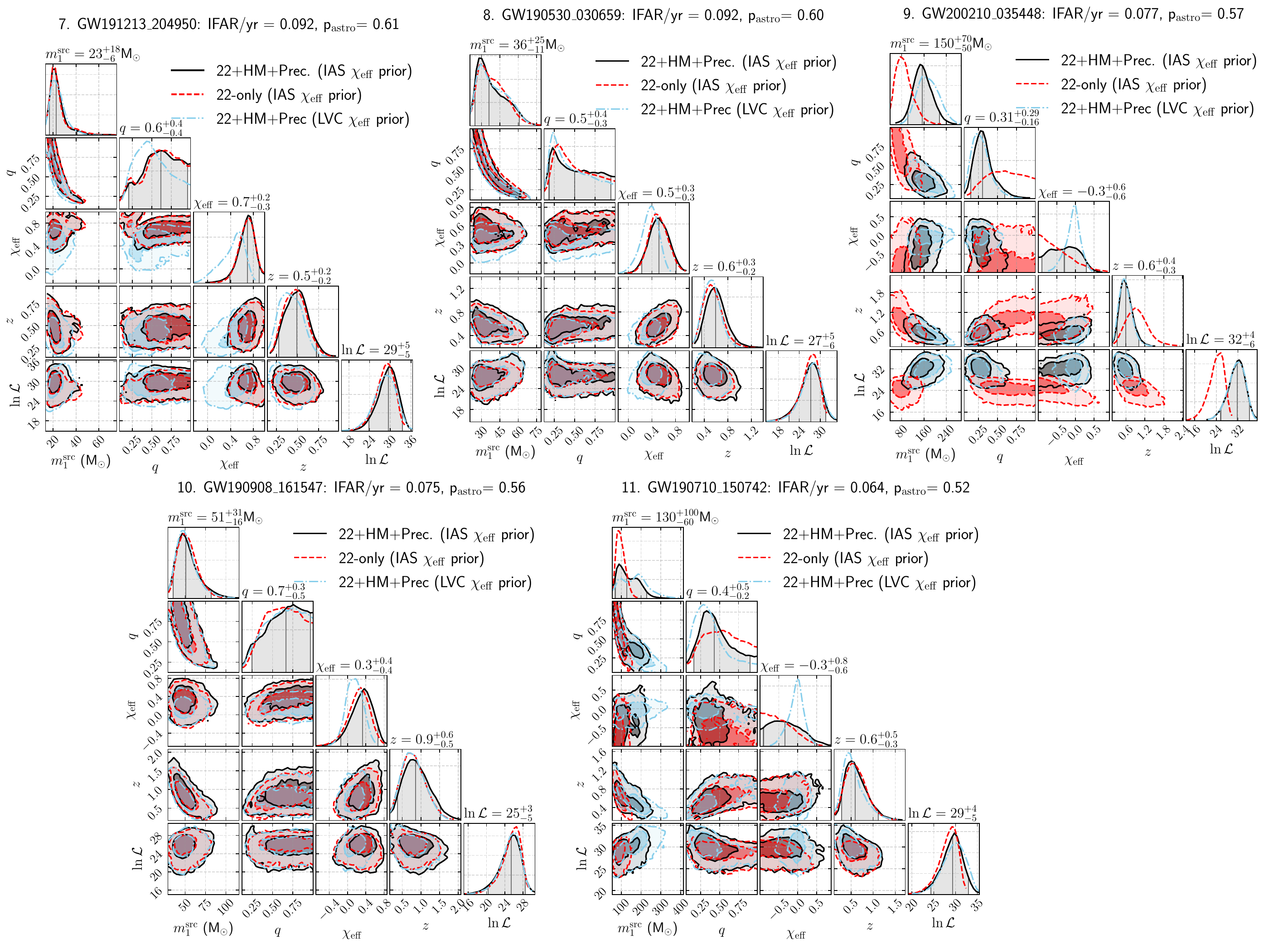}
        \caption{Same as Fig.~\ref{fig:PE_Events_1} but for events numbered 7--11 in the Table~\ref{tab:NewEvents}.}
    \label{fig:PE_Events_2}
    \end{figure*}
    
\clearpage

\bibliographystyle{apsrev4-1-etal}
\bibliography{GW}

\begin{thebibliography}{142}%
\makeatletter
\providecommand \@ifxundefined [1]{%
 \@ifx{#1\undefined}
}%
\providecommand \@ifnum [1]{%
 \ifnum #1\expandafter \@firstoftwo
 \else \expandafter \@secondoftwo
 \fi
}%
\providecommand \@ifx [1]{%
 \ifx #1\expandafter \@firstoftwo
 \else \expandafter \@secondoftwo
 \fi
}%
\providecommand \natexlab [1]{#1}%
\providecommand \enquote  [1]{``#1''}%
\providecommand \bibnamefont  [1]{#1}%
\providecommand \bibfnamefont [1]{#1}%
\providecommand \citenamefont [1]{#1}%
\providecommand \href@noop [0]{\@secondoftwo}%
\providecommand \href [0]{\begingroup \@sanitize@url \@href}%
\providecommand \@href[1]{\@@startlink{#1}\@@href}%
\providecommand \@@href[1]{\endgroup#1\@@endlink}%
\providecommand \@sanitize@url [0]{\catcode `\\12\catcode `\$12\catcode
  `\&12\catcode `\#12\catcode `\^12\catcode `\_12\catcode `\%12\relax}%
\providecommand \@@startlink[1]{}%
\providecommand \@@endlink[0]{}%
\providecommand \url  [0]{\begingroup\@sanitize@url \@url }%
\providecommand \@url [1]{\endgroup\@href {#1}{\urlprefix }}%
\providecommand \urlprefix  [0]{URL }%
\providecommand \Eprint [0]{\href }%
\providecommand \doibase [0]{http://dx.doi.org/}%
\providecommand \selectlanguage [0]{\@gobble}%
\providecommand \bibinfo  [0]{\@secondoftwo}%
\providecommand \bibfield  [0]{\@secondoftwo}%
\providecommand \translation [1]{[#1]}%
\providecommand \BibitemOpen [0]{}%
\providecommand \bibitemStop [0]{}%
\providecommand \bibitemNoStop [0]{.\EOS\space}%
\providecommand \EOS [0]{\spacefactor3000\relax}%
\providecommand \BibitemShut  [1]{\csname bibitem#1\endcsname}%
\let\auto@bib@innerbib\@empty
\bibitem [{\citenamefont {{Wadekar}}\ \emph {et~al.}(2023)\citenamefont
  {{Wadekar}}, \citenamefont {{Venumadhav}}, \citenamefont {{Mehta}},
  \citenamefont {{Roulet}}, \citenamefont {{Olsen}}, \citenamefont {{Mushkin}},
  \citenamefont {{Zackay}},\ and\ \citenamefont
  {{Zaldarriaga}}}]{Wad23_TemplateBanks}%
  \BibitemOpen
  \bibfield  {author} {\bibinfo {author} {\bibfnamefont {D.}~\bibnamefont
  {{Wadekar}}}, \bibinfo {author} {\bibfnamefont {T.}~\bibnamefont
  {{Venumadhav}}}, \bibinfo {author} {\bibfnamefont {A.~K.}\ \bibnamefont
  {{Mehta}}}, \bibinfo {author} {\bibfnamefont {J.}~\bibnamefont {{Roulet}}},
  \emph {et~al.},\ }\href@noop {} {\bibfield  {journal} {\bibinfo  {journal}
  {arXiv e-prints}\ ,\ \bibinfo {eid} {arXiv:2310.15233}} (\bibinfo {year}
  {2023})},\ \Eprint {http://arxiv.org/abs/2310.15233} {arXiv:2310.15233
  [gr-qc]} \BibitemShut {NoStop}%
\bibitem [{\citenamefont {{Wadekar}}\ \emph {et~al.}(2024)\citenamefont
  {{Wadekar}}, \citenamefont {{Venumadhav}}, \citenamefont {{Roulet}},
  \citenamefont {{Mehta}}, \citenamefont {{Zackay}}, \citenamefont
  {{Mushkin}},\ and\ \citenamefont {{Zaldarriaga}}}]{Wad23_Pipeline}%
  \BibitemOpen
  \bibfield  {author} {\bibinfo {author} {\bibfnamefont {D.}~\bibnamefont
  {{Wadekar}}}, \bibinfo {author} {\bibfnamefont {T.}~\bibnamefont
  {{Venumadhav}}}, \bibinfo {author} {\bibfnamefont {J.}~\bibnamefont
  {{Roulet}}}, \bibinfo {author} {\bibfnamefont {A.~K.}\ \bibnamefont
  {{Mehta}}}, \bibinfo {author} {\bibfnamefont {B.}~\bibnamefont {{Zackay}}},
  \bibinfo {author} {\bibfnamefont {J.}~\bibnamefont {{Mushkin}}}, \ and\
  \bibinfo {author} {\bibfnamefont {M.}~\bibnamefont {{Zaldarriaga}}},\ }\href
  {\doibase 10.1103/PhysRevD.110.044063} {\bibfield  {journal} {\bibinfo
  {journal} {\prd}\ }\textbf {\bibinfo {volume} {110}},\ \bibinfo {eid}
  {044063} (\bibinfo {year} {2024})},\ \Eprint
  {http://arxiv.org/abs/2405.17400} {arXiv:2405.17400 [gr-qc]} \BibitemShut
  {NoStop}%
\bibitem [{\citenamefont {Abbott}\ \emph
  {et~al.}(2016{\natexlab{a}})\citenamefont {Abbott} \emph
  {et~al.}}]{O1catalog_LVC2016}%
  \BibitemOpen
  \bibfield  {author} {\bibinfo {author} {\bibfnamefont {B.~P.}\ \bibnamefont
  {Abbott}} \emph {et~al.} (\bibinfo {collaboration} {LIGO Scientific
  Collaboration and Virgo Collaboration}),\ }\href {\doibase
  10.1103/PhysRevX.6.041015} {\bibfield  {journal} {\bibinfo  {journal} {Phys.
  Rev. X}\ }\textbf {\bibinfo {volume} {6}},\ \bibinfo {pages} {041015}
  (\bibinfo {year} {2016}{\natexlab{a}})}\BibitemShut {NoStop}%
\bibitem [{\citenamefont {Abbott}\ \emph {et~al.}(2019)\citenamefont {Abbott}
  \emph {et~al.}}]{gwtc1_o2catalog_LVC2018}%
  \BibitemOpen
  \bibfield  {author} {\bibinfo {author} {\bibfnamefont {B.~P.}\ \bibnamefont
  {Abbott}} \emph {et~al.} (\bibinfo {collaboration} {LIGO Scientific,
  Virgo}),\ }\href {\doibase 10.1103/PhysRevX.9.031040} {\bibfield  {journal}
  {\bibinfo  {journal} {Phys. Rev. X}\ }\textbf {\bibinfo {volume} {9}},\
  \bibinfo {pages} {031040} (\bibinfo {year} {2019})},\ \Eprint
  {http://arxiv.org/abs/1811.12907} {arXiv:1811.12907 [astro-ph.HE]}
  \BibitemShut {NoStop}%
\bibitem [{\citenamefont {{Abbott}}\ \emph {et~al.}(2021)\citenamefont
  {{Abbott}} \emph {et~al.}}]{lvc_o3a_gwtc2_catalog_2021}%
  \BibitemOpen
  \bibfield  {author} {\bibinfo {author} {\bibfnamefont {R.}~\bibnamefont
  {{Abbott}}} \emph {et~al.},\ }\href {\doibase 10.1103/PhysRevX.11.021053}
  {\bibfield  {journal} {\bibinfo  {journal} {Physical Review X}\ }\textbf
  {\bibinfo {volume} {11}},\ \bibinfo {eid} {021053} (\bibinfo {year}
  {2021})},\ \Eprint {http://arxiv.org/abs/2010.14527} {arXiv:2010.14527
  [gr-qc]} \BibitemShut {NoStop}%
\bibitem [{\citenamefont {{The LIGO Scientific Collaboration}}\ \emph
  {et~al.}(2021)\citenamefont {{The LIGO Scientific Collaboration}},
  \citenamefont {{the Virgo Collaboration}} \emph
  {et~al.}}]{lvc_o3a_deep_gwtc2_1_update_2021}%
  \BibitemOpen
  \bibfield  {author} {\bibinfo {author} {\bibnamefont {{The LIGO Scientific
  Collaboration}}}, \bibinfo {author} {\bibnamefont {{the Virgo
  Collaboration}}},  \emph {et~al.},\ }\href
  {https://ui.adsabs.harvard.edu/abs/2021arXiv210801045T} {\bibfield  {journal}
  {\bibinfo  {journal} {arXiv e-prints}\ ,\ \bibinfo {eid} {arXiv:2108.01045}}
  (\bibinfo {year} {2021})},\ \Eprint {http://arxiv.org/abs/2108.01045}
  {arXiv:2108.01045 [gr-qc]} \BibitemShut {NoStop}%
\bibitem [{\citenamefont {Abbott}\ \emph {et~al.}(2021)\citenamefont {Abbott}
  \emph {et~al.}}]{lvc_gwtc3_o3_ab_catalog_2021}%
  \BibitemOpen
  \bibfield  {author} {\bibinfo {author} {\bibfnamefont {R.}~\bibnamefont
  {Abbott}} \emph {et~al.} (\bibinfo {collaboration} {LIGO Scientific, VIRGO,
  KAGRA}),\ }\href@noop {} {\  (\bibinfo {year} {2021})},\ \Eprint
  {http://arxiv.org/abs/2111.03606} {arXiv:2111.03606 [gr-qc]} \BibitemShut
  {NoStop}%
\bibitem [{\citenamefont {Abbott}\ \emph {et~al.}(2022)\citenamefont {Abbott}
  \emph {et~al.}}]{LVK_O3_IMBH_search}%
  \BibitemOpen
  \bibfield  {author} {\bibinfo {author} {\bibfnamefont {R.}~\bibnamefont
  {Abbott}} \emph {et~al.} (\bibinfo {collaboration} {LIGO Scientific, VIRGO,
  KAGRA}),\ }\href {\doibase 10.1051/0004-6361/202141452} {\bibfield  {journal}
  {\bibinfo  {journal} {Astron. Astrophys.}\ }\textbf {\bibinfo {volume}
  {659}},\ \bibinfo {pages} {A84} (\bibinfo {year} {2022})},\ \Eprint
  {http://arxiv.org/abs/2105.15120} {arXiv:2105.15120 [astro-ph.HE]}
  \BibitemShut {NoStop}%
\bibitem [{\citenamefont {{Kumar}}\ and\ \citenamefont {{Dent}}(2024)}]{Kum24}%
  \BibitemOpen
  \bibfield  {author} {\bibinfo {author} {\bibfnamefont {P.}~\bibnamefont
  {{Kumar}}}\ and\ \bibinfo {author} {\bibfnamefont {T.}~\bibnamefont
  {{Dent}}},\ }\href {\doibase 10.48550/arXiv.2403.10439} {\bibfield  {journal}
  {\bibinfo  {journal} {arXiv e-prints}\ ,\ \bibinfo {eid} {arXiv:2403.10439}}
  (\bibinfo {year} {2024})},\ \Eprint {http://arxiv.org/abs/2403.10439}
  {arXiv:2403.10439 [gr-qc]} \BibitemShut {NoStop}%
\bibitem [{\citenamefont {Venumadhav}\ \emph {et~al.}(2019)\citenamefont
  {Venumadhav}, \citenamefont {Zackay}, \citenamefont {Roulet}, \citenamefont
  {Dai},\ and\ \citenamefont
  {Zaldarriaga}}]{ias_pipeline_o1_catalog_new_search_prd2019}%
  \BibitemOpen
  \bibfield  {author} {\bibinfo {author} {\bibfnamefont {T.}~\bibnamefont
  {Venumadhav}}, \bibinfo {author} {\bibfnamefont {B.}~\bibnamefont {Zackay}},
  \bibinfo {author} {\bibfnamefont {J.}~\bibnamefont {Roulet}}, \bibinfo
  {author} {\bibfnamefont {L.}~\bibnamefont {Dai}}, \ and\ \bibinfo {author}
  {\bibfnamefont {M.}~\bibnamefont {Zaldarriaga}},\ }\href {\doibase
  10.1103/PhysRevD.100.023011} {\bibfield  {journal} {\bibinfo  {journal}
  {Phys. Rev.}\ }\textbf {\bibinfo {volume} {D100}},\ \bibinfo {pages} {023011}
  (\bibinfo {year} {2019})},\ \Eprint {http://arxiv.org/abs/1902.10341}
  {arXiv:1902.10341 [astro-ph.IM]} \BibitemShut {NoStop}%
\bibitem [{\citenamefont {{Venumadhav}}\ \emph {et~al.}(2020)\citenamefont
  {{Venumadhav}}, \citenamefont {{Zackay}}, \citenamefont {{Roulet}},
  \citenamefont {{Dai}},\ and\ \citenamefont
  {{Zaldarriaga}}}]{ias_o2_pipeline_new_events_prd2020}%
  \BibitemOpen
  \bibfield  {author} {\bibinfo {author} {\bibfnamefont {T.}~\bibnamefont
  {{Venumadhav}}}, \bibinfo {author} {\bibfnamefont {B.}~\bibnamefont
  {{Zackay}}}, \bibinfo {author} {\bibfnamefont {J.}~\bibnamefont {{Roulet}}},
  \bibinfo {author} {\bibfnamefont {L.}~\bibnamefont {{Dai}}}, \ and\ \bibinfo
  {author} {\bibfnamefont {M.}~\bibnamefont {{Zaldarriaga}}},\ }\href {\doibase
  10.1103/PhysRevD.101.083030} {\bibfield  {journal} {\bibinfo  {journal}
  {\prd}\ }\textbf {\bibinfo {volume} {101}},\ \bibinfo {eid} {083030}
  (\bibinfo {year} {2020})},\ \Eprint {http://arxiv.org/abs/1904.07214}
  {arXiv:1904.07214 [astro-ph.HE]} \BibitemShut {NoStop}%
\bibitem [{\citenamefont {Olsen}\ \emph {et~al.}(2022)\citenamefont {Olsen},
  \citenamefont {Venumadhav}, \citenamefont {Mushkin}, \citenamefont {Roulet},
  \citenamefont {Zackay},\ and\ \citenamefont {Zaldarriaga}}]{Ols22_ias_o3a}%
  \BibitemOpen
  \bibfield  {author} {\bibinfo {author} {\bibfnamefont {S.}~\bibnamefont
  {Olsen}}, \bibinfo {author} {\bibfnamefont {T.}~\bibnamefont {Venumadhav}},
  \bibinfo {author} {\bibfnamefont {J.}~\bibnamefont {Mushkin}}, \bibinfo
  {author} {\bibfnamefont {J.}~\bibnamefont {Roulet}}, \bibinfo {author}
  {\bibfnamefont {B.}~\bibnamefont {Zackay}}, \ and\ \bibinfo {author}
  {\bibfnamefont {M.}~\bibnamefont {Zaldarriaga}},\ }\href {\doibase
  10.1103/PhysRevD.106.043009} {\bibfield  {journal} {\bibinfo  {journal}
  {Phys. Rev. D}\ }\textbf {\bibinfo {volume} {106}},\ \bibinfo {pages}
  {043009} (\bibinfo {year} {2022})}\BibitemShut {NoStop}%
\bibitem [{\citenamefont {Nitz}\ \emph {et~al.}(2019)\citenamefont {Nitz},
  \citenamefont {Capano}, \citenamefont {Nielsen}, \citenamefont {Reyes},
  \citenamefont {White}, \citenamefont {Brown},\ and\ \citenamefont
  {Krishnan}}]{NitzCatalog_1-OGC_o1_2018}%
  \BibitemOpen
  \bibfield  {author} {\bibinfo {author} {\bibfnamefont {A.~H.}\ \bibnamefont
  {Nitz}}, \bibinfo {author} {\bibfnamefont {C.}~\bibnamefont {Capano}},
  \bibinfo {author} {\bibfnamefont {A.~B.}\ \bibnamefont {Nielsen}}, \bibinfo
  {author} {\bibfnamefont {S.}~\bibnamefont {Reyes}}, \bibinfo {author}
  {\bibfnamefont {R.}~\bibnamefont {White}}, \bibinfo {author} {\bibfnamefont
  {D.~A.}\ \bibnamefont {Brown}}, \ and\ \bibinfo {author} {\bibfnamefont
  {B.}~\bibnamefont {Krishnan}},\ }\href {\doibase 10.3847/1538-4357/ab0108}
  {\bibfield  {journal} {\bibinfo  {journal} {The Astrophysical Journal}\
  }\textbf {\bibinfo {volume} {872}},\ \bibinfo {pages} {195} (\bibinfo {year}
  {2019})}\BibitemShut {NoStop}%
\bibitem [{\citenamefont {Nitz}\ \emph {et~al.}(2020)\citenamefont {Nitz},
  \citenamefont {Dent}, \citenamefont {Davies}, \citenamefont {Kumar},
  \citenamefont {Capano}, \citenamefont {Harry}, \citenamefont {Mozzon},
  \citenamefont {Nuttall}, \citenamefont {Lundgren},\ and\ \citenamefont
  {T\'apai}}]{NitzCatalog_2-OGC_o2_2020}%
  \BibitemOpen
  \bibfield  {author} {\bibinfo {author} {\bibfnamefont {A.~H.}\ \bibnamefont
  {Nitz}}, \bibinfo {author} {\bibfnamefont {T.}~\bibnamefont {Dent}}, \bibinfo
  {author} {\bibfnamefont {G.~S.}\ \bibnamefont {Davies}}, \bibinfo {author}
  {\bibfnamefont {S.}~\bibnamefont {Kumar}},  \emph {et~al.},\ }\href {\doibase
  10.3847/1538-4357/ab733f} {\bibfield  {journal} {\bibinfo  {journal}
  {Astrophys. J.}\ }\textbf {\bibinfo {volume} {891}},\ \bibinfo {pages} {123}
  (\bibinfo {year} {2020})},\ \Eprint {http://arxiv.org/abs/1910.05331}
  {arXiv:1910.05331 [astro-ph.HE]} \BibitemShut {NoStop}%
\bibitem [{\citenamefont {{Nitz}}\ \emph {et~al.}(2021)\citenamefont {{Nitz}},
  \citenamefont {{Capano}}, \citenamefont {{Kumar}}, \citenamefont {{Wang}},
  \citenamefont {{Kastha}}, \citenamefont {{Sch{\"a}fer}}, \citenamefont
  {{Dhurkunde}},\ and\ \citenamefont {{Cabero}}}]{nitz_o3a_3ogc_catalog_2021}%
  \BibitemOpen
  \bibfield  {author} {\bibinfo {author} {\bibfnamefont {A.~H.}\ \bibnamefont
  {{Nitz}}}, \bibinfo {author} {\bibfnamefont {C.~D.}\ \bibnamefont
  {{Capano}}}, \bibinfo {author} {\bibfnamefont {S.}~\bibnamefont {{Kumar}}},
  \bibinfo {author} {\bibfnamefont {Y.-F.}\ \bibnamefont {{Wang}}},  \emph
  {et~al.},\ }\href {https://ui.adsabs.harvard.edu/abs/2021arXiv210509151N}
  {\bibfield  {journal} {\bibinfo  {journal} {arXiv e-prints}\ ,\ \bibinfo
  {eid} {arXiv:2105.09151}} (\bibinfo {year} {2021})},\ \Eprint
  {http://arxiv.org/abs/2105.09151} {arXiv:2105.09151 [astro-ph.HE]}
  \BibitemShut {NoStop}%
\bibitem [{\citenamefont {Nitz}\ \emph {et~al.}(2021)\citenamefont {Nitz},
  \citenamefont {Kumar}, \citenamefont {Wang}, \citenamefont {Kastha},
  \citenamefont {Wu}, \citenamefont {Schäfer}, \citenamefont {Dhurkunde},\
  and\ \citenamefont {Capano}}]{nitz_4ogc_o3_ab_catalog_2021}%
  \BibitemOpen
  \bibfield  {author} {\bibinfo {author} {\bibfnamefont {A.~H.}\ \bibnamefont
  {Nitz}}, \bibinfo {author} {\bibfnamefont {S.}~\bibnamefont {Kumar}},
  \bibinfo {author} {\bibfnamefont {Y.-F.}\ \bibnamefont {Wang}}, \bibinfo
  {author} {\bibfnamefont {S.}~\bibnamefont {Kastha}},  \emph {et~al.},\
  }\href@noop {} {\enquote {\bibinfo {title} {4-{OGC}: Catalog of gravitational
  waves from compact-binary mergers},}\ } (\bibinfo {year} {2021}),\ \Eprint
  {http://arxiv.org/abs/2112.06878} {arXiv:2112.06878 [astro-ph.HE]}
  \BibitemShut {NoStop}%
\bibitem [{\citenamefont {{Chia}}\ \emph {et~al.}(2024)\citenamefont {{Chia}},
  \citenamefont {{Edwards}}, \citenamefont {{Wadekar}}, \citenamefont
  {{Zimmerman}}, \citenamefont {{Olsen}}, \citenamefont {{Roulet}},
  \citenamefont {{Venumadhav}}, \citenamefont {{Zackay}},\ and\ \citenamefont
  {{Zaldarriaga}}}]{Chi23}%
  \BibitemOpen
  \bibfield  {author} {\bibinfo {author} {\bibfnamefont {H.~S.}\ \bibnamefont
  {{Chia}}}, \bibinfo {author} {\bibfnamefont {T.~D.~P.}\ \bibnamefont
  {{Edwards}}}, \bibinfo {author} {\bibfnamefont {D.}~\bibnamefont
  {{Wadekar}}}, \bibinfo {author} {\bibfnamefont {A.}~\bibnamefont
  {{Zimmerman}}},  \emph {et~al.},\ }\href {\doibase
  10.1103/PhysRevD.110.063007} {\bibfield  {journal} {\bibinfo  {journal}
  {\prd}\ }\textbf {\bibinfo {volume} {110}},\ \bibinfo {eid} {063007}
  (\bibinfo {year} {2024})},\ \Eprint {http://arxiv.org/abs/2306.00050}
  {arXiv:2306.00050 [gr-qc]} \BibitemShut {NoStop}%
\bibitem [{\citenamefont {{Mehta}}\ \emph {et~al.}(2023)\citenamefont
  {{Mehta}}, \citenamefont {{Olsen}}, \citenamefont {{Wadekar}}, \citenamefont
  {{Roulet}}, \citenamefont {{Venumadhav}}, \citenamefont {{Mushkin}},
  \citenamefont {{Zackay}},\ and\ \citenamefont
  {{Zaldarriaga}}}]{Meh23_ias_o3b}%
  \BibitemOpen
  \bibfield  {author} {\bibinfo {author} {\bibfnamefont {A.~K.}\ \bibnamefont
  {{Mehta}}}, \bibinfo {author} {\bibfnamefont {S.}~\bibnamefont {{Olsen}}},
  \bibinfo {author} {\bibfnamefont {D.}~\bibnamefont {{Wadekar}}}, \bibinfo
  {author} {\bibfnamefont {J.}~\bibnamefont {{Roulet}}},  \emph {et~al.},\
  }\href {\doibase 10.48550/arXiv.2311.06061} {\bibfield  {journal} {\bibinfo
  {journal} {arXiv e-prints}\ ,\ \bibinfo {eid} {arXiv:2311.06061}} (\bibinfo
  {year} {2023})},\ \Eprint {http://arxiv.org/abs/2311.06061} {arXiv:2311.06061
  [gr-qc]} \BibitemShut {NoStop}%
\bibitem [{\citenamefont {{Koloniari}}\ \emph {et~al.}(2024)\citenamefont
  {{Koloniari}}, \citenamefont {{Koursoumpa}}, \citenamefont {{Nousi}},
  \citenamefont {{Lampropoulos}}, \citenamefont {{Passalis}}, \citenamefont
  {{Tefas}},\ and\ \citenamefont {{Stergioulas}}}]{Kol24_Ares_ML_Search}%
  \BibitemOpen
  \bibfield  {author} {\bibinfo {author} {\bibfnamefont {A.~E.}\ \bibnamefont
  {{Koloniari}}}, \bibinfo {author} {\bibfnamefont {E.~C.}\ \bibnamefont
  {{Koursoumpa}}}, \bibinfo {author} {\bibfnamefont {P.}~\bibnamefont
  {{Nousi}}}, \bibinfo {author} {\bibfnamefont {P.}~\bibnamefont
  {{Lampropoulos}}}, \bibinfo {author} {\bibfnamefont {N.}~\bibnamefont
  {{Passalis}}}, \bibinfo {author} {\bibfnamefont {A.}~\bibnamefont {{Tefas}}},
  \ and\ \bibinfo {author} {\bibfnamefont {N.}~\bibnamefont {{Stergioulas}}},\
  }\href@noop {} {\bibfield  {journal} {\bibinfo  {journal} {arXiv e-prints}\
  ,\ \bibinfo {eid} {arXiv:2407.07820}} (\bibinfo {year} {2024})},\ \Eprint
  {http://arxiv.org/abs/2407.07820} {arXiv:2407.07820 [gr-qc]} \BibitemShut
  {NoStop}%
\bibitem [{\citenamefont {{Vallisneri}}\ \emph {et~al.}(2015)\citenamefont
  {{Vallisneri}}, \citenamefont {{Kanner}}, \citenamefont {{Williams}},
  \citenamefont {{Weinstein}},\ and\ \citenamefont {{Stephens}}}]{GWOSC}%
  \BibitemOpen
  \bibfield  {author} {\bibinfo {author} {\bibfnamefont {M.}~\bibnamefont
  {{Vallisneri}}}, \bibinfo {author} {\bibfnamefont {J.}~\bibnamefont
  {{Kanner}}}, \bibinfo {author} {\bibfnamefont {R.}~\bibnamefont
  {{Williams}}}, \bibinfo {author} {\bibfnamefont {A.}~\bibnamefont
  {{Weinstein}}}, \ and\ \bibinfo {author} {\bibfnamefont {B.}~\bibnamefont
  {{Stephens}}},\ }in\ \href {\doibase 10.1088/1742-6596/610/1/012021} {\emph
  {\bibinfo {booktitle} {Journal of Physics Conference Series}}},\ \bibinfo
  {series} {Journal of Physics Conference Series}, Vol.\ \bibinfo {volume}
  {610}\ (\bibinfo {year} {2015})\ p.\ \bibinfo {pages} {012021},\ \Eprint
  {http://arxiv.org/abs/1410.4839} {arXiv:1410.4839 [gr-qc]} \BibitemShut
  {NoStop}%
\bibitem [{\citenamefont {{Roulet}}\ \emph {et~al.}(2022)\citenamefont
  {{Roulet}}, \citenamefont {{Olsen}}, \citenamefont {{Mushkin}}, \citenamefont
  {{Islam}}, \citenamefont {{Venumadhav}}, \citenamefont {{Zackay}},\ and\
  \citenamefont {{Zaldarriaga}}}]{Rou22_cogwheel}%
  \BibitemOpen
  \bibfield  {author} {\bibinfo {author} {\bibfnamefont {J.}~\bibnamefont
  {{Roulet}}}, \bibinfo {author} {\bibfnamefont {S.}~\bibnamefont {{Olsen}}},
  \bibinfo {author} {\bibfnamefont {J.}~\bibnamefont {{Mushkin}}}, \bibinfo
  {author} {\bibfnamefont {T.}~\bibnamefont {{Islam}}}, \bibinfo {author}
  {\bibfnamefont {T.}~\bibnamefont {{Venumadhav}}}, \bibinfo {author}
  {\bibfnamefont {B.}~\bibnamefont {{Zackay}}}, \ and\ \bibinfo {author}
  {\bibfnamefont {M.}~\bibnamefont {{Zaldarriaga}}},\ }\href {\doibase
  10.1103/PhysRevD.106.123015} {\bibfield  {journal} {\bibinfo  {journal}
  {\prd}\ }\textbf {\bibinfo {volume} {106}},\ \bibinfo {eid} {123015}
  (\bibinfo {year} {2022})},\ \Eprint {http://arxiv.org/abs/2207.03508}
  {arXiv:2207.03508 [gr-qc]} \BibitemShut {NoStop}%
\bibitem [{\citenamefont {Thorne}(1980)}]{Tho80}%
  \BibitemOpen
  \bibfield  {author} {\bibinfo {author} {\bibfnamefont {K.~S.}\ \bibnamefont
  {Thorne}},\ }\href {\doibase 10.1103/RevModPhys.52.299} {\bibfield  {journal}
  {\bibinfo  {journal} {Rev. Mod. Phys.}\ }\textbf {\bibinfo {volume} {52}},\
  \bibinfo {pages} {299} (\bibinfo {year} {1980})}\BibitemShut {NoStop}%
\bibitem [{\citenamefont {{Pan}}\ \emph {et~al.}(2011)\citenamefont {{Pan}},
  \citenamefont {{Buonanno}}, \citenamefont {{Boyle}}, \citenamefont
  {{Buchman}}, \citenamefont {{Kidder}}, \citenamefont {{Pfeiffer}},\ and\
  \citenamefont {{Scheel}}}]{Pan11}%
  \BibitemOpen
  \bibfield  {author} {\bibinfo {author} {\bibfnamefont {Y.}~\bibnamefont
  {{Pan}}}, \bibinfo {author} {\bibfnamefont {A.}~\bibnamefont {{Buonanno}}},
  \bibinfo {author} {\bibfnamefont {M.}~\bibnamefont {{Boyle}}}, \bibinfo
  {author} {\bibfnamefont {L.~T.}\ \bibnamefont {{Buchman}}}, \bibinfo {author}
  {\bibfnamefont {L.~E.}\ \bibnamefont {{Kidder}}}, \bibinfo {author}
  {\bibfnamefont {H.~P.}\ \bibnamefont {{Pfeiffer}}}, \ and\ \bibinfo {author}
  {\bibfnamefont {M.~A.}\ \bibnamefont {{Scheel}}},\ }\href {\doibase
  10.1103/PhysRevD.84.124052} {\bibfield  {journal} {\bibinfo  {journal}
  {\prd}\ }\textbf {\bibinfo {volume} {84}},\ \bibinfo {eid} {124052} (\bibinfo
  {year} {2011})},\ \Eprint {http://arxiv.org/abs/1106.1021} {arXiv:1106.1021
  [gr-qc]} \BibitemShut {NoStop}%
\bibitem [{\citenamefont {{Mills}}\ and\ \citenamefont
  {{Fairhurst}}(2021)}]{Mil21}%
  \BibitemOpen
  \bibfield  {author} {\bibinfo {author} {\bibfnamefont {C.}~\bibnamefont
  {{Mills}}}\ and\ \bibinfo {author} {\bibfnamefont {S.}~\bibnamefont
  {{Fairhurst}}},\ }\href {\doibase 10.1103/PhysRevD.103.024042} {\bibfield
  {journal} {\bibinfo  {journal} {\prd}\ }\textbf {\bibinfo {volume} {103}},\
  \bibinfo {eid} {024042} (\bibinfo {year} {2021})},\ \Eprint
  {http://arxiv.org/abs/2007.04313} {arXiv:2007.04313 [gr-qc]} \BibitemShut
  {NoStop}%
\bibitem [{\citenamefont {{Varma}}\ \emph {et~al.}(2014)\citenamefont
  {{Varma}}, \citenamefont {{Ajith}}, \citenamefont {{Husa}}, \citenamefont
  {{Bustillo}}, \citenamefont {{Hannam}},\ and\ \citenamefont
  {{P{\"u}rrer}}}]{Var14}%
  \BibitemOpen
  \bibfield  {author} {\bibinfo {author} {\bibfnamefont {V.}~\bibnamefont
  {{Varma}}}, \bibinfo {author} {\bibfnamefont {P.}~\bibnamefont {{Ajith}}},
  \bibinfo {author} {\bibfnamefont {S.}~\bibnamefont {{Husa}}}, \bibinfo
  {author} {\bibfnamefont {J.~C.}\ \bibnamefont {{Bustillo}}}, \bibinfo
  {author} {\bibfnamefont {M.}~\bibnamefont {{Hannam}}}, \ and\ \bibinfo
  {author} {\bibfnamefont {M.}~\bibnamefont {{P{\"u}rrer}}},\ }\href {\doibase
  10.1103/PhysRevD.90.124004} {\bibfield  {journal} {\bibinfo  {journal}
  {\prd}\ }\textbf {\bibinfo {volume} {90}},\ \bibinfo {eid} {124004} (\bibinfo
  {year} {2014})},\ \Eprint {http://arxiv.org/abs/1409.2349} {arXiv:1409.2349
  [gr-qc]} \BibitemShut {NoStop}%
\bibitem [{\citenamefont {{Abbott}}\ \emph
  {et~al.}(2020{\natexlab{a}})\citenamefont {{Abbott}}, \citenamefont {{LIGO
  Scientific Collaboration}},\ and\ \citenamefont {{Virgo
  Collaboration}}}]{GW190412}%
  \BibitemOpen
  \bibfield  {author} {\bibinfo {author} {\bibfnamefont {R.}~\bibnamefont
  {{Abbott}}}, \bibinfo {author} {\bibnamefont {{LIGO Scientific
  Collaboration}}}, \ and\ \bibinfo {author} {\bibnamefont {{Virgo
  Collaboration}}},\ }\href {\doibase 10.1103/PhysRevD.102.043015} {\bibfield
  {journal} {\bibinfo  {journal} {\prd}\ }\textbf {\bibinfo {volume} {102}},\
  \bibinfo {eid} {043015} (\bibinfo {year} {2020}{\natexlab{a}})},\ \Eprint
  {http://arxiv.org/abs/2004.08342} {arXiv:2004.08342 [astro-ph.HE]}
  \BibitemShut {NoStop}%
\bibitem [{\citenamefont {{Abbott}}\ \emph
  {et~al.}(2020{\natexlab{b}})\citenamefont {{Abbott}} \emph
  {et~al.}}]{GW190814}%
  \BibitemOpen
  \bibfield  {author} {\bibinfo {author} {\bibfnamefont {R.}~\bibnamefont
  {{Abbott}}} \emph {et~al.},\ }\href {\doibase 10.3847/2041-8213/ab960f}
  {\bibfield  {journal} {\bibinfo  {journal} {\apjl}\ }\textbf {\bibinfo
  {volume} {896}},\ \bibinfo {eid} {L44} (\bibinfo {year}
  {2020}{\natexlab{b}})},\ \Eprint {http://arxiv.org/abs/2006.12611}
  {arXiv:2006.12611 [astro-ph.HE]} \BibitemShut {NoStop}%
\bibitem [{\citenamefont {{Kastha}}\ \emph {et~al.}(2018)\citenamefont
  {{Kastha}}, \citenamefont {{Gupta}}, \citenamefont {{Arun}}, \citenamefont
  {{Sathyaprakash}},\ and\ \citenamefont {{Van Den Broeck}}}]{Kas18}%
  \BibitemOpen
  \bibfield  {author} {\bibinfo {author} {\bibfnamefont {S.}~\bibnamefont
  {{Kastha}}}, \bibinfo {author} {\bibfnamefont {A.}~\bibnamefont {{Gupta}}},
  \bibinfo {author} {\bibfnamefont {K.~G.}\ \bibnamefont {{Arun}}}, \bibinfo
  {author} {\bibfnamefont {B.~S.}\ \bibnamefont {{Sathyaprakash}}}, \ and\
  \bibinfo {author} {\bibfnamefont {C.}~\bibnamefont {{Van Den Broeck}}},\
  }\href {\doibase 10.1103/PhysRevD.98.124033} {\bibfield  {journal} {\bibinfo
  {journal} {\prd}\ }\textbf {\bibinfo {volume} {98}},\ \bibinfo {eid} {124033}
  (\bibinfo {year} {2018})},\ \Eprint {http://arxiv.org/abs/1809.10465}
  {arXiv:1809.10465 [gr-qc]} \BibitemShut {NoStop}%
\bibitem [{\citenamefont {{Kastha}}\ \emph {et~al.}(2019)\citenamefont
  {{Kastha}}, \citenamefont {{Gupta}}, \citenamefont {{Arun}}, \citenamefont
  {{Sathyaprakash}},\ and\ \citenamefont {{Van Den Broeck}}}]{Kas19}%
  \BibitemOpen
  \bibfield  {author} {\bibinfo {author} {\bibfnamefont {S.}~\bibnamefont
  {{Kastha}}}, \bibinfo {author} {\bibfnamefont {A.}~\bibnamefont {{Gupta}}},
  \bibinfo {author} {\bibfnamefont {K.~G.}\ \bibnamefont {{Arun}}}, \bibinfo
  {author} {\bibfnamefont {B.~S.}\ \bibnamefont {{Sathyaprakash}}}, \ and\
  \bibinfo {author} {\bibfnamefont {C.}~\bibnamefont {{Van Den Broeck}}},\
  }\href {\doibase 10.1103/PhysRevD.100.044007} {\bibfield  {journal} {\bibinfo
   {journal} {\prd}\ }\textbf {\bibinfo {volume} {100}},\ \bibinfo {eid}
  {044007} (\bibinfo {year} {2019})},\ \Eprint
  {http://arxiv.org/abs/1905.07277} {arXiv:1905.07277 [gr-qc]} \BibitemShut
  {NoStop}%
\bibitem [{\citenamefont {{Dhanpal}}\ \emph {et~al.}(2019)\citenamefont
  {{Dhanpal}}, \citenamefont {{Ghosh}}, \citenamefont {{Mehta}}, \citenamefont
  {{Ajith}},\ and\ \citenamefont {{Sathyaprakash}}}]{Dha19}%
  \BibitemOpen
  \bibfield  {author} {\bibinfo {author} {\bibfnamefont {S.}~\bibnamefont
  {{Dhanpal}}}, \bibinfo {author} {\bibfnamefont {A.}~\bibnamefont {{Ghosh}}},
  \bibinfo {author} {\bibfnamefont {A.~K.}\ \bibnamefont {{Mehta}}}, \bibinfo
  {author} {\bibfnamefont {P.}~\bibnamefont {{Ajith}}}, \ and\ \bibinfo
  {author} {\bibfnamefont {B.~S.}\ \bibnamefont {{Sathyaprakash}}},\ }\href
  {\doibase 10.1103/PhysRevD.99.104056} {\bibfield  {journal} {\bibinfo
  {journal} {\prd}\ }\textbf {\bibinfo {volume} {99}},\ \bibinfo {eid} {104056}
  (\bibinfo {year} {2019})},\ \Eprint {http://arxiv.org/abs/1804.03297}
  {arXiv:1804.03297 [gr-qc]} \BibitemShut {NoStop}%
\bibitem [{\citenamefont {{Islam}}\ \emph {et~al.}(2020)\citenamefont
  {{Islam}}, \citenamefont {{Mehta}}, \citenamefont {{Ghosh}}, \citenamefont
  {{Varma}}, \citenamefont {{Ajith}},\ and\ \citenamefont
  {{Sathyaprakash}}}]{Isl20}%
  \BibitemOpen
  \bibfield  {author} {\bibinfo {author} {\bibfnamefont {T.}~\bibnamefont
  {{Islam}}}, \bibinfo {author} {\bibfnamefont {A.~K.}\ \bibnamefont
  {{Mehta}}}, \bibinfo {author} {\bibfnamefont {A.}~\bibnamefont {{Ghosh}}},
  \bibinfo {author} {\bibfnamefont {V.}~\bibnamefont {{Varma}}}, \bibinfo
  {author} {\bibfnamefont {P.}~\bibnamefont {{Ajith}}}, \ and\ \bibinfo
  {author} {\bibfnamefont {B.~S.}\ \bibnamefont {{Sathyaprakash}}},\ }\href
  {\doibase 10.1103/PhysRevD.101.024032} {\bibfield  {journal} {\bibinfo
  {journal} {\prd}\ }\textbf {\bibinfo {volume} {101}},\ \bibinfo {eid}
  {024032} (\bibinfo {year} {2020})},\ \Eprint
  {http://arxiv.org/abs/1910.14259} {arXiv:1910.14259 [gr-qc]} \BibitemShut
  {NoStop}%
\bibitem [{\citenamefont {Capano}\ \emph {et~al.}(2021)\citenamefont {Capano},
  \citenamefont {Cabero}, \citenamefont {Westerweck}, \citenamefont {Abedi},
  \citenamefont {Kastha}, \citenamefont {Nitz}, \citenamefont {Nielsen},\ and\
  \citenamefont {Krishnan}}]{hm_observation_qn_ringdown_capano2021}%
  \BibitemOpen
  \bibfield  {author} {\bibinfo {author} {\bibfnamefont {C.~D.}\ \bibnamefont
  {Capano}}, \bibinfo {author} {\bibfnamefont {M.}~\bibnamefont {Cabero}},
  \bibinfo {author} {\bibfnamefont {J.}~\bibnamefont {Westerweck}}, \bibinfo
  {author} {\bibfnamefont {J.}~\bibnamefont {Abedi}},  \emph {et~al.},\ }\href
  {https://arxiv.org/abs/2105.05238} {\enquote {\bibinfo {title} {Observation
  of a multimode quasi-normal spectrum from a perturbed black hole},}\ }
  (\bibinfo {year} {2021}),\ \Eprint {http://arxiv.org/abs/2105.05238}
  {arXiv:2105.05238 [gr-qc]} \BibitemShut {NoStop}%
\bibitem [{\citenamefont {{Islam}}\ \emph
  {et~al.}(2021{\natexlab{a}})\citenamefont {{Islam}}, \citenamefont {{Field}},
  \citenamefont {{Haster}},\ and\ \citenamefont {{Smith}}}]{Isl21_IMRI}%
  \BibitemOpen
  \bibfield  {author} {\bibinfo {author} {\bibfnamefont {T.}~\bibnamefont
  {{Islam}}}, \bibinfo {author} {\bibfnamefont {S.~E.}\ \bibnamefont
  {{Field}}}, \bibinfo {author} {\bibfnamefont {C.-J.}\ \bibnamefont
  {{Haster}}}, \ and\ \bibinfo {author} {\bibfnamefont {R.}~\bibnamefont
  {{Smith}}},\ }\href {\doibase 10.1103/PhysRevD.104.084068} {\bibfield
  {journal} {\bibinfo  {journal} {\prd}\ }\textbf {\bibinfo {volume} {104}},\
  \bibinfo {eid} {084068} (\bibinfo {year} {2021}{\natexlab{a}})},\ \Eprint
  {http://arxiv.org/abs/2105.04422} {arXiv:2105.04422 [gr-qc]} \BibitemShut
  {NoStop}%
\bibitem [{\citenamefont {{Islam}}\ \emph
  {et~al.}(2021{\natexlab{b}})\citenamefont {{Islam}}, \citenamefont {{Field}},
  \citenamefont {{Haster}},\ and\ \citenamefont {{Smith}}}]{Isl21_190412}%
  \BibitemOpen
  \bibfield  {author} {\bibinfo {author} {\bibfnamefont {T.}~\bibnamefont
  {{Islam}}}, \bibinfo {author} {\bibfnamefont {S.~E.}\ \bibnamefont
  {{Field}}}, \bibinfo {author} {\bibfnamefont {C.-J.}\ \bibnamefont
  {{Haster}}}, \ and\ \bibinfo {author} {\bibfnamefont {R.}~\bibnamefont
  {{Smith}}},\ }\href {\doibase 10.1103/PhysRevD.103.104027} {\bibfield
  {journal} {\bibinfo  {journal} {\prd}\ }\textbf {\bibinfo {volume} {103}},\
  \bibinfo {eid} {104027} (\bibinfo {year} {2021}{\natexlab{b}})},\ \Eprint
  {http://arxiv.org/abs/2010.04848} {arXiv:2010.04848 [gr-qc]} \BibitemShut
  {NoStop}%
\bibitem [{\citenamefont {Calder\'on~Bustillo}\ \emph
  {et~al.}(2018)\citenamefont {Calder\'on~Bustillo}, \citenamefont {Clark},
  \citenamefont {Laguna},\ and\ \citenamefont {Shoemaker}}]{Cal18_HM_BH_kicks}%
  \BibitemOpen
  \bibfield  {author} {\bibinfo {author} {\bibfnamefont {J.}~\bibnamefont
  {Calder\'on~Bustillo}}, \bibinfo {author} {\bibfnamefont {J.~A.}\
  \bibnamefont {Clark}}, \bibinfo {author} {\bibfnamefont {P.}~\bibnamefont
  {Laguna}}, \ and\ \bibinfo {author} {\bibfnamefont {D.}~\bibnamefont
  {Shoemaker}},\ }\href {\doibase 10.1103/PhysRevLett.121.191102} {\bibfield
  {journal} {\bibinfo  {journal} {Phys. Rev. Lett.}\ }\textbf {\bibinfo
  {volume} {121}},\ \bibinfo {pages} {191102} (\bibinfo {year} {2018})},\
  \Eprint {http://arxiv.org/abs/1806.11160} {arXiv:1806.11160 [gr-qc]}
  \BibitemShut {NoStop}%
\bibitem [{\citenamefont {{London}}\ \emph {et~al.}(2018)\citenamefont
  {{London}}, \citenamefont {{Khan}}, \citenamefont {{Fauchon-Jones}},
  \citenamefont {{Garc{\'\i}a}}, \citenamefont {{Hannam}}, \citenamefont
  {{Husa}}, \citenamefont {{Jim{\'e}nez-Forteza}}, \citenamefont
  {{Kalaghatgi}}, \citenamefont {{Ohme}},\ and\ \citenamefont
  {{Pannarale}}}]{Lon18}%
  \BibitemOpen
  \bibfield  {author} {\bibinfo {author} {\bibfnamefont {L.}~\bibnamefont
  {{London}}}, \bibinfo {author} {\bibfnamefont {S.}~\bibnamefont {{Khan}}},
  \bibinfo {author} {\bibfnamefont {E.}~\bibnamefont {{Fauchon-Jones}}},
  \bibinfo {author} {\bibfnamefont {C.}~\bibnamefont {{Garc{\'\i}a}}},  \emph
  {et~al.},\ }\href {\doibase 10.1103/PhysRevLett.120.161102} {\bibfield
  {journal} {\bibinfo  {journal} {\prl}\ }\textbf {\bibinfo {volume} {120}},\
  \bibinfo {eid} {161102} (\bibinfo {year} {2018})},\ \Eprint
  {http://arxiv.org/abs/1708.00404} {arXiv:1708.00404 [gr-qc]} \BibitemShut
  {NoStop}%
\bibitem [{\citenamefont {{Garc{\'\i}a-Quir{\'o}s}}\ \emph
  {et~al.}(2020)\citenamefont {{Garc{\'\i}a-Quir{\'o}s}}, \citenamefont
  {{Colleoni}}, \citenamefont {{Husa}}, \citenamefont {{Estell{\'e}s}},
  \citenamefont {{Pratten}}, \citenamefont {{Ramos-Buades}}, \citenamefont
  {{Mateu-Lucena}},\ and\ \citenamefont {{Jaume}}}]{Gar20}%
  \BibitemOpen
  \bibfield  {author} {\bibinfo {author} {\bibfnamefont {C.}~\bibnamefont
  {{Garc{\'\i}a-Quir{\'o}s}}}, \bibinfo {author} {\bibfnamefont
  {M.}~\bibnamefont {{Colleoni}}}, \bibinfo {author} {\bibfnamefont
  {S.}~\bibnamefont {{Husa}}}, \bibinfo {author} {\bibfnamefont
  {H.}~\bibnamefont {{Estell{\'e}s}}},  \emph {et~al.},\ }\href {\doibase
  10.1103/PhysRevD.102.064002} {\bibfield  {journal} {\bibinfo  {journal}
  {\prd}\ }\textbf {\bibinfo {volume} {102}},\ \bibinfo {eid} {064002}
  (\bibinfo {year} {2020})},\ \Eprint {http://arxiv.org/abs/2001.10914}
  {arXiv:2001.10914 [gr-qc]} \BibitemShut {NoStop}%
\bibitem [{\citenamefont {{Mehta}}\ \emph {et~al.}(2017)\citenamefont
  {{Mehta}}, \citenamefont {{Mishra}}, \citenamefont {{Varma}},\ and\
  \citenamefont {{Ajith}}}]{Meh17}%
  \BibitemOpen
  \bibfield  {author} {\bibinfo {author} {\bibfnamefont {A.~K.}\ \bibnamefont
  {{Mehta}}}, \bibinfo {author} {\bibfnamefont {C.~K.}\ \bibnamefont
  {{Mishra}}}, \bibinfo {author} {\bibfnamefont {V.}~\bibnamefont {{Varma}}}, \
  and\ \bibinfo {author} {\bibfnamefont {P.}~\bibnamefont {{Ajith}}},\ }\href
  {\doibase 10.1103/PhysRevD.96.124010} {\bibfield  {journal} {\bibinfo
  {journal} {\prd}\ }\textbf {\bibinfo {volume} {96}},\ \bibinfo {eid} {124010}
  (\bibinfo {year} {2017})},\ \Eprint {http://arxiv.org/abs/1708.03501}
  {arXiv:1708.03501 [gr-qc]} \BibitemShut {NoStop}%
\bibitem [{\citenamefont {{Ramos-Buades}}\ \emph {et~al.}(2023)\citenamefont
  {{Ramos-Buades}}, \citenamefont {{Buonanno}}, \citenamefont {{Estell{\'e}s}},
  \citenamefont {{Khalil}}, \citenamefont {{Mihaylov}}, \citenamefont
  {{Ossokine}}, \citenamefont {{Pompili}},\ and\ \citenamefont
  {{Shiferaw}}}]{Ram24_SEOBv5PHM}%
  \BibitemOpen
  \bibfield  {author} {\bibinfo {author} {\bibfnamefont {A.}~\bibnamefont
  {{Ramos-Buades}}}, \bibinfo {author} {\bibfnamefont {A.}~\bibnamefont
  {{Buonanno}}}, \bibinfo {author} {\bibfnamefont {H.}~\bibnamefont
  {{Estell{\'e}s}}}, \bibinfo {author} {\bibfnamefont {M.}~\bibnamefont
  {{Khalil}}},  \emph {et~al.},\ }\href {\doibase 10.1103/PhysRevD.108.124037}
  {\bibfield  {journal} {\bibinfo  {journal} {\prd}\ }\textbf {\bibinfo
  {volume} {108}},\ \bibinfo {eid} {124037} (\bibinfo {year} {2023})},\ \Eprint
  {http://arxiv.org/abs/2303.18046} {arXiv:2303.18046 [gr-qc]} \BibitemShut
  {NoStop}%
\bibitem [{\citenamefont {{Chandra}}\ \emph {et~al.}(2022)\citenamefont
  {{Chandra}}, \citenamefont {{Calder{\'o}n Bustillo}}, \citenamefont {{Pai}},\
  and\ \citenamefont {{Harry}}}]{Cha22}%
  \BibitemOpen
  \bibfield  {author} {\bibinfo {author} {\bibfnamefont {K.}~\bibnamefont
  {{Chandra}}}, \bibinfo {author} {\bibfnamefont {J.}~\bibnamefont
  {{Calder{\'o}n Bustillo}}}, \bibinfo {author} {\bibfnamefont
  {A.}~\bibnamefont {{Pai}}}, \ and\ \bibinfo {author} {\bibfnamefont
  {I.}~\bibnamefont {{Harry}}},\ }\href@noop {} {\bibfield  {journal} {\bibinfo
   {journal} {arXiv e-prints}\ ,\ \bibinfo {eid} {arXiv:2207.01654}} (\bibinfo
  {year} {2022})},\ \Eprint {http://arxiv.org/abs/2207.01654} {arXiv:2207.01654
  [gr-qc]} \BibitemShut {NoStop}%
\bibitem [{\citenamefont {Harry}\ \emph {et~al.}(2018)\citenamefont {Harry},
  \citenamefont {Bustillo},\ and\ \citenamefont {Nitz}}]{Har18}%
  \BibitemOpen
  \bibfield  {author} {\bibinfo {author} {\bibfnamefont {I.}~\bibnamefont
  {Harry}}, \bibinfo {author} {\bibfnamefont {J.~C.}\ \bibnamefont {Bustillo}},
  \ and\ \bibinfo {author} {\bibfnamefont {A.}~\bibnamefont {Nitz}},\ }\href
  {\doibase 10.1103/PhysRevD.97.023004} {\bibfield  {journal} {\bibinfo
  {journal} {Phys. Rev. D}\ }\textbf {\bibinfo {volume} {97}},\ \bibinfo
  {pages} {023004} (\bibinfo {year} {2018})}\BibitemShut {NoStop}%
\bibitem [{\citenamefont {{Schmidt}}\ \emph {et~al.}(2023)\citenamefont
  {{Schmidt}}, \citenamefont {{Gadre}},\ and\ \citenamefont
  {{Caudill}}}]{Sch23_NF_TemplateBank}%
  \BibitemOpen
  \bibfield  {author} {\bibinfo {author} {\bibfnamefont {S.}~\bibnamefont
  {{Schmidt}}}, \bibinfo {author} {\bibfnamefont {B.}~\bibnamefont {{Gadre}}},
  \ and\ \bibinfo {author} {\bibfnamefont {S.}~\bibnamefont {{Caudill}}},\
  }\href {\doibase 10.48550/arXiv.2302.00436} {\bibfield  {journal} {\bibinfo
  {journal} {arXiv e-prints}\ ,\ \bibinfo {eid} {arXiv:2302.00436}} (\bibinfo
  {year} {2023})},\ \Eprint {http://arxiv.org/abs/2302.00436} {arXiv:2302.00436
  [gr-qc]} \BibitemShut {NoStop}%
\bibitem [{\citenamefont {Capano}\ \emph {et~al.}(2014)\citenamefont {Capano},
  \citenamefont {Pan},\ and\ \citenamefont {Buonanno}}]{Cap14}%
  \BibitemOpen
  \bibfield  {author} {\bibinfo {author} {\bibfnamefont {C.}~\bibnamefont
  {Capano}}, \bibinfo {author} {\bibfnamefont {Y.}~\bibnamefont {Pan}}, \ and\
  \bibinfo {author} {\bibfnamefont {A.}~\bibnamefont {Buonanno}},\ }\href
  {\doibase 10.1103/physrevd.89.102003} {\bibfield  {journal} {\bibinfo
  {journal} {Physical Review D}\ }\textbf {\bibinfo {volume} {89}} (\bibinfo
  {year} {2014}),\ 10.1103/physrevd.89.102003}\BibitemShut {NoStop}%
\bibitem [{\citenamefont {Neyman}\ and\ \citenamefont
  {Pearson}(1933)}]{neymanpearson}%
  \BibitemOpen
  \bibfield  {author} {\bibinfo {author} {\bibfnamefont {J.}~\bibnamefont
  {Neyman}}\ and\ \bibinfo {author} {\bibfnamefont {E.~S.}\ \bibnamefont
  {Pearson}},\ }\href@noop {} {\bibfield  {journal} {\bibinfo  {journal}
  {Philosophical Transactions of the Royal Society of London. Series A,
  Containing Papers of a Mathematical or Physical Character}\ }\textbf
  {\bibinfo {volume} {231}},\ \bibinfo {pages} {289} (\bibinfo {year}
  {1933})}\BibitemShut {NoStop}%
\bibitem [{\citenamefont {{Nitz}}\ \emph {et~al.}(2017)\citenamefont {{Nitz}},
  \citenamefont {{Dent}}, \citenamefont {{Dal Canton}}, \citenamefont
  {{Fairhurst}},\ and\ \citenamefont {{Brown}}}]{CoherentScore}%
  \BibitemOpen
  \bibfield  {author} {\bibinfo {author} {\bibfnamefont {A.~H.}\ \bibnamefont
  {{Nitz}}}, \bibinfo {author} {\bibfnamefont {T.}~\bibnamefont {{Dent}}},
  \bibinfo {author} {\bibfnamefont {T.}~\bibnamefont {{Dal Canton}}}, \bibinfo
  {author} {\bibfnamefont {S.}~\bibnamefont {{Fairhurst}}}, \ and\ \bibinfo
  {author} {\bibfnamefont {D.~A.}\ \bibnamefont {{Brown}}},\ }\href {\doibase
  10.3847/1538-4357/aa8f50} {\bibfield  {journal} {\bibinfo  {journal} {\apj}\
  }\textbf {\bibinfo {volume} {849}},\ \bibinfo {eid} {118} (\bibinfo {year}
  {2017})},\ \Eprint {http://arxiv.org/abs/1705.01513} {arXiv:1705.01513
  [gr-qc]} \BibitemShut {NoStop}%
\bibitem [{\citenamefont {{Roulet}}\ \emph {et~al.}(2024)\citenamefont
  {{Roulet}}, \citenamefont {{Mushkin}}, \citenamefont {{Wadekar}},
  \citenamefont {{Venumadhav}}, \citenamefont {{Zackay}},\ and\ \citenamefont
  {{Zaldarriaga}}}]{Rou23_CoherentScore}%
  \BibitemOpen
  \bibfield  {author} {\bibinfo {author} {\bibfnamefont {J.}~\bibnamefont
  {{Roulet}}}, \bibinfo {author} {\bibfnamefont {J.}~\bibnamefont {{Mushkin}}},
  \bibinfo {author} {\bibfnamefont {D.}~\bibnamefont {{Wadekar}}}, \bibinfo
  {author} {\bibfnamefont {T.}~\bibnamefont {{Venumadhav}}}, \bibinfo {author}
  {\bibfnamefont {B.}~\bibnamefont {{Zackay}}}, \ and\ \bibinfo {author}
  {\bibfnamefont {M.}~\bibnamefont {{Zaldarriaga}}},\ }\href {\doibase
  10.48550/arXiv.2404.02435} {\bibfield  {journal} {\bibinfo  {journal} {arXiv
  e-prints}\ ,\ \bibinfo {eid} {arXiv:2404.02435}} (\bibinfo {year} {2024})},\
  \Eprint {http://arxiv.org/abs/2404.02435} {arXiv:2404.02435 [gr-qc]}
  \BibitemShut {NoStop}%
\bibitem [{\citenamefont {Leslie}\ \emph {et~al.}(2021)\citenamefont {Leslie},
  \citenamefont {Dai},\ and\ \citenamefont
  {Pratten}}]{relative_binning_higher_modes}%
  \BibitemOpen
  \bibfield  {author} {\bibinfo {author} {\bibfnamefont {N.}~\bibnamefont
  {Leslie}}, \bibinfo {author} {\bibfnamefont {L.}~\bibnamefont {Dai}}, \ and\
  \bibinfo {author} {\bibfnamefont {G.}~\bibnamefont {Pratten}},\ }\href@noop
  {} {\enquote {\bibinfo {title} {Mode-by-mode relative binning: Fast
  likelihood estimation for gravitational waveforms with spin-orbit precession
  and multiple harmonics},}\ } (\bibinfo {year} {2021}),\ \Eprint
  {http://arxiv.org/abs/2109.09872} {arXiv:2109.09872 [astro-ph.IM]}
  \BibitemShut {NoStop}%
\bibitem [{\citenamefont {Wadekar}\ \emph {et~al.}({\natexlab{a}})\citenamefont
  {Wadekar} \emph {et~al.}}]{Wad25_TIER_GW}%
  \BibitemOpen
  \bibfield  {author} {\bibinfo {author} {\bibfnamefont {D.}~\bibnamefont
  {Wadekar}} \emph {et~al.},\ }\href@noop {} {\bibfield  {journal} {\bibinfo
  {journal} {in preparation}\ } ({\natexlab{a}})}\BibitemShut {NoStop}%
\bibitem [{\citenamefont {Zackay}\ \emph {et~al.}(2021)\citenamefont {Zackay},
  \citenamefont {Venumadhav}, \citenamefont {Roulet}, \citenamefont {Dai},\
  and\ \citenamefont {Zaldarriaga}}]{psd_drift}%
  \BibitemOpen
  \bibfield  {author} {\bibinfo {author} {\bibfnamefont {B.}~\bibnamefont
  {Zackay}}, \bibinfo {author} {\bibfnamefont {T.}~\bibnamefont {Venumadhav}},
  \bibinfo {author} {\bibfnamefont {J.}~\bibnamefont {Roulet}}, \bibinfo
  {author} {\bibfnamefont {L.}~\bibnamefont {Dai}}, \ and\ \bibinfo {author}
  {\bibfnamefont {M.}~\bibnamefont {Zaldarriaga}},\ }\href {\doibase
  10.1103/PhysRevD.104.063034} {\bibfield  {journal} {\bibinfo  {journal}
  {Phys. Rev. D}\ }\textbf {\bibinfo {volume} {104}},\ \bibinfo {pages}
  {063034} (\bibinfo {year} {2021})}\BibitemShut {NoStop}%
\bibitem [{\citenamefont {Abbott}\ \emph
  {et~al.}(2016{\natexlab{b}})\citenamefont {Abbott} \emph {et~al.}}]{LIGO_O1}%
  \BibitemOpen
  \bibfield  {author} {\bibinfo {author} {\bibfnamefont {B.~P.}\ \bibnamefont
  {Abbott}} \emph {et~al.} (\bibinfo {collaboration} {LIGO Scientific,
  Virgo}),\ }\href {\doibase 10.1103/PhysRevD.93.122003} {\bibfield  {journal}
  {\bibinfo  {journal} {Phys. Rev. D}\ }\textbf {\bibinfo {volume} {93}},\
  \bibinfo {pages} {122003} (\bibinfo {year} {2016}{\natexlab{b}})},\ \Eprint
  {http://arxiv.org/abs/1602.03839} {arXiv:1602.03839 [gr-qc]} \BibitemShut
  {NoStop}%
\bibitem [{\citenamefont {Pratten}\ \emph {et~al.}(2020)\citenamefont
  {Pratten}, \citenamefont {Husa}, \citenamefont {Garcia-Quiros}, \citenamefont
  {Colleoni}, \citenamefont {Ramos-Buades}, \citenamefont {Estelles},\ and\
  \citenamefont {Jaume}}]{Pratten:2020fqn}%
  \BibitemOpen
  \bibfield  {author} {\bibinfo {author} {\bibfnamefont {G.}~\bibnamefont
  {Pratten}}, \bibinfo {author} {\bibfnamefont {S.}~\bibnamefont {Husa}},
  \bibinfo {author} {\bibfnamefont {C.}~\bibnamefont {Garcia-Quiros}}, \bibinfo
  {author} {\bibfnamefont {M.}~\bibnamefont {Colleoni}}, \bibinfo {author}
  {\bibfnamefont {A.}~\bibnamefont {Ramos-Buades}}, \bibinfo {author}
  {\bibfnamefont {H.}~\bibnamefont {Estelles}}, \ and\ \bibinfo {author}
  {\bibfnamefont {R.}~\bibnamefont {Jaume}},\ }\href {\doibase
  10.1103/PhysRevD.102.064001} {\bibfield  {journal} {\bibinfo  {journal}
  {Phys. Rev. D}\ }\textbf {\bibinfo {volume} {102}},\ \bibinfo {pages}
  {064001} (\bibinfo {year} {2020})},\ \Eprint
  {http://arxiv.org/abs/2001.11412} {arXiv:2001.11412 [gr-qc]} \BibitemShut
  {NoStop}%
\bibitem [{\citenamefont {Roulet}\ \emph {et~al.}(2020)\citenamefont {Roulet},
  \citenamefont {Venumadhav}, \citenamefont {Zackay}, \citenamefont {Dai},\
  and\ \citenamefont {Zaldarriaga}}]{ias_popO2_Roulet_2020}%
  \BibitemOpen
  \bibfield  {author} {\bibinfo {author} {\bibfnamefont {J.}~\bibnamefont
  {Roulet}}, \bibinfo {author} {\bibfnamefont {T.}~\bibnamefont {Venumadhav}},
  \bibinfo {author} {\bibfnamefont {B.}~\bibnamefont {Zackay}}, \bibinfo
  {author} {\bibfnamefont {L.}~\bibnamefont {Dai}}, \ and\ \bibinfo {author}
  {\bibfnamefont {M.}~\bibnamefont {Zaldarriaga}},\ }\href {\doibase
  10.1103/physrevd.102.123022} {\bibfield  {journal} {\bibinfo  {journal}
  {Physical Review D}\ }\textbf {\bibinfo {volume} {102}} (\bibinfo {year}
  {2020}),\ 10.1103/physrevd.102.123022}\BibitemShut {NoStop}%
\bibitem [{\citenamefont {{Fishbach}}\ and\ \citenamefont
  {{Holz}}(2017)}]{Fis17}%
  \BibitemOpen
  \bibfield  {author} {\bibinfo {author} {\bibfnamefont {M.}~\bibnamefont
  {{Fishbach}}}\ and\ \bibinfo {author} {\bibfnamefont {D.~E.}\ \bibnamefont
  {{Holz}}},\ }\href {\doibase 10.3847/2041-8213/aa9bf6} {\bibfield  {journal}
  {\bibinfo  {journal} {\apjl}\ }\textbf {\bibinfo {volume} {851}},\ \bibinfo
  {eid} {L25} (\bibinfo {year} {2017})},\ \Eprint
  {http://arxiv.org/abs/1709.08584} {arXiv:1709.08584 [astro-ph.HE]}
  \BibitemShut {NoStop}%
\bibitem [{\citenamefont {Wadekar}\ \emph {et~al.}({\natexlab{b}})\citenamefont
  {Wadekar} \emph {et~al.}}]{Wad25_NormFlow_HM_search}%
  \BibitemOpen
  \bibfield  {author} {\bibinfo {author} {\bibfnamefont {D.}~\bibnamefont
  {Wadekar}} \emph {et~al.},\ }\href@noop {} {\bibfield  {journal} {\bibinfo
  {journal} {in preparation}\ } ({\natexlab{b}})}\BibitemShut {NoStop}%
\bibitem [{\citenamefont {Lange}(2023)}]{Lange2023}%
  \BibitemOpen
  \bibfield  {author} {\bibinfo {author} {\bibfnamefont {J.~U.}\ \bibnamefont
  {Lange}},\ }\href@noop {} {\enquote {\bibinfo {title} {Nautilus: boosting
  bayesian importance nested sampling with deep learning},}\ } (\bibinfo {year}
  {2023}),\ \Eprint {http://arxiv.org/abs/2306.16923} {arXiv:2306.16923
  [astro-ph.IM]} \BibitemShut {NoStop}%
\bibitem [{\citenamefont {Zackay}\ \emph {et~al.}(2018)\citenamefont {Zackay},
  \citenamefont {Dai},\ and\ \citenamefont {Venumadhav}}]{relative_binning}%
  \BibitemOpen
  \bibfield  {author} {\bibinfo {author} {\bibfnamefont {B.}~\bibnamefont
  {Zackay}}, \bibinfo {author} {\bibfnamefont {L.}~\bibnamefont {Dai}}, \ and\
  \bibinfo {author} {\bibfnamefont {T.}~\bibnamefont {Venumadhav}},\
  }\href@noop {} {\  (\bibinfo {year} {2018})},\ \Eprint
  {http://arxiv.org/abs/1806.08792} {arXiv:1806.08792 [astro-ph.IM]}
  \BibitemShut {NoStop}%
\bibitem [{\citenamefont {Ade}\ \emph {et~al.}(2016)\citenamefont {Ade} \emph
  {et~al.}}]{cosmology_planck2015}%
  \BibitemOpen
  \bibfield  {author} {\bibinfo {author} {\bibfnamefont {P.~A.~R.}\
  \bibnamefont {Ade}} \emph {et~al.} (\bibinfo {collaboration} {Planck}),\
  }\href {\doibase 10.1051/0004-6361/201525830} {\bibfield  {journal} {\bibinfo
   {journal} {Astron. Astrophys.}\ }\textbf {\bibinfo {volume} {594}},\
  \bibinfo {pages} {A13} (\bibinfo {year} {2016})},\ \Eprint
  {http://arxiv.org/abs/1502.01589} {arXiv:1502.01589 [astro-ph.CO]}
  \BibitemShut {NoStop}%
\bibitem [{\citenamefont {{Yu}}\ \emph {et~al.}(2023)\citenamefont {{Yu}},
  \citenamefont {{Roulet}}, \citenamefont {{Venumadhav}}, \citenamefont
  {{Zackay}},\ and\ \citenamefont {{Zaldarriaga}}}]{Yu23_XODE}%
  \BibitemOpen
  \bibfield  {author} {\bibinfo {author} {\bibfnamefont {H.}~\bibnamefont
  {{Yu}}}, \bibinfo {author} {\bibfnamefont {J.}~\bibnamefont {{Roulet}}},
  \bibinfo {author} {\bibfnamefont {T.}~\bibnamefont {{Venumadhav}}}, \bibinfo
  {author} {\bibfnamefont {B.}~\bibnamefont {{Zackay}}}, \ and\ \bibinfo
  {author} {\bibfnamefont {M.}~\bibnamefont {{Zaldarriaga}}},\ }\href {\doibase
  10.1103/PhysRevD.108.064059} {\bibfield  {journal} {\bibinfo  {journal}
  {\prd}\ }\textbf {\bibinfo {volume} {108}},\ \bibinfo {eid} {064059}
  (\bibinfo {year} {2023})},\ \Eprint {http://arxiv.org/abs/2306.08774}
  {arXiv:2306.08774 [gr-qc]} \BibitemShut {NoStop}%
\bibitem [{\citenamefont {Mishra}\ \emph {et~al.}(2025)\citenamefont {Mishra},
  \citenamefont {Bhaumik}, \citenamefont {Gayathri}, \citenamefont
  {Szczepa{\'n}czyk}, \citenamefont {Bartos},\ and\ \citenamefont
  {Klimenko}}]{Mis24_CWB}%
  \BibitemOpen
  \bibfield  {author} {\bibinfo {author} {\bibfnamefont {T.}~\bibnamefont
  {Mishra}}, \bibinfo {author} {\bibfnamefont {S.}~\bibnamefont {Bhaumik}},
  \bibinfo {author} {\bibfnamefont {V.}~\bibnamefont {Gayathri}}, \bibinfo
  {author} {\bibfnamefont {M.~J.}\ \bibnamefont {Szczepa{\'n}czyk}}, \bibinfo
  {author} {\bibfnamefont {I.}~\bibnamefont {Bartos}}, \ and\ \bibinfo {author}
  {\bibfnamefont {S.}~\bibnamefont {Klimenko}},\ }\href {\doibase
  10.1103/PhysRevD.111.023054} {\bibfield  {journal} {\bibinfo  {journal}
  {Phys. Rev. D}\ }\textbf {\bibinfo {volume} {111}},\ \bibinfo {pages}
  {023054} (\bibinfo {year} {2025})},\ \Eprint
  {http://arxiv.org/abs/2410.15191} {arXiv:2410.15191 [astro-ph.HE]}
  \BibitemShut {NoStop}%
\bibitem [{\citenamefont {Pratten}\ \emph {et~al.}(2021)\citenamefont
  {Pratten}, \citenamefont {Garc\'{\i}a-Quir\'os}, \citenamefont {Colleoni},
  \citenamefont {Ramos-Buades}, \citenamefont {Estell\'es}, \citenamefont
  {Mateu-Lucena}, \citenamefont {Jaume}, \citenamefont {Haney}, \citenamefont
  {Keitel}, \citenamefont {Thompson},\ and\ \citenamefont {Husa}}]{Pra21_XPHM}%
  \BibitemOpen
  \bibfield  {author} {\bibinfo {author} {\bibfnamefont {G.}~\bibnamefont
  {Pratten}}, \bibinfo {author} {\bibfnamefont {C.}~\bibnamefont
  {Garc\'{\i}a-Quir\'os}}, \bibinfo {author} {\bibfnamefont {M.}~\bibnamefont
  {Colleoni}}, \bibinfo {author} {\bibfnamefont {A.}~\bibnamefont
  {Ramos-Buades}},  \emph {et~al.},\ }\href {\doibase
  10.1103/PhysRevD.103.104056} {\bibfield  {journal} {\bibinfo  {journal}
  {Phys. Rev. D}\ }\textbf {\bibinfo {volume} {103}},\ \bibinfo {pages}
  {104056} (\bibinfo {year} {2021})}\BibitemShut {NoStop}%
\bibitem [{\citenamefont {{Chatziioannou}}\ \emph {et~al.}(2017)\citenamefont
  {{Chatziioannou}}, \citenamefont {{Klein}}, \citenamefont {{Yunes}},\ and\
  \citenamefont {{Cornish}}}]{Cha17_MSA}%
  \BibitemOpen
  \bibfield  {author} {\bibinfo {author} {\bibfnamefont {K.}~\bibnamefont
  {{Chatziioannou}}}, \bibinfo {author} {\bibfnamefont {A.}~\bibnamefont
  {{Klein}}}, \bibinfo {author} {\bibfnamefont {N.}~\bibnamefont {{Yunes}}}, \
  and\ \bibinfo {author} {\bibfnamefont {N.}~\bibnamefont {{Cornish}}},\ }\href
  {\doibase 10.1103/PhysRevD.95.104004} {\bibfield  {journal} {\bibinfo
  {journal} {\prd}\ }\textbf {\bibinfo {volume} {95}},\ \bibinfo {eid} {104004}
  (\bibinfo {year} {2017})},\ \Eprint {http://arxiv.org/abs/1703.03967}
  {arXiv:1703.03967 [gr-qc]} \BibitemShut {NoStop}%
\bibitem [{\citenamefont {{Islam}}\ \emph {et~al.}(2023)\citenamefont
  {{Islam}}, \citenamefont {{Vajpeyi}}, \citenamefont {{Shaik}}, \citenamefont
  {{Haster}}, \citenamefont {{Varma}}, \citenamefont {{Field}}, \citenamefont
  {{Lange}}, \citenamefont {{O'Shaughnessy}},\ and\ \citenamefont
  {{Smith}}}]{Isl23_NRsurrogate}%
  \BibitemOpen
  \bibfield  {author} {\bibinfo {author} {\bibfnamefont {T.}~\bibnamefont
  {{Islam}}}, \bibinfo {author} {\bibfnamefont {A.}~\bibnamefont {{Vajpeyi}}},
  \bibinfo {author} {\bibfnamefont {F.~H.}\ \bibnamefont {{Shaik}}}, \bibinfo
  {author} {\bibfnamefont {C.-J.}\ \bibnamefont {{Haster}}},  \emph {et~al.},\
  }\href {\doibase 10.48550/arXiv.2309.14473} {\bibfield  {journal} {\bibinfo
  {journal} {arXiv e-prints}\ ,\ \bibinfo {eid} {arXiv:2309.14473}} (\bibinfo
  {year} {2023})},\ \Eprint {http://arxiv.org/abs/2309.14473} {arXiv:2309.14473
  [gr-qc]} \BibitemShut {NoStop}%
\bibitem [{\citenamefont {Sachdev}\ \emph {et~al.}(2019)\citenamefont
  {Sachdev}, \citenamefont {Caudill} \emph {et~al.}}]{gstlal}%
  \BibitemOpen
  \bibfield  {author} {\bibinfo {author} {\bibfnamefont {S.}~\bibnamefont
  {Sachdev}}, \bibinfo {author} {\bibfnamefont {S.}~\bibnamefont {Caudill}},
  \emph {et~al.},\ }\href {http://arxiv.org/abs/1901.08580v1} {\  (\bibinfo
  {year} {2019})},\ \Eprint {http://arxiv.org/abs/1901.08580v1}
  {arXiv:1901.08580v1 [gr-qc]} \BibitemShut {NoStop}%
\bibitem [{\citenamefont {{Usman}}\ \emph {et~al.}(2016)\citenamefont
  {{Usman}}, \citenamefont {{Nitz}}, \citenamefont {{Harry}} \emph
  {et~al.}}]{PYCBCPipeline}%
  \BibitemOpen
  \bibfield  {author} {\bibinfo {author} {\bibfnamefont {S.~A.}\ \bibnamefont
  {{Usman}}}, \bibinfo {author} {\bibfnamefont {A.~H.}\ \bibnamefont {{Nitz}}},
  \bibinfo {author} {\bibfnamefont {I.~W.}\ \bibnamefont {{Harry}}},  \emph
  {et~al.},\ }\href {\doibase 10.1088/0264-9381/33/21/215004} {\bibfield
  {journal} {\bibinfo  {journal} {Classical and Quantum Gravity}\ }\textbf
  {\bibinfo {volume} {33}},\ \bibinfo {eid} {215004} (\bibinfo {year}
  {2016})},\ \Eprint {http://arxiv.org/abs/1508.02357} {arXiv:1508.02357
  [gr-qc]} \BibitemShut {NoStop}%
\bibitem [{\citenamefont {Andres}\ \emph {et~al.}(2022)\citenamefont {Andres},
  \citenamefont {Assiduo}, \citenamefont {Aubin}, \citenamefont {Chierici},
  \citenamefont {Estevez}, \citenamefont {Faedi}, \citenamefont {Guidi},
  \citenamefont {Juste}, \citenamefont {Marion}, \citenamefont {Mours},
  \citenamefont {Nitoglia},\ and\ \citenamefont
  {Sordini}}]{mbta_o3a_pastro_andres2022}%
  \BibitemOpen
  \bibfield  {author} {\bibinfo {author} {\bibfnamefont {N.}~\bibnamefont
  {Andres}}, \bibinfo {author} {\bibfnamefont {M.}~\bibnamefont {Assiduo}},
  \bibinfo {author} {\bibfnamefont {F.}~\bibnamefont {Aubin}}, \bibinfo
  {author} {\bibfnamefont {R.}~\bibnamefont {Chierici}},  \emph {et~al.},\
  }\href {\doibase 10.1088/1361-6382/ac482a} {\bibfield  {journal} {\bibinfo
  {journal} {Classical and Quantum Gravity}\ }\textbf {\bibinfo {volume}
  {39}},\ \bibinfo {pages} {055002} (\bibinfo {year} {2022})}\BibitemShut
  {NoStop}%
\bibitem [{\citenamefont {Klimenko}\ \emph {et~al.}(2016)\citenamefont
  {Klimenko}, \citenamefont {Vedovato}, \citenamefont {Drago}, \citenamefont
  {Salemi}, \citenamefont {Tiwari}, \citenamefont {Prodi}, \citenamefont
  {Lazzaro}, \citenamefont {Ackley}, \citenamefont {Tiwari}, \citenamefont
  {Da~Silva},\ and\ \citenamefont {Mitselmakher}}]{Klimenko:2016}%
  \BibitemOpen
  \bibfield  {author} {\bibinfo {author} {\bibfnamefont {S.}~\bibnamefont
  {Klimenko}}, \bibinfo {author} {\bibfnamefont {G.}~\bibnamefont {Vedovato}},
  \bibinfo {author} {\bibfnamefont {M.}~\bibnamefont {Drago}}, \bibinfo
  {author} {\bibfnamefont {F.}~\bibnamefont {Salemi}},  \emph {et~al.},\ }\href
  {\doibase 10.1103/PhysRevD.93.042004} {\bibfield  {journal} {\bibinfo
  {journal} {Phys. Rev. D}\ }\textbf {\bibinfo {volume} {93}},\ \bibinfo
  {pages} {042004} (\bibinfo {year} {2016})}\BibitemShut {NoStop}%
\bibitem [{\citenamefont {Coleman~Miller}\ and\ \citenamefont
  {Hamilton}(2002)}]{imbh_from_globular_clusters2002}%
  \BibitemOpen
  \bibfield  {author} {\bibinfo {author} {\bibfnamefont {M.}~\bibnamefont
  {Coleman~Miller}}\ and\ \bibinfo {author} {\bibfnamefont {D.~P.}\
  \bibnamefont {Hamilton}},\ }\href {\doibase 10.1046/j.1365-8711.2002.05112.x}
  {\bibfield  {journal} {\bibinfo  {journal} {Monthly Notices of the Royal
  Astronomical Society}\ }\textbf {\bibinfo {volume} {330}},\ \bibinfo {pages}
  {232} (\bibinfo {year} {2002})},\ \Eprint
  {http://arxiv.org/abs/https://academic.oup.com/mnras/article-pdf/330/1/232/18413718/330-1-232.pdf}
  {https://academic.oup.com/mnras/article-pdf/330/1/232/18413718/330-1-232.pdf}
  \BibitemShut {NoStop}%
\bibitem [{\citenamefont {{Franciolini}}\ \emph {et~al.}(2024)\citenamefont
  {{Franciolini}}, \citenamefont {{Kritos}}, \citenamefont {{Reali}},
  \citenamefont {{Broekgaarden}},\ and\ \citenamefont {{Berti}}}]{Fra24_UMG}%
  \BibitemOpen
  \bibfield  {author} {\bibinfo {author} {\bibfnamefont {G.}~\bibnamefont
  {{Franciolini}}}, \bibinfo {author} {\bibfnamefont {K.}~\bibnamefont
  {{Kritos}}}, \bibinfo {author} {\bibfnamefont {L.}~\bibnamefont {{Reali}}},
  \bibinfo {author} {\bibfnamefont {F.}~\bibnamefont {{Broekgaarden}}}, \ and\
  \bibinfo {author} {\bibfnamefont {E.}~\bibnamefont {{Berti}}},\ }\href
  {\doibase 10.48550/arXiv.2401.13038} {\bibfield  {journal} {\bibinfo
  {journal} {arXiv e-prints}\ ,\ \bibinfo {eid} {arXiv:2401.13038}} (\bibinfo
  {year} {2024})},\ \Eprint {http://arxiv.org/abs/2401.13038} {arXiv:2401.13038
  [astro-ph.HE]} \BibitemShut {NoStop}%
\bibitem [{\citenamefont {{Greene}}\ \emph {et~al.}(2020)\citenamefont
  {{Greene}}, \citenamefont {{Strader}},\ and\ \citenamefont {{Ho}}}]{Gre20}%
  \BibitemOpen
  \bibfield  {author} {\bibinfo {author} {\bibfnamefont {J.~E.}\ \bibnamefont
  {{Greene}}}, \bibinfo {author} {\bibfnamefont {J.}~\bibnamefont {{Strader}}},
  \ and\ \bibinfo {author} {\bibfnamefont {L.~C.}\ \bibnamefont {{Ho}}},\
  }\href {\doibase 10.1146/annurev-astro-032620-021835} {\bibfield  {journal}
  {\bibinfo  {journal} {\araa}\ }\textbf {\bibinfo {volume} {58}},\ \bibinfo
  {pages} {257} (\bibinfo {year} {2020})},\ \Eprint
  {http://arxiv.org/abs/1911.09678} {arXiv:1911.09678 [astro-ph.GA]}
  \BibitemShut {NoStop}%
\bibitem [{\citenamefont {Abbott}\ \emph {et~al.}(2020)\citenamefont {Abbott}
  \emph {et~al.}}]{GW190521}%
  \BibitemOpen
  \bibfield  {author} {\bibinfo {author} {\bibfnamefont {R.}~\bibnamefont
  {Abbott}} \emph {et~al.} (\bibinfo {collaboration} {{LIGO Scientific
  Collaboration and Virgo Collaboration}}),\ }\href {\doibase
  10.1103/PhysRevLett.125.101102} {\bibfield  {journal} {\bibinfo  {journal}
  {Phys. Rev. Lett.}\ }\textbf {\bibinfo {volume} {125}},\ \bibinfo {pages}
  {101102} (\bibinfo {year} {2020})}\BibitemShut {NoStop}%
\bibitem [{\citenamefont {{Fragione}}\ \emph {et~al.}(2018)\citenamefont
  {{Fragione}}, \citenamefont {{Ginsburg}},\ and\ \citenamefont
  {{Kocsis}}}]{Fra18_IMBH_LISA}%
  \BibitemOpen
  \bibfield  {author} {\bibinfo {author} {\bibfnamefont {G.}~\bibnamefont
  {{Fragione}}}, \bibinfo {author} {\bibfnamefont {I.}~\bibnamefont
  {{Ginsburg}}}, \ and\ \bibinfo {author} {\bibfnamefont {B.}~\bibnamefont
  {{Kocsis}}},\ }\href {\doibase 10.3847/1538-4357/aab368} {\bibfield
  {journal} {\bibinfo  {journal} {\apj}\ }\textbf {\bibinfo {volume} {856}},\
  \bibinfo {eid} {92} (\bibinfo {year} {2018})},\ \Eprint
  {http://arxiv.org/abs/1711.00483} {arXiv:1711.00483 [astro-ph.GA]}
  \BibitemShut {NoStop}%
\bibitem [{\citenamefont {Marchant}\ and\ \citenamefont
  {Moriya}(2020)}]{Marchant:2020haw}%
  \BibitemOpen
  \bibfield  {author} {\bibinfo {author} {\bibfnamefont {P.}~\bibnamefont
  {Marchant}}\ and\ \bibinfo {author} {\bibfnamefont {T.}~\bibnamefont
  {Moriya}},\ }\href {\doibase 10.1051/0004-6361/202038902} {\bibfield
  {journal} {\bibinfo  {journal} {Astron. Astrophys.}\ }\textbf {\bibinfo
  {volume} {640}},\ \bibinfo {pages} {L18} (\bibinfo {year} {2020})},\ \Eprint
  {http://arxiv.org/abs/2007.06220} {arXiv:2007.06220 [astro-ph.HE]}
  \BibitemShut {NoStop}%
\bibitem [{\citenamefont {Farmer}\ \emph {et~al.}(2019)\citenamefont {Farmer},
  \citenamefont {Renzo}, \citenamefont {de~Mink}, \citenamefont {Marchant},\
  and\ \citenamefont {Justham}}]{Farmer:2019jed}%
  \BibitemOpen
  \bibfield  {author} {\bibinfo {author} {\bibfnamefont {R.}~\bibnamefont
  {Farmer}}, \bibinfo {author} {\bibfnamefont {M.}~\bibnamefont {Renzo}},
  \bibinfo {author} {\bibfnamefont {S.~E.}\ \bibnamefont {de~Mink}}, \bibinfo
  {author} {\bibfnamefont {P.}~\bibnamefont {Marchant}}, \ and\ \bibinfo
  {author} {\bibfnamefont {S.}~\bibnamefont {Justham}},\ }\href {\doibase
  10.3847/1538-4357/ab518b} {\  (\bibinfo {year} {2019}),\
  10.3847/1538-4357/ab518b},\ \Eprint {http://arxiv.org/abs/1910.12874}
  {arXiv:1910.12874 [astro-ph.SR]} \BibitemShut {NoStop}%
\bibitem [{\citenamefont {{van Son}}\ \emph {et~al.}(2020)\citenamefont {{van
  Son}}, \citenamefont {{De Mink}}, \citenamefont {{Broekgaarden}},
  \citenamefont {{Renzo}}, \citenamefont {{Justham}}, \citenamefont
  {{Laplace}}, \citenamefont {{Mor{\'a}n-Fraile}}, \citenamefont {{Hendriks}},\
  and\ \citenamefont {{Farmer}}}]{VanSon20_UMG_Pollution}%
  \BibitemOpen
  \bibfield  {author} {\bibinfo {author} {\bibfnamefont {L.~A.~C.}\
  \bibnamefont {{van Son}}}, \bibinfo {author} {\bibfnamefont {S.~E.}\
  \bibnamefont {{De Mink}}}, \bibinfo {author} {\bibfnamefont {F.~S.}\
  \bibnamefont {{Broekgaarden}}}, \bibinfo {author} {\bibfnamefont
  {M.}~\bibnamefont {{Renzo}}},  \emph {et~al.},\ }\href {\doibase
  10.3847/1538-4357/ab9809} {\bibfield  {journal} {\bibinfo  {journal} {\apj}\
  }\textbf {\bibinfo {volume} {897}},\ \bibinfo {eid} {100} (\bibinfo {year}
  {2020})},\ \Eprint {http://arxiv.org/abs/2004.05187} {arXiv:2004.05187
  [astro-ph.HE]} \BibitemShut {NoStop}%
\bibitem [{\citenamefont {Mehta}\ \emph {et~al.}(2022)\citenamefont {Mehta},
  \citenamefont {Buonanno}, \citenamefont {Gair}, \citenamefont {Miller},
  \citenamefont {Farag}, \citenamefont {deBoer}, \citenamefont {Wiescher},\
  and\ \citenamefont {Timmes}}]{Mehta:2021fgz}%
  \BibitemOpen
  \bibfield  {author} {\bibinfo {author} {\bibfnamefont {A.~K.}\ \bibnamefont
  {Mehta}}, \bibinfo {author} {\bibfnamefont {A.}~\bibnamefont {Buonanno}},
  \bibinfo {author} {\bibfnamefont {J.}~\bibnamefont {Gair}}, \bibinfo {author}
  {\bibfnamefont {M.~C.}\ \bibnamefont {Miller}},  \emph {et~al.},\ }\href
  {\doibase 10.3847/1538-4357/ac3130} {\bibfield  {journal} {\bibinfo
  {journal} {Astrophys. J.}\ }\textbf {\bibinfo {volume} {924}},\ \bibinfo
  {pages} {39} (\bibinfo {year} {2022})},\ \Eprint
  {http://arxiv.org/abs/2105.06366} {arXiv:2105.06366 [gr-qc]} \BibitemShut
  {NoStop}%
\bibitem [{\citenamefont {{Hendriks}}\ \emph {et~al.}(2023)\citenamefont
  {{Hendriks}}, \citenamefont {{van Son}}, \citenamefont {{Renzo}},
  \citenamefont {{Izzard}},\ and\ \citenamefont {{Farmer}}}]{Hen23_UMG}%
  \BibitemOpen
  \bibfield  {author} {\bibinfo {author} {\bibfnamefont {D.~D.}\ \bibnamefont
  {{Hendriks}}}, \bibinfo {author} {\bibfnamefont {L.~A.~C.}\ \bibnamefont
  {{van Son}}}, \bibinfo {author} {\bibfnamefont {M.}~\bibnamefont {{Renzo}}},
  \bibinfo {author} {\bibfnamefont {R.~G.}\ \bibnamefont {{Izzard}}}, \ and\
  \bibinfo {author} {\bibfnamefont {R.}~\bibnamefont {{Farmer}}},\ }\href
  {\doibase 10.1093/mnras/stad2857} {\bibfield  {journal} {\bibinfo  {journal}
  {\mnras}\ } (\bibinfo {year} {2023}),\ 10.1093/mnras/stad2857},\ \Eprint
  {http://arxiv.org/abs/2309.09339} {arXiv:2309.09339 [astro-ph.HE]}
  \BibitemShut {NoStop}%
\bibitem [{\citenamefont {{Golomb}}\ \emph {et~al.}(2023)\citenamefont
  {{Golomb}}, \citenamefont {{Isi}},\ and\ \citenamefont {{Farr}}}]{Gol23_UMG}%
  \BibitemOpen
  \bibfield  {author} {\bibinfo {author} {\bibfnamefont {J.}~\bibnamefont
  {{Golomb}}}, \bibinfo {author} {\bibfnamefont {M.}~\bibnamefont {{Isi}}}, \
  and\ \bibinfo {author} {\bibfnamefont {W.}~\bibnamefont {{Farr}}},\ }\href
  {\doibase 10.48550/arXiv.2312.03973} {\bibfield  {journal} {\bibinfo
  {journal} {arXiv e-prints}\ ,\ \bibinfo {eid} {arXiv:2312.03973}} (\bibinfo
  {year} {2023})},\ \Eprint {http://arxiv.org/abs/2312.03973} {arXiv:2312.03973
  [astro-ph.HE]} \BibitemShut {NoStop}%
\bibitem [{\citenamefont {Abbott}\ \emph {et~al.}(2023)\citenamefont {Abbott}
  \emph {et~al.}}]{LVKO3bpopulation}%
  \BibitemOpen
  \bibfield  {author} {\bibinfo {author} {\bibfnamefont {R.}~\bibnamefont
  {Abbott}} \emph {et~al.} (\bibinfo {collaboration} {KAGRA, VIRGO, LIGO
  Scientific}),\ }\href {\doibase 10.1103/PhysRevX.13.011048} {\bibfield
  {journal} {\bibinfo  {journal} {Phys. Rev. X}\ }\textbf {\bibinfo {volume}
  {13}},\ \bibinfo {pages} {011048} (\bibinfo {year} {2023})},\ \Eprint
  {http://arxiv.org/abs/2111.03634} {arXiv:2111.03634 [astro-ph.HE]}
  \BibitemShut {NoStop}%
\bibitem [{\citenamefont {Roulet}\ \emph {et~al.}(2021)\citenamefont {Roulet},
  \citenamefont {Chia}, \citenamefont {Olsen}, \citenamefont {Dai},
  \citenamefont {Venumadhav}, \citenamefont {Zackay},\ and\ \citenamefont
  {Zaldarriaga}}]{ias_o3a_population_analysis_prd2021roulet}%
  \BibitemOpen
  \bibfield  {author} {\bibinfo {author} {\bibfnamefont {J.}~\bibnamefont
  {Roulet}}, \bibinfo {author} {\bibfnamefont {H.~S.}\ \bibnamefont {Chia}},
  \bibinfo {author} {\bibfnamefont {S.}~\bibnamefont {Olsen}}, \bibinfo
  {author} {\bibfnamefont {L.}~\bibnamefont {Dai}}, \bibinfo {author}
  {\bibfnamefont {T.}~\bibnamefont {Venumadhav}}, \bibinfo {author}
  {\bibfnamefont {B.}~\bibnamefont {Zackay}}, \ and\ \bibinfo {author}
  {\bibfnamefont {M.}~\bibnamefont {Zaldarriaga}},\ }\href {\doibase
  10.1103/PhysRevD.104.083010} {\bibfield  {journal} {\bibinfo  {journal}
  {Phys. Rev. D}\ }\textbf {\bibinfo {volume} {104}},\ \bibinfo {pages}
  {083010} (\bibinfo {year} {2021})}\BibitemShut {NoStop}%
\bibitem [{\citenamefont {Anagnostou}\ \emph {et~al.}(2020)\citenamefont
  {Anagnostou}, \citenamefont {Trenti},\ and\ \citenamefont
  {Melatos}}]{hierarchical_7merger_scenario2020b}%
  \BibitemOpen
  \bibfield  {author} {\bibinfo {author} {\bibfnamefont {O.}~\bibnamefont
  {Anagnostou}}, \bibinfo {author} {\bibfnamefont {M.}~\bibnamefont {Trenti}},
  \ and\ \bibinfo {author} {\bibfnamefont {A.}~\bibnamefont {Melatos}},\
  }\href@noop {} {\  (\bibinfo {year} {2020})},\ \Eprint
  {http://arxiv.org/abs/2010.06161} {arXiv:2010.06161 [astro-ph.HE]}
  \BibitemShut {NoStop}%
\bibitem [{\citenamefont {Fragione}\ \emph {et~al.}(2020)\citenamefont
  {Fragione}, \citenamefont {Loeb},\ and\ \citenamefont
  {Rasio}}]{hierarchical_from_dynamical_in_any_star_cluster2020b}%
  \BibitemOpen
  \bibfield  {author} {\bibinfo {author} {\bibfnamefont {G.}~\bibnamefont
  {Fragione}}, \bibinfo {author} {\bibfnamefont {A.}~\bibnamefont {Loeb}}, \
  and\ \bibinfo {author} {\bibfnamefont {F.~A.}\ \bibnamefont {Rasio}},\ }\href
  {\doibase 10.3847/2041-8213/abbc0a} {\bibfield  {journal} {\bibinfo
  {journal} {Astrophys. J. Lett.}\ }\textbf {\bibinfo {volume} {902}},\
  \bibinfo {pages} {L26} (\bibinfo {year} {2020})},\ \Eprint
  {http://arxiv.org/abs/2009.05065} {arXiv:2009.05065 [astro-ph.GA]}
  \BibitemShut {NoStop}%
\bibitem [{\citenamefont {Veske}\ \emph {et~al.}(2021)\citenamefont {Veske},
  \citenamefont {Sullivan}, \citenamefont {M\'arka}, \citenamefont {Bartos},
  \citenamefont {Corley}, \citenamefont {Samsing}, \citenamefont {Buscicchio},\
  and\ \citenamefont
  {M\'arka}}]{hierarchical_mergerFamily_dynamical_mass_dist_matters2021}%
  \BibitemOpen
  \bibfield  {author} {\bibinfo {author} {\bibfnamefont {D.}~\bibnamefont
  {Veske}}, \bibinfo {author} {\bibfnamefont {A.~G.}\ \bibnamefont {Sullivan}},
  \bibinfo {author} {\bibfnamefont {Z.}~\bibnamefont {M\'arka}}, \bibinfo
  {author} {\bibfnamefont {I.}~\bibnamefont {Bartos}},  \emph {et~al.},\ }\href
  {\doibase 10.3847/2041-8213/abd721} {\bibfield  {journal} {\bibinfo
  {journal} {Astrophys. J. Lett.}\ }\textbf {\bibinfo {volume} {907}},\
  \bibinfo {pages} {L48} (\bibinfo {year} {2021})},\ \Eprint
  {http://arxiv.org/abs/2011.06591} {arXiv:2011.06591 [astro-ph.HE]}
  \BibitemShut {NoStop}%
\bibitem [{\citenamefont {Fragione}\ \emph {et~al.}(2021)\citenamefont
  {Fragione}, \citenamefont {Kocsis}, \citenamefont {Rasio},\ and\
  \citenamefont
  {Silk}}]{hierarchical_rate_sensitive_to_natal_spins_Fragione2021kocsis}%
  \BibitemOpen
  \bibfield  {author} {\bibinfo {author} {\bibfnamefont {G.}~\bibnamefont
  {Fragione}}, \bibinfo {author} {\bibfnamefont {B.}~\bibnamefont {Kocsis}},
  \bibinfo {author} {\bibfnamefont {F.~A.}\ \bibnamefont {Rasio}}, \ and\
  \bibinfo {author} {\bibfnamefont {J.}~\bibnamefont {Silk}},\ }\href@noop {}
  {\  (\bibinfo {year} {2021})},\ \Eprint {http://arxiv.org/abs/2107.04639}
  {arXiv:2107.04639 [astro-ph.GA]} \BibitemShut {NoStop}%
\bibitem [{\citenamefont {{Kremer}}\ \emph {et~al.}(2020)\citenamefont
  {{Kremer}}, \citenamefont {{Spera}}, \citenamefont {{Becker}}, \citenamefont
  {{Chatterjee}}, \citenamefont {{Di Carlo}}, \citenamefont {{Fragione}},
  \citenamefont {{Rodriguez}}, \citenamefont {{Ye}},\ and\ \citenamefont
  {{Rasio}}}]{UMG_Kre20_StellarCollisions}%
  \BibitemOpen
  \bibfield  {author} {\bibinfo {author} {\bibfnamefont {K.}~\bibnamefont
  {{Kremer}}}, \bibinfo {author} {\bibfnamefont {M.}~\bibnamefont {{Spera}}},
  \bibinfo {author} {\bibfnamefont {D.}~\bibnamefont {{Becker}}}, \bibinfo
  {author} {\bibfnamefont {S.}~\bibnamefont {{Chatterjee}}},  \emph {et~al.},\
  }\href {\doibase 10.3847/1538-4357/abb945} {\bibfield  {journal} {\bibinfo
  {journal} {\apj}\ }\textbf {\bibinfo {volume} {903}},\ \bibinfo {eid} {45}
  (\bibinfo {year} {2020})},\ \Eprint {http://arxiv.org/abs/2006.10771}
  {arXiv:2006.10771 [astro-ph.HE]} \BibitemShut {NoStop}%
\bibitem [{\citenamefont {Yang}\ \emph {et~al.}(2019)\citenamefont {Yang},
  \citenamefont {Bartos}, \citenamefont {Gayathri}, \citenamefont {Ford},
  \citenamefont {Haiman}, \citenamefont {Klimenko}, \citenamefont {Kocsis},
  \citenamefont {M\'arka}, \citenamefont {M\'arka}, \citenamefont {McKernan},\
  and\ \citenamefont {O'Shaughnessy}}]{hierarchical_mergers_agn_kocsis2019}%
  \BibitemOpen
  \bibfield  {author} {\bibinfo {author} {\bibfnamefont {Y.}~\bibnamefont
  {Yang}}, \bibinfo {author} {\bibfnamefont {I.}~\bibnamefont {Bartos}},
  \bibinfo {author} {\bibfnamefont {V.}~\bibnamefont {Gayathri}}, \bibinfo
  {author} {\bibfnamefont {K.~E.~S.}\ \bibnamefont {Ford}},  \emph {et~al.},\
  }\href {\doibase 10.1103/PhysRevLett.123.181101} {\bibfield  {journal}
  {\bibinfo  {journal} {Phys. Rev. Lett.}\ }\textbf {\bibinfo {volume} {123}},\
  \bibinfo {pages} {181101} (\bibinfo {year} {2019})}\BibitemShut {NoStop}%
\bibitem [{\citenamefont {McKernan}\ \emph {et~al.}(2020)\citenamefont
  {McKernan}, \citenamefont {Ford}, \citenamefont {O’Shaugnessy},\ and\
  \citenamefont
  {Wysocki}}]{agn_bbh_population_chieff_q_simulation_mckernan_ford2019}%
  \BibitemOpen
  \bibfield  {author} {\bibinfo {author} {\bibfnamefont {B.}~\bibnamefont
  {McKernan}}, \bibinfo {author} {\bibfnamefont {K.~E.~S.}\ \bibnamefont
  {Ford}}, \bibinfo {author} {\bibfnamefont {R.}~\bibnamefont
  {O’Shaugnessy}}, \ and\ \bibinfo {author} {\bibfnamefont {D.}~\bibnamefont
  {Wysocki}},\ }\href {\doibase 10.1093/mnras/staa740} {\bibfield  {journal}
  {\bibinfo  {journal} {Monthly Notices of the Royal Astronomical Society}\
  }\textbf {\bibinfo {volume} {494}},\ \bibinfo {pages} {1203} (\bibinfo {year}
  {2020})},\ \Eprint
  {http://arxiv.org/abs/https://academic.oup.com/mnras/article-pdf/494/1/1203/33029648/staa740.pdf}
  {https://academic.oup.com/mnras/article-pdf/494/1/1203/33029648/staa740.pdf}
  \BibitemShut {NoStop}%
\bibitem [{\citenamefont {Tagawa}\ \emph {et~al.}(2020)\citenamefont {Tagawa},
  \citenamefont {Haiman}, \citenamefont {Bartos},\ and\ \citenamefont
  {Kocsis}}]{bbh_spin_evolution_agn_Tagawa_2020a}%
  \BibitemOpen
  \bibfield  {author} {\bibinfo {author} {\bibfnamefont {H.}~\bibnamefont
  {Tagawa}}, \bibinfo {author} {\bibfnamefont {Z.}~\bibnamefont {Haiman}},
  \bibinfo {author} {\bibfnamefont {I.}~\bibnamefont {Bartos}}, \ and\ \bibinfo
  {author} {\bibfnamefont {B.}~\bibnamefont {Kocsis}},\ }\href {\doibase
  10.3847/1538-4357/aba2cc} {\bibfield  {journal} {\bibinfo  {journal} {The
  Astrophysical Journal}\ }\textbf {\bibinfo {volume} {899}},\ \bibinfo {pages}
  {26} (\bibinfo {year} {2020})}\BibitemShut {NoStop}%
\bibitem [{\citenamefont {{Ishibashi, W.}}\ and\ \citenamefont {{Gr\"obner,
  M.}}(2020)}]{bbh_evolution_agn_merger_timescale_ishibashi2020a}%
  \BibitemOpen
  \bibfield  {author} {\bibinfo {author} {\bibnamefont {{Ishibashi, W.}}}\ and\
  \bibinfo {author} {\bibnamefont {{Gr\"obner, M.}}},\ }\href {\doibase
  10.1051/0004-6361/202037799} {\bibfield  {journal} {\bibinfo  {journal}
  {A\&A}\ }\textbf {\bibinfo {volume} {639}},\ \bibinfo {pages} {A108}
  (\bibinfo {year} {2020})}\BibitemShut {NoStop}%
\bibitem [{\citenamefont {Secunda}\ \emph {et~al.}(2020)\citenamefont
  {Secunda}, \citenamefont {Bellovary}, \citenamefont {Mac~Low}, \citenamefont
  {Ford}, \citenamefont {McKernan}, \citenamefont {Leigh}, \citenamefont
  {Lyra}, \citenamefont {Sandor},\ and\ \citenamefont
  {Adorno}}]{migration_traps_spins_rates_mckernan_ford2020a}%
  \BibitemOpen
  \bibfield  {author} {\bibinfo {author} {\bibfnamefont {A.}~\bibnamefont
  {Secunda}}, \bibinfo {author} {\bibfnamefont {J.}~\bibnamefont {Bellovary}},
  \bibinfo {author} {\bibfnamefont {M.-M.}\ \bibnamefont {Mac~Low}}, \bibinfo
  {author} {\bibfnamefont {K.~E.~S.}\ \bibnamefont {Ford}},  \emph {et~al.},\
  }\href {\doibase 10.3847/1538-4357/abbc1d} {\bibfield  {journal} {\bibinfo
  {journal} {Astrophys. J.}\ }\textbf {\bibinfo {volume} {903}},\ \bibinfo
  {pages} {133} (\bibinfo {year} {2020})},\ \Eprint
  {http://arxiv.org/abs/2004.11936} {arXiv:2004.11936 [astro-ph.HE]}
  \BibitemShut {NoStop}%
\bibitem [{\citenamefont {Tagawa}\ \emph {et~al.}(2021)\citenamefont {Tagawa},
  \citenamefont {Kocsis}, \citenamefont {Haiman}, \citenamefont {Bartos},
  \citenamefont {Omukai},\ and\ \citenamefont
  {Samsing}}]{mass_gap_agn_bbh_mergers2021}%
  \BibitemOpen
  \bibfield  {author} {\bibinfo {author} {\bibfnamefont {H.}~\bibnamefont
  {Tagawa}}, \bibinfo {author} {\bibfnamefont {B.}~\bibnamefont {Kocsis}},
  \bibinfo {author} {\bibfnamefont {Z.}~\bibnamefont {Haiman}}, \bibinfo
  {author} {\bibfnamefont {I.}~\bibnamefont {Bartos}}, \bibinfo {author}
  {\bibfnamefont {K.}~\bibnamefont {Omukai}}, \ and\ \bibinfo {author}
  {\bibfnamefont {J.}~\bibnamefont {Samsing}},\ }\href {\doibase
  10.3847/1538-4357/abd555} {\bibfield  {journal} {\bibinfo  {journal}
  {Astrophys. J.}\ }\textbf {\bibinfo {volume} {908}},\ \bibinfo {pages} {194}
  (\bibinfo {year} {2021})},\ \Eprint {http://arxiv.org/abs/2012.00011}
  {arXiv:2012.00011 [astro-ph.HE]} \BibitemShut {NoStop}%
\bibitem [{\citenamefont {Fabj}\ \emph {et~al.}(2020)\citenamefont {Fabj},
  \citenamefont {Nasim}, \citenamefont {Caban}, \citenamefont {Ford},
  \citenamefont {McKernan},\ and\ \citenamefont
  {Bellovary}}]{agn_accretion_disk_merger_population2020a}%
  \BibitemOpen
  \bibfield  {author} {\bibinfo {author} {\bibfnamefont {G.}~\bibnamefont
  {Fabj}}, \bibinfo {author} {\bibfnamefont {S.~S.}\ \bibnamefont {Nasim}},
  \bibinfo {author} {\bibfnamefont {F.}~\bibnamefont {Caban}}, \bibinfo
  {author} {\bibfnamefont {K.~E.~S.}\ \bibnamefont {Ford}}, \bibinfo {author}
  {\bibfnamefont {B.}~\bibnamefont {McKernan}}, \ and\ \bibinfo {author}
  {\bibfnamefont {J.~M.}\ \bibnamefont {Bellovary}},\ }\href {\doibase
  10.1093/mnras/staa3004} {\bibfield  {journal} {\bibinfo  {journal} {Mon. Not.
  Roy. Astron. Soc.}\ }\textbf {\bibinfo {volume} {499}},\ \bibinfo {pages}
  {2608} (\bibinfo {year} {2020})},\ \Eprint {http://arxiv.org/abs/2006.11229}
  {arXiv:2006.11229 [astro-ph.GA]} \BibitemShut {NoStop}%
\bibitem [{\citenamefont {{Tagawa}}\ \emph {et~al.}(2020)\citenamefont
  {{Tagawa}}, \citenamefont {{Haiman}}, \citenamefont {{Bartos}},\ and\
  \citenamefont {{Kocsis}}}]{Tag20_AGN}%
  \BibitemOpen
  \bibfield  {author} {\bibinfo {author} {\bibfnamefont {H.}~\bibnamefont
  {{Tagawa}}}, \bibinfo {author} {\bibfnamefont {Z.}~\bibnamefont {{Haiman}}},
  \bibinfo {author} {\bibfnamefont {I.}~\bibnamefont {{Bartos}}}, \ and\
  \bibinfo {author} {\bibfnamefont {B.}~\bibnamefont {{Kocsis}}},\ }\href
  {\doibase 10.3847/1538-4357/aba2cc} {\bibfield  {journal} {\bibinfo
  {journal} {\apj}\ }\textbf {\bibinfo {volume} {899}},\ \bibinfo {eid} {26}
  (\bibinfo {year} {2020})},\ \Eprint {http://arxiv.org/abs/2004.11914}
  {arXiv:2004.11914 [astro-ph.HE]} \BibitemShut {NoStop}%
\bibitem [{\citenamefont {{Samsing}}\ \emph {et~al.}(2022)\citenamefont
  {{Samsing}}, \citenamefont {{Bartos}}, \citenamefont {{D'Orazio}},
  \citenamefont {{Haiman}}, \citenamefont {{Kocsis}}, \citenamefont {{Leigh}},
  \citenamefont {{Liu}}, \citenamefont {{Pessah}},\ and\ \citenamefont
  {{Tagawa}}}]{Sam22_AGN}%
  \BibitemOpen
  \bibfield  {author} {\bibinfo {author} {\bibfnamefont {J.}~\bibnamefont
  {{Samsing}}}, \bibinfo {author} {\bibfnamefont {I.}~\bibnamefont {{Bartos}}},
  \bibinfo {author} {\bibfnamefont {D.~J.}\ \bibnamefont {{D'Orazio}}},
  \bibinfo {author} {\bibfnamefont {Z.}~\bibnamefont {{Haiman}}},  \emph
  {et~al.},\ }\href {\doibase 10.1038/s41586-021-04333-1} {\bibfield  {journal}
  {\bibinfo  {journal} {\nat}\ }\textbf {\bibinfo {volume} {603}},\ \bibinfo
  {pages} {237} (\bibinfo {year} {2022})},\ \Eprint
  {http://arxiv.org/abs/2010.09765} {arXiv:2010.09765 [astro-ph.HE]}
  \BibitemShut {NoStop}%
\bibitem [{\citenamefont {Dai}\ \emph {et~al.}(2017)\citenamefont {Dai},
  \citenamefont {Venumadhav},\ and\ \citenamefont {Sigurdson}}]{Dai16_Lensing}%
  \BibitemOpen
  \bibfield  {author} {\bibinfo {author} {\bibfnamefont {L.}~\bibnamefont
  {Dai}}, \bibinfo {author} {\bibfnamefont {T.}~\bibnamefont {Venumadhav}}, \
  and\ \bibinfo {author} {\bibfnamefont {K.}~\bibnamefont {Sigurdson}},\ }\href
  {\doibase 10.1103/PhysRevD.95.044011} {\bibfield  {journal} {\bibinfo
  {journal} {Phys. Rev. D}\ }\textbf {\bibinfo {volume} {95}},\ \bibinfo
  {pages} {044011} (\bibinfo {year} {2017})},\ \Eprint
  {http://arxiv.org/abs/1605.09398} {arXiv:1605.09398 [astro-ph.CO]}
  \BibitemShut {NoStop}%
\bibitem [{\citenamefont {{Fishbach}}\ \emph {et~al.}(2018)\citenamefont
  {{Fishbach}}, \citenamefont {{Holz}},\ and\ \citenamefont
  {{Farr}}}]{Fis18_merger_redshift}%
  \BibitemOpen
  \bibfield  {author} {\bibinfo {author} {\bibfnamefont {M.}~\bibnamefont
  {{Fishbach}}}, \bibinfo {author} {\bibfnamefont {D.~E.}\ \bibnamefont
  {{Holz}}}, \ and\ \bibinfo {author} {\bibfnamefont {W.~M.}\ \bibnamefont
  {{Farr}}},\ }\href {\doibase 10.3847/2041-8213/aad800} {\bibfield  {journal}
  {\bibinfo  {journal} {\apjl}\ }\textbf {\bibinfo {volume} {863}},\ \bibinfo
  {eid} {L41} (\bibinfo {year} {2018})},\ \Eprint
  {http://arxiv.org/abs/1805.10270} {arXiv:1805.10270 [astro-ph.HE]}
  \BibitemShut {NoStop}%
\bibitem [{\citenamefont {{Madau}}\ and\ \citenamefont
  {{Dickinson}}(2014)}]{Mad14_SFR}%
  \BibitemOpen
  \bibfield  {author} {\bibinfo {author} {\bibfnamefont {P.}~\bibnamefont
  {{Madau}}}\ and\ \bibinfo {author} {\bibfnamefont {M.}~\bibnamefont
  {{Dickinson}}},\ }\href {\doibase 10.1146/annurev-astro-081811-125615}
  {\bibfield  {journal} {\bibinfo  {journal} {\araa}\ }\textbf {\bibinfo
  {volume} {52}},\ \bibinfo {pages} {415} (\bibinfo {year} {2014})},\ \Eprint
  {http://arxiv.org/abs/1403.0007} {arXiv:1403.0007 [astro-ph.CO]} \BibitemShut
  {NoStop}%
\bibitem [{\citenamefont {{Mapelli}}\ \emph {et~al.}(2022)\citenamefont
  {{Mapelli}}, \citenamefont {{Bouffanais}}, \citenamefont {{Santoliquido}},
  \citenamefont {{Arca Sedda}},\ and\ \citenamefont
  {{Artale}}}]{Map22_MergerRate}%
  \BibitemOpen
  \bibfield  {author} {\bibinfo {author} {\bibfnamefont {M.}~\bibnamefont
  {{Mapelli}}}, \bibinfo {author} {\bibfnamefont {Y.}~\bibnamefont
  {{Bouffanais}}}, \bibinfo {author} {\bibfnamefont {F.}~\bibnamefont
  {{Santoliquido}}}, \bibinfo {author} {\bibfnamefont {M.}~\bibnamefont {{Arca
  Sedda}}}, \ and\ \bibinfo {author} {\bibfnamefont {M.~C.}\ \bibnamefont
  {{Artale}}},\ }\href {\doibase 10.1093/mnras/stac422} {\bibfield  {journal}
  {\bibinfo  {journal} {\mnras}\ }\textbf {\bibinfo {volume} {511}},\ \bibinfo
  {pages} {5797} (\bibinfo {year} {2022})},\ \Eprint
  {http://arxiv.org/abs/2109.06222} {arXiv:2109.06222 [astro-ph.HE]}
  \BibitemShut {NoStop}%
\bibitem [{\citenamefont {{van Son}}\ \emph {et~al.}(2022)\citenamefont {{van
  Son}}, \citenamefont {{de Mink}}, \citenamefont {{Callister}}, \citenamefont
  {{Justham}}, \citenamefont {{Renzo}}, \citenamefont {{Wagg}}, \citenamefont
  {{Broekgaarden}}, \citenamefont {{Kummer}}, \citenamefont {{Pakmor}},\ and\
  \citenamefont {{Mandel}}}]{Van22_MergerRate}%
  \BibitemOpen
  \bibfield  {author} {\bibinfo {author} {\bibfnamefont {L.~A.~C.}\
  \bibnamefont {{van Son}}}, \bibinfo {author} {\bibfnamefont {S.~E.}\
  \bibnamefont {{de Mink}}}, \bibinfo {author} {\bibfnamefont {T.}~\bibnamefont
  {{Callister}}}, \bibinfo {author} {\bibfnamefont {S.}~\bibnamefont
  {{Justham}}},  \emph {et~al.},\ }\href {\doibase 10.3847/1538-4357/ac64a3}
  {\bibfield  {journal} {\bibinfo  {journal} {\apj}\ }\textbf {\bibinfo
  {volume} {931}},\ \bibinfo {eid} {17} (\bibinfo {year} {2022})},\ \Eprint
  {http://arxiv.org/abs/2110.01634} {arXiv:2110.01634 [astro-ph.HE]}
  \BibitemShut {NoStop}%
\bibitem [{\citenamefont {{Boesky}}\ \emph
  {et~al.}(2024{\natexlab{a}})\citenamefont {{Boesky}}, \citenamefont
  {{Broekgaarden}},\ and\ \citenamefont {{Berger}}}]{Boe24_DelayTime_COMPASS}%
  \BibitemOpen
  \bibfield  {author} {\bibinfo {author} {\bibfnamefont {A.}~\bibnamefont
  {{Boesky}}}, \bibinfo {author} {\bibfnamefont {F.~S.}\ \bibnamefont
  {{Broekgaarden}}}, \ and\ \bibinfo {author} {\bibfnamefont {E.}~\bibnamefont
  {{Berger}}},\ }\href {\doibase 10.48550/arXiv.2405.01623} {\bibfield
  {journal} {\bibinfo  {journal} {arXiv e-prints}\ ,\ \bibinfo {eid}
  {arXiv:2405.01623}} (\bibinfo {year} {2024}{\natexlab{a}})},\ \Eprint
  {http://arxiv.org/abs/2405.01623} {arXiv:2405.01623 [astro-ph.HE]}
  \BibitemShut {NoStop}%
\bibitem [{\citenamefont {{Boesky}}\ \emph
  {et~al.}(2024{\natexlab{b}})\citenamefont {{Boesky}}, \citenamefont
  {{Broekgaarden}},\ and\ \citenamefont
  {{Berger}}}]{Boe24_RedshiftEvolution_COMPASS}%
  \BibitemOpen
  \bibfield  {author} {\bibinfo {author} {\bibfnamefont {A.}~\bibnamefont
  {{Boesky}}}, \bibinfo {author} {\bibfnamefont {F.~S.}\ \bibnamefont
  {{Broekgaarden}}}, \ and\ \bibinfo {author} {\bibfnamefont {E.}~\bibnamefont
  {{Berger}}},\ }\href {\doibase 10.48550/arXiv.2405.01630} {\bibfield
  {journal} {\bibinfo  {journal} {arXiv e-prints}\ ,\ \bibinfo {eid}
  {arXiv:2405.01630}} (\bibinfo {year} {2024}{\natexlab{b}})},\ \Eprint
  {http://arxiv.org/abs/2405.01630} {arXiv:2405.01630 [astro-ph.HE]}
  \BibitemShut {NoStop}%
\bibitem [{\citenamefont {Mandel}\ and\ \citenamefont
  {de~Mink}(2016)}]{Man16_CHE}%
  \BibitemOpen
  \bibfield  {author} {\bibinfo {author} {\bibfnamefont {I.}~\bibnamefont
  {Mandel}}\ and\ \bibinfo {author} {\bibfnamefont {S.~E.}\ \bibnamefont
  {de~Mink}},\ }\href {\doibase 10.1093/mnras/stw379} {\bibfield  {journal}
  {\bibinfo  {journal} {Monthly Notices of the Royal Astronomical Society}\
  }\textbf {\bibinfo {volume} {458}},\ \bibinfo {pages} {2634} (\bibinfo {year}
  {2016})}\BibitemShut {NoStop}%
\bibitem [{\citenamefont {{Mapelli}}\ \emph {et~al.}(2019)\citenamefont
  {{Mapelli}}, \citenamefont {{Giacobbo}}, \citenamefont {{Santoliquido}},\
  and\ \citenamefont {{Artale}}}]{Map19_MergerRate}%
  \BibitemOpen
  \bibfield  {author} {\bibinfo {author} {\bibfnamefont {M.}~\bibnamefont
  {{Mapelli}}}, \bibinfo {author} {\bibfnamefont {N.}~\bibnamefont
  {{Giacobbo}}}, \bibinfo {author} {\bibfnamefont {F.}~\bibnamefont
  {{Santoliquido}}}, \ and\ \bibinfo {author} {\bibfnamefont {M.~C.}\
  \bibnamefont {{Artale}}},\ }\href {\doibase 10.1093/mnras/stz1150} {\bibfield
   {journal} {\bibinfo  {journal} {\mnras}\ }\textbf {\bibinfo {volume}
  {487}},\ \bibinfo {pages} {2} (\bibinfo {year} {2019})},\ \Eprint
  {http://arxiv.org/abs/1902.01419} {arXiv:1902.01419 [astro-ph.HE]}
  \BibitemShut {NoStop}%
\bibitem [{\citenamefont {{Biscoveanu}}\ \emph {et~al.}(2022)\citenamefont
  {{Biscoveanu}}, \citenamefont {{Callister}}, \citenamefont {{Haster}},
  \citenamefont {{Ng}}, \citenamefont {{Vitale}},\ and\ \citenamefont
  {{Farr}}}]{Bis22}%
  \BibitemOpen
  \bibfield  {author} {\bibinfo {author} {\bibfnamefont {S.}~\bibnamefont
  {{Biscoveanu}}}, \bibinfo {author} {\bibfnamefont {T.~A.}\ \bibnamefont
  {{Callister}}}, \bibinfo {author} {\bibfnamefont {C.-J.}\ \bibnamefont
  {{Haster}}}, \bibinfo {author} {\bibfnamefont {K.~K.~Y.}\ \bibnamefont
  {{Ng}}}, \bibinfo {author} {\bibfnamefont {S.}~\bibnamefont {{Vitale}}}, \
  and\ \bibinfo {author} {\bibfnamefont {W.~M.}\ \bibnamefont {{Farr}}},\
  }\href {\doibase 10.3847/2041-8213/ac71a8} {\bibfield  {journal} {\bibinfo
  {journal} {\apjl}\ }\textbf {\bibinfo {volume} {932}},\ \bibinfo {eid} {L19}
  (\bibinfo {year} {2022})},\ \Eprint {http://arxiv.org/abs/2204.01578}
  {arXiv:2204.01578 [astro-ph.HE]} \BibitemShut {NoStop}%
\bibitem [{\citenamefont {{Kushnir}}\ \emph {et~al.}(2016)\citenamefont
  {{Kushnir}}, \citenamefont {{Zaldarriaga}}, \citenamefont {{Kollmeier}},\
  and\ \citenamefont {{Waldman}}}]{Kus16_MergerTime_150914}%
  \BibitemOpen
  \bibfield  {author} {\bibinfo {author} {\bibfnamefont {D.}~\bibnamefont
  {{Kushnir}}}, \bibinfo {author} {\bibfnamefont {M.}~\bibnamefont
  {{Zaldarriaga}}}, \bibinfo {author} {\bibfnamefont {J.~A.}\ \bibnamefont
  {{Kollmeier}}}, \ and\ \bibinfo {author} {\bibfnamefont {R.}~\bibnamefont
  {{Waldman}}},\ }\href {\doibase 10.1093/mnras/stw1684} {\bibfield  {journal}
  {\bibinfo  {journal} {\mnras}\ }\textbf {\bibinfo {volume} {462}},\ \bibinfo
  {pages} {844} (\bibinfo {year} {2016})},\ \Eprint
  {http://arxiv.org/abs/1605.03839} {arXiv:1605.03839 [astro-ph.HE]}
  \BibitemShut {NoStop}%
\bibitem [{\citenamefont {{Zaldarriaga}}\ \emph {et~al.}(2018)\citenamefont
  {{Zaldarriaga}}, \citenamefont {{Kushnir}},\ and\ \citenamefont
  {{Kollmeier}}}]{chieff_isolated_field_binary_matias2018}%
  \BibitemOpen
  \bibfield  {author} {\bibinfo {author} {\bibfnamefont {M.}~\bibnamefont
  {{Zaldarriaga}}}, \bibinfo {author} {\bibfnamefont {D.}~\bibnamefont
  {{Kushnir}}}, \ and\ \bibinfo {author} {\bibfnamefont {J.~A.}\ \bibnamefont
  {{Kollmeier}}},\ }\href {\doibase 10.1093/mnras/stx2577} {\bibfield
  {journal} {\bibinfo  {journal} {\mnras}\ }\textbf {\bibinfo {volume} {473}},\
  \bibinfo {pages} {4174} (\bibinfo {year} {2018})},\ \Eprint
  {http://arxiv.org/abs/1702.00885} {arXiv:1702.00885 [astro-ph.HE]}
  \BibitemShut {NoStop}%
\bibitem [{\citenamefont {{Rodriguez}}\ \emph {et~al.}(2016)\citenamefont
  {{Rodriguez}}, \citenamefont {{Chatterjee}},\ and\ \citenamefont
  {{Rasio}}}]{Rod16_Dynamical}%
  \BibitemOpen
  \bibfield  {author} {\bibinfo {author} {\bibfnamefont {C.~L.}\ \bibnamefont
  {{Rodriguez}}}, \bibinfo {author} {\bibfnamefont {S.}~\bibnamefont
  {{Chatterjee}}}, \ and\ \bibinfo {author} {\bibfnamefont {F.~A.}\
  \bibnamefont {{Rasio}}},\ }\href {\doibase 10.1103/PhysRevD.93.084029}
  {\bibfield  {journal} {\bibinfo  {journal} {\prd}\ }\textbf {\bibinfo
  {volume} {93}},\ \bibinfo {eid} {084029} (\bibinfo {year} {2016})},\ \Eprint
  {http://arxiv.org/abs/1602.02444} {arXiv:1602.02444 [astro-ph.HE]}
  \BibitemShut {NoStop}%
\bibitem [{\citenamefont {{Rodriguez}}\ \emph {et~al.}(2019)\citenamefont
  {{Rodriguez}}, \citenamefont {{Zevin}}, \citenamefont {{Amaro-Seoane}},
  \citenamefont {{Chatterjee}}, \citenamefont {{Kremer}}, \citenamefont
  {{Rasio}},\ and\ \citenamefont {{Ye}}}]{Rodriquez_2019_2g}%
  \BibitemOpen
  \bibfield  {author} {\bibinfo {author} {\bibfnamefont {C.~L.}\ \bibnamefont
  {{Rodriguez}}}, \bibinfo {author} {\bibfnamefont {M.}~\bibnamefont
  {{Zevin}}}, \bibinfo {author} {\bibfnamefont {P.}~\bibnamefont
  {{Amaro-Seoane}}}, \bibinfo {author} {\bibfnamefont {S.}~\bibnamefont
  {{Chatterjee}}}, \bibinfo {author} {\bibfnamefont {K.}~\bibnamefont
  {{Kremer}}}, \bibinfo {author} {\bibfnamefont {F.~A.}\ \bibnamefont
  {{Rasio}}}, \ and\ \bibinfo {author} {\bibfnamefont {C.~S.}\ \bibnamefont
  {{Ye}}},\ }\href {\doibase 10.1103/PhysRevD.100.043027} {\bibfield  {journal}
  {\bibinfo  {journal} {\prd}\ }\textbf {\bibinfo {volume} {100}},\ \bibinfo
  {eid} {043027} (\bibinfo {year} {2019})},\ \Eprint
  {http://arxiv.org/abs/1906.10260} {arXiv:1906.10260 [astro-ph.HE]}
  \BibitemShut {NoStop}%
\bibitem [{\citenamefont {{Ivanova}}\ \emph {et~al.}(2013)\citenamefont
  {{Ivanova}}, \citenamefont {{Justham}}, \citenamefont {{Chen}}, \citenamefont
  {{De Marco}}, \citenamefont {{Fryer}}, \citenamefont {{Gaburov}},
  \citenamefont {{Ge}}, \citenamefont {{Glebbeek}}, \citenamefont {{Han}},
  \citenamefont {{Li}}, \citenamefont {{Lu}}, \citenamefont {{Marsh}},
  \citenamefont {{Podsiadlowski}}, \citenamefont {{Potter}}, \citenamefont
  {{Soker}}, \citenamefont {{Taam}}, \citenamefont {{Tauris}}, \citenamefont
  {{van den Heuvel}},\ and\ \citenamefont {{Webbink}}}]{Iva13_common_envelope}%
  \BibitemOpen
  \bibfield  {author} {\bibinfo {author} {\bibfnamefont {N.}~\bibnamefont
  {{Ivanova}}}, \bibinfo {author} {\bibfnamefont {S.}~\bibnamefont
  {{Justham}}}, \bibinfo {author} {\bibfnamefont {X.}~\bibnamefont {{Chen}}},
  \bibinfo {author} {\bibfnamefont {O.}~\bibnamefont {{De Marco}}},  \emph
  {et~al.},\ }\href {\doibase 10.1007/s00159-013-0059-2} {\bibfield  {journal}
  {\bibinfo  {journal} {\aapr}\ }\textbf {\bibinfo {volume} {21}},\ \bibinfo
  {eid} {59} (\bibinfo {year} {2013})},\ \Eprint
  {http://arxiv.org/abs/1209.4302} {arXiv:1209.4302 [astro-ph.HE]} \BibitemShut
  {NoStop}%
\bibitem [{\citenamefont {{Postnov}}\ and\ \citenamefont
  {{Yungelson}}(2014)}]{Pos14_mass_transfer}%
  \BibitemOpen
  \bibfield  {author} {\bibinfo {author} {\bibfnamefont {K.~A.}\ \bibnamefont
  {{Postnov}}}\ and\ \bibinfo {author} {\bibfnamefont {L.~R.}\ \bibnamefont
  {{Yungelson}}},\ }\href {\doibase 10.12942/lrr-2014-3} {\bibfield  {journal}
  {\bibinfo  {journal} {Living Reviews in Relativity}\ }\textbf {\bibinfo
  {volume} {17}},\ \bibinfo {eid} {3} (\bibinfo {year} {2014})},\ \Eprint
  {http://arxiv.org/abs/1403.4754} {arXiv:1403.4754 [astro-ph.HE]} \BibitemShut
  {NoStop}%
\bibitem [{\citenamefont {{Gerosa}}\ \emph {et~al.}(2013)\citenamefont
  {{Gerosa}}, \citenamefont {{Kesden}}, \citenamefont {{Berti}}, \citenamefont
  {{O'Shaughnessy}},\ and\ \citenamefont
  {{Sperhake}}}]{Ger13_MassRatioReversal}%
  \BibitemOpen
  \bibfield  {author} {\bibinfo {author} {\bibfnamefont {D.}~\bibnamefont
  {{Gerosa}}}, \bibinfo {author} {\bibfnamefont {M.}~\bibnamefont {{Kesden}}},
  \bibinfo {author} {\bibfnamefont {E.}~\bibnamefont {{Berti}}}, \bibinfo
  {author} {\bibfnamefont {R.}~\bibnamefont {{O'Shaughnessy}}}, \ and\ \bibinfo
  {author} {\bibfnamefont {U.}~\bibnamefont {{Sperhake}}},\ }\href {\doibase
  10.1103/PhysRevD.87.104028} {\bibfield  {journal} {\bibinfo  {journal}
  {\prd}\ }\textbf {\bibinfo {volume} {87}},\ \bibinfo {eid} {104028} (\bibinfo
  {year} {2013})},\ \Eprint {http://arxiv.org/abs/1302.4442} {arXiv:1302.4442
  [gr-qc]} \BibitemShut {NoStop}%
\bibitem [{\citenamefont {{Broekgaarden}}\ \emph {et~al.}(2022)\citenamefont
  {{Broekgaarden}}, \citenamefont {{Stevenson}},\ and\ \citenamefont
  {{Thrane}}}]{Broekgaarden_2022_reversal}%
  \BibitemOpen
  \bibfield  {author} {\bibinfo {author} {\bibfnamefont {F.~S.}\ \bibnamefont
  {{Broekgaarden}}}, \bibinfo {author} {\bibfnamefont {S.}~\bibnamefont
  {{Stevenson}}}, \ and\ \bibinfo {author} {\bibfnamefont {E.}~\bibnamefont
  {{Thrane}}},\ }\href {\doibase 10.3847/1538-4357/ac8879} {\bibfield
  {journal} {\bibinfo  {journal} {\apj}\ }\textbf {\bibinfo {volume} {938}},\
  \bibinfo {eid} {45} (\bibinfo {year} {2022})},\ \Eprint
  {http://arxiv.org/abs/2205.01693} {arXiv:2205.01693 [astro-ph.HE]}
  \BibitemShut {NoStop}%
\bibitem [{\citenamefont {{Mehta}}\ \emph {et~al.}(2025)\citenamefont
  {{Mehta}}, \citenamefont {{Wadekar}}, \citenamefont {{Roulet}}, \citenamefont
  {{Anantpurkar}}, \citenamefont {{Venumadhav}}, \citenamefont {{Mushkin}},
  \citenamefont {{Zackay}}, \citenamefont {{Zaldarriaga}},\ and\ \citenamefont
  {{Islam}}}]{Meh25_VT_HM}%
  \BibitemOpen
  \bibfield  {author} {\bibinfo {author} {\bibfnamefont {A.~K.}\ \bibnamefont
  {{Mehta}}}, \bibinfo {author} {\bibfnamefont {D.}~\bibnamefont {{Wadekar}}},
  \bibinfo {author} {\bibfnamefont {J.}~\bibnamefont {{Roulet}}}, \bibinfo
  {author} {\bibfnamefont {I.}~\bibnamefont {{Anantpurkar}}},  \emph {et~al.},\
  }\href {\doibase 10.48550/arXiv.2501.17939} {\bibfield  {journal} {\bibinfo
  {journal} {arXiv e-prints}\ ,\ \bibinfo {eid} {arXiv:2501.17939}} (\bibinfo
  {year} {2025})},\ \Eprint {http://arxiv.org/abs/2501.17939} {arXiv:2501.17939
  [gr-qc]} \BibitemShut {NoStop}%
\bibitem [{\citenamefont {LIGO Scientific~Collaboration}(2021)}]{zenodoLVK}%
  \BibitemOpen
  \bibfield  {author} {\bibinfo {author} {\bibfnamefont {K.~C.}\ \bibnamefont
  {LIGO Scientific~Collaboration}, \bibfnamefont {Virgo~Collaboration}},\
  }\href {\doibase 10.5281/zenodo.5546675} {\  (\bibinfo {year} {2021}),\
  10.5281/zenodo.5546675}\BibitemShut {NoStop}%
\bibitem [{\citenamefont {{McIsaac}}\ \emph {et~al.}(2023)\citenamefont
  {{McIsaac}}, \citenamefont {{Hoy}},\ and\ \citenamefont {{Harry}}}]{McI23}%
  \BibitemOpen
  \bibfield  {author} {\bibinfo {author} {\bibfnamefont {C.}~\bibnamefont
  {{McIsaac}}}, \bibinfo {author} {\bibfnamefont {C.}~\bibnamefont {{Hoy}}}, \
  and\ \bibinfo {author} {\bibfnamefont {I.}~\bibnamefont {{Harry}}},\ }\href
  {\doibase 10.48550/arXiv.2303.17364} {\bibfield  {journal} {\bibinfo
  {journal} {arXiv e-prints}\ ,\ \bibinfo {eid} {arXiv:2303.17364}} (\bibinfo
  {year} {2023})},\ \Eprint {http://arxiv.org/abs/2303.17364} {arXiv:2303.17364
  [gr-qc]} \BibitemShut {NoStop}%
\bibitem [{\citenamefont {{Schmidt}}\ \emph
  {et~al.}(2024{\natexlab{a}})\citenamefont {{Schmidt}}, \citenamefont
  {{Caudill}} \emph {et~al.}}]{Sch24_Precession_Pipeline}%
  \BibitemOpen
  \bibfield  {author} {\bibinfo {author} {\bibfnamefont {S.}~\bibnamefont
  {{Schmidt}}}, \bibinfo {author} {\bibfnamefont {S.}~\bibnamefont
  {{Caudill}}},  \emph {et~al.},\ }\href {\doibase 10.48550/arXiv.2403.17186}
  {\bibfield  {journal} {\bibinfo  {journal} {arXiv e-prints}\ ,\ \bibinfo
  {eid} {arXiv:2403.17186}} (\bibinfo {year} {2024}{\natexlab{a}})},\ \Eprint
  {http://arxiv.org/abs/2403.17186} {arXiv:2403.17186 [gr-qc]} \BibitemShut
  {NoStop}%
\bibitem [{\citenamefont {{Schmidt}}\ \emph
  {et~al.}(2024{\natexlab{b}})\citenamefont {{Schmidt}}, \citenamefont
  {{Caudill}} \emph {et~al.}}]{Sch24_Precession_O3_Search}%
  \BibitemOpen
  \bibfield  {author} {\bibinfo {author} {\bibfnamefont {S.}~\bibnamefont
  {{Schmidt}}}, \bibinfo {author} {\bibfnamefont {S.}~\bibnamefont
  {{Caudill}}},  \emph {et~al.},\ }\href {\doibase 10.48550/arXiv.2406.17832}
  {\bibfield  {journal} {\bibinfo  {journal} {arXiv e-prints}\ ,\ \bibinfo
  {eid} {arXiv:2406.17832}} (\bibinfo {year} {2024}{\natexlab{b}})},\ \Eprint
  {http://arxiv.org/abs/2406.17832} {arXiv:2406.17832 [gr-qc]} \BibitemShut
  {NoStop}%
\bibitem [{\citenamefont {{Dhurkunde}}\ and\ \citenamefont
  {{Nitz}}(2022)}]{Dhu22}%
  \BibitemOpen
  \bibfield  {author} {\bibinfo {author} {\bibfnamefont {R.}~\bibnamefont
  {{Dhurkunde}}}\ and\ \bibinfo {author} {\bibfnamefont {A.~H.}\ \bibnamefont
  {{Nitz}}},\ }\href {\doibase 10.1103/PhysRevD.106.103035} {\bibfield
  {journal} {\bibinfo  {journal} {\prd}\ }\textbf {\bibinfo {volume} {106}},\
  \bibinfo {eid} {103035} (\bibinfo {year} {2022})},\ \Eprint
  {http://arxiv.org/abs/2207.14645} {arXiv:2207.14645 [astro-ph.IM]}
  \BibitemShut {NoStop}%
\bibitem [{\citenamefont {{Dhurkunde}}\ and\ \citenamefont
  {{Nitz}}(2023)}]{Dhu23_eccentricity_search}%
  \BibitemOpen
  \bibfield  {author} {\bibinfo {author} {\bibfnamefont {R.}~\bibnamefont
  {{Dhurkunde}}}\ and\ \bibinfo {author} {\bibfnamefont {A.~H.}\ \bibnamefont
  {{Nitz}}},\ }\href {\doibase 10.48550/arXiv.2311.00242} {\bibfield  {journal}
  {\bibinfo  {journal} {arXiv e-prints}\ ,\ \bibinfo {eid} {arXiv:2311.00242}}
  (\bibinfo {year} {2023})},\ \Eprint {http://arxiv.org/abs/2311.00242}
  {arXiv:2311.00242 [astro-ph.HE]} \BibitemShut {NoStop}%
\bibitem [{\citenamefont {{Nitz}}\ \emph {et~al.}(2020)\citenamefont {{Nitz}},
  \citenamefont {{Lenon}},\ and\ \citenamefont
  {{Brown}}}]{Nit20_eccentricity_search}%
  \BibitemOpen
  \bibfield  {author} {\bibinfo {author} {\bibfnamefont {A.~H.}\ \bibnamefont
  {{Nitz}}}, \bibinfo {author} {\bibfnamefont {A.}~\bibnamefont {{Lenon}}}, \
  and\ \bibinfo {author} {\bibfnamefont {D.~A.}\ \bibnamefont {{Brown}}},\
  }\href {\doibase 10.3847/1538-4357/ab6611} {\bibfield  {journal} {\bibinfo
  {journal} {\apj}\ }\textbf {\bibinfo {volume} {890}},\ \bibinfo {eid} {1}
  (\bibinfo {year} {2020})},\ \Eprint {http://arxiv.org/abs/1912.05464}
  {arXiv:1912.05464 [astro-ph.HE]} \BibitemShut {NoStop}%
\bibitem [{\citenamefont {Zhou}\ \emph {et~al.}()\citenamefont {Zhou} \emph
  {et~al.}}]{Zho25_Prec_TemplateBank}%
  \BibitemOpen
  \bibfield  {author} {\bibinfo {author} {\bibfnamefont {Z.}~\bibnamefont
  {Zhou}} \emph {et~al.},\ }\href@noop {} {\bibinfo  {journal} {in
  preparation}\ }\BibitemShut {NoStop}%
\bibitem [{\citenamefont {Islam}\ \emph {et~al.}()\citenamefont {Islam} \emph
  {et~al.}}]{Isl25_Eccentricity_TemplateBank}%
  \BibitemOpen
\bibfield  {journal} {  }\bibfield  {author} {\bibinfo {author} {\bibfnamefont
  {T.}~\bibnamefont {Islam}} \emph {et~al.},\ }\href@noop {} {\bibinfo
  {journal} {in preparation}\ }\BibitemShut {NoStop}%
\bibitem [{\citenamefont {Payne}\ \emph {et~al.}(2019)\citenamefont {Payne},
  \citenamefont {Talbot},\ and\ \citenamefont
  {Thrane}}]{reweighting_importance_sampling_thrane2019}%
  \BibitemOpen
\bibfield  {journal} {  }\bibfield  {author} {\bibinfo {author} {\bibfnamefont
  {E.}~\bibnamefont {Payne}}, \bibinfo {author} {\bibfnamefont
  {C.}~\bibnamefont {Talbot}}, \ and\ \bibinfo {author} {\bibfnamefont
  {E.}~\bibnamefont {Thrane}},\ }\href {\doibase 10.1103/PhysRevD.100.123017}
  {\bibfield  {journal} {\bibinfo  {journal} {Phys. Rev. D}\ }\textbf {\bibinfo
  {volume} {100}},\ \bibinfo {pages} {123017} (\bibinfo {year}
  {2019})}\BibitemShut {NoStop}%
\bibitem [{\citenamefont {Zackay}\ \emph {et~al.}(2019)\citenamefont {Zackay},
  \citenamefont {Venumadhav}, \citenamefont {Dai}, \citenamefont {Roulet},\
  and\ \citenamefont
  {Zaldarriaga}}]{flat_chieff_prior_o1event1216_formation_channels2019}%
  \BibitemOpen
  \bibfield  {author} {\bibinfo {author} {\bibfnamefont {B.}~\bibnamefont
  {Zackay}}, \bibinfo {author} {\bibfnamefont {T.}~\bibnamefont {Venumadhav}},
  \bibinfo {author} {\bibfnamefont {L.}~\bibnamefont {Dai}}, \bibinfo {author}
  {\bibfnamefont {J.}~\bibnamefont {Roulet}}, \ and\ \bibinfo {author}
  {\bibfnamefont {M.}~\bibnamefont {Zaldarriaga}},\ }\href {\doibase
  10.1103/PhysRevD.100.023007} {\bibfield  {journal} {\bibinfo  {journal}
  {Phys. Rev. D}\ }\textbf {\bibinfo {volume} {100}},\ \bibinfo {pages}
  {023007} (\bibinfo {year} {2019})},\ \Eprint
  {http://arxiv.org/abs/1902.10331} {arXiv:1902.10331 [astro-ph.HE]}
  \BibitemShut {NoStop}%
\bibitem [{\citenamefont {{Gaebel}}\ \emph {et~al.}(2019)\citenamefont
  {{Gaebel}}, \citenamefont {{Veitch}}, \citenamefont {{Dent}},\ and\
  \citenamefont {{Farr}}}]{Gae19_Pop_Subthreshold}%
  \BibitemOpen
  \bibfield  {author} {\bibinfo {author} {\bibfnamefont {S.~M.}\ \bibnamefont
  {{Gaebel}}}, \bibinfo {author} {\bibfnamefont {J.}~\bibnamefont {{Veitch}}},
  \bibinfo {author} {\bibfnamefont {T.}~\bibnamefont {{Dent}}}, \ and\ \bibinfo
  {author} {\bibfnamefont {W.~M.}\ \bibnamefont {{Farr}}},\ }\href {\doibase
  10.1093/mnras/stz225} {\bibfield  {journal} {\bibinfo  {journal} {\mnras}\
  }\textbf {\bibinfo {volume} {484}},\ \bibinfo {pages} {4008} (\bibinfo {year}
  {2019})},\ \Eprint {http://arxiv.org/abs/1809.03815} {arXiv:1809.03815
  [astro-ph.IM]} \BibitemShut {NoStop}%
\bibitem [{\citenamefont {Mehta}\ \emph {et~al.}()\citenamefont {Mehta} \emph
  {et~al.}}]{Meh25_Population_HM}%
  \BibitemOpen
  \bibfield  {author} {\bibinfo {author} {\bibfnamefont {A.~K.}\ \bibnamefont
  {Mehta}} \emph {et~al.},\ }\href@noop {} {\bibinfo  {journal} {in
  preparation}\ }\BibitemShut {NoStop}%
\bibitem [{\citenamefont {Ng}\ \emph {et~al.}(2023)\citenamefont {Ng} \emph
  {et~al.}}]{Ng22_ET_HM}%
  \BibitemOpen
\bibfield  {journal} {  }\bibfield  {author} {\bibinfo {author} {\bibfnamefont
  {K.~K.~Y.}\ \bibnamefont {Ng}} \emph {et~al.},\ }\href {\doibase
  10.1103/PhysRevD.107.024041} {\bibfield  {journal} {\bibinfo  {journal}
  {Phys. Rev. D}\ }\textbf {\bibinfo {volume} {107}},\ \bibinfo {pages}
  {024041} (\bibinfo {year} {2023})},\ \Eprint
  {http://arxiv.org/abs/2210.03132} {arXiv:2210.03132 [astro-ph.CO]}
  \BibitemShut {NoStop}%
\bibitem [{\citenamefont {{Fairhurst}}\ \emph {et~al.}(2023)\citenamefont
  {{Fairhurst}}, \citenamefont {{Mills}}, \citenamefont {{Colpi}},
  \citenamefont {{Schneider}}, \citenamefont {{Sesana}}, \citenamefont
  {{Trinca}},\ and\ \citenamefont {{Valiante}}}]{Fai23}%
  \BibitemOpen
  \bibfield  {author} {\bibinfo {author} {\bibfnamefont {S.}~\bibnamefont
  {{Fairhurst}}}, \bibinfo {author} {\bibfnamefont {C.}~\bibnamefont
  {{Mills}}}, \bibinfo {author} {\bibfnamefont {M.}~\bibnamefont {{Colpi}}},
  \bibinfo {author} {\bibfnamefont {R.}~\bibnamefont {{Schneider}}}, \bibinfo
  {author} {\bibfnamefont {A.}~\bibnamefont {{Sesana}}}, \bibinfo {author}
  {\bibfnamefont {A.}~\bibnamefont {{Trinca}}}, \ and\ \bibinfo {author}
  {\bibfnamefont {R.}~\bibnamefont {{Valiante}}},\ }\href {\doibase
  10.48550/arXiv.2310.18158} {\bibfield  {journal} {\bibinfo  {journal} {arXiv
  e-prints}\ ,\ \bibinfo {eid} {arXiv:2310.18158}} (\bibinfo {year} {2023})},\
  \Eprint {http://arxiv.org/abs/2310.18158} {arXiv:2310.18158 [astro-ph.HE]}
  \BibitemShut {NoStop}%
\bibitem [{\citenamefont {Maggiore}\ \emph {et~al.}(2020)\citenamefont
  {Maggiore} \emph {et~al.}}]{einsteinTelescope_Maggiore:2019uih}%
  \BibitemOpen
  \bibfield  {author} {\bibinfo {author} {\bibfnamefont {M.}~\bibnamefont
  {Maggiore}} \emph {et~al.},\ }\href {\doibase 10.1088/1475-7516/2020/03/050}
  {\bibfield  {journal} {\bibinfo  {journal} {JCAP}\ }\textbf {\bibinfo
  {volume} {03}},\ \bibinfo {pages} {050} (\bibinfo {year} {2020})},\ \Eprint
  {http://arxiv.org/abs/1912.02622} {arXiv:1912.02622 [astro-ph.CO]}
  \BibitemShut {NoStop}%
\bibitem [{\citenamefont {{Evans}}\ \emph {et~al.}(2023)\citenamefont {{Evans}}
  \emph {et~al.}}]{Eva23}%
  \BibitemOpen
  \bibfield  {author} {\bibinfo {author} {\bibfnamefont {M.}~\bibnamefont
  {{Evans}}} \emph {et~al.},\ }\href {\doibase 10.48550/arXiv.2306.13745}
  {\bibfield  {journal} {\bibinfo  {journal} {arXiv e-prints}\ ,\ \bibinfo
  {eid} {arXiv:2306.13745}} (\bibinfo {year} {2023})},\ \Eprint
  {http://arxiv.org/abs/2306.13745} {arXiv:2306.13745 [astro-ph.IM]}
  \BibitemShut {NoStop}%
\bibitem [{\citenamefont {Kawamura}\ \emph {et~al.}(2020)\citenamefont
  {Kawamura} \emph {et~al.}}]{decigo_spaceDetector_kawamura2020}%
  \BibitemOpen
  \bibfield  {author} {\bibinfo {author} {\bibfnamefont {S.}~\bibnamefont
  {Kawamura}} \emph {et~al.},\ }\href {\doibase 10.48550/ARXIV.2006.13545} {\
  (\bibinfo {year} {2020}),\ 10.48550/ARXIV.2006.13545}\BibitemShut {NoStop}%
\bibitem [{\citenamefont {{Amaro-Seoane}}\ \emph {et~al.}(2017)\citenamefont
  {{Amaro-Seoane}} \emph {et~al.}}]{LISA}%
  \BibitemOpen
  \bibfield  {author} {\bibinfo {author} {\bibfnamefont {P.}~\bibnamefont
  {{Amaro-Seoane}}} \emph {et~al.},\ }\href {\doibase
  10.48550/arXiv.1702.00786} {\bibfield  {journal} {\bibinfo  {journal} {arXiv
  e-prints}\ ,\ \bibinfo {eid} {arXiv:1702.00786}} (\bibinfo {year} {2017})},\
  \Eprint {http://arxiv.org/abs/1702.00786} {arXiv:1702.00786 [astro-ph.IM]}
  \BibitemShut {NoStop}%
\bibitem [{\citenamefont {Akutsu}\ \emph {et~al.}(2021)\citenamefont {Akutsu}
  \emph {et~al.}}]{KAGRA:2020tym}%
  \BibitemOpen
  \bibfield  {author} {\bibinfo {author} {\bibfnamefont {T.}~\bibnamefont
  {Akutsu}} \emph {et~al.} (\bibinfo {collaboration} {KAGRA}),\ }\href
  {\doibase 10.1093/ptep/ptaa125} {\bibfield  {journal} {\bibinfo  {journal}
  {PTEP}\ }\textbf {\bibinfo {volume} {2021}},\ \bibinfo {pages} {05A101}
  (\bibinfo {year} {2021})},\ \Eprint {http://arxiv.org/abs/2005.05574}
  {arXiv:2005.05574 [physics.ins-det]} \BibitemShut {NoStop}%
\bibitem [{\citenamefont {{Arun}}\ \emph
  {et~al.}(2007{\natexlab{a}})\citenamefont {{Arun}}, \citenamefont {{Iyer}},
  \citenamefont {{Sathyaprakash}}, \citenamefont {{Sinha}},\ and\ \citenamefont
  {{van den Broeck}}}]{Aru07_LISA}%
  \BibitemOpen
  \bibfield  {author} {\bibinfo {author} {\bibfnamefont {K.~G.}\ \bibnamefont
  {{Arun}}}, \bibinfo {author} {\bibfnamefont {B.~R.}\ \bibnamefont {{Iyer}}},
  \bibinfo {author} {\bibfnamefont {B.~S.}\ \bibnamefont {{Sathyaprakash}}},
  \bibinfo {author} {\bibfnamefont {S.}~\bibnamefont {{Sinha}}}, \ and\
  \bibinfo {author} {\bibfnamefont {C.}~\bibnamefont {{van den Broeck}}},\
  }\href {\doibase 10.1103/PhysRevD.76.104016} {\bibfield  {journal} {\bibinfo
  {journal} {\prd}\ }\textbf {\bibinfo {volume} {76}},\ \bibinfo {eid} {104016}
  (\bibinfo {year} {2007}{\natexlab{a}})},\ \Eprint
  {http://arxiv.org/abs/0707.3920} {arXiv:0707.3920 [astro-ph]} \BibitemShut
  {NoStop}%
\bibitem [{\citenamefont {{Arun}}\ \emph
  {et~al.}(2007{\natexlab{b}})\citenamefont {{Arun}}, \citenamefont {{Iyer}},
  \citenamefont {{Sathyaprakash}},\ and\ \citenamefont
  {{Sinha}}}]{Aru07_LISA_2}%
  \BibitemOpen
  \bibfield  {author} {\bibinfo {author} {\bibfnamefont {K.~G.}\ \bibnamefont
  {{Arun}}}, \bibinfo {author} {\bibfnamefont {B.~R.}\ \bibnamefont {{Iyer}}},
  \bibinfo {author} {\bibfnamefont {B.~S.}\ \bibnamefont {{Sathyaprakash}}}, \
  and\ \bibinfo {author} {\bibfnamefont {S.}~\bibnamefont {{Sinha}}},\ }\href
  {\doibase 10.1103/PhysRevD.75.124002} {\bibfield  {journal} {\bibinfo
  {journal} {\prd}\ }\textbf {\bibinfo {volume} {75}},\ \bibinfo {eid} {124002}
  (\bibinfo {year} {2007}{\natexlab{b}})},\ \Eprint
  {http://arxiv.org/abs/0704.1086} {arXiv:0704.1086 [gr-qc]} \BibitemShut
  {NoStop}%
\bibitem [{\citenamefont {{Porter}}\ and\ \citenamefont
  {{Cornish}}(2008)}]{Por08_LISA}%
  \BibitemOpen
  \bibfield  {author} {\bibinfo {author} {\bibfnamefont {E.~K.}\ \bibnamefont
  {{Porter}}}\ and\ \bibinfo {author} {\bibfnamefont {N.~J.}\ \bibnamefont
  {{Cornish}}},\ }\href {\doibase 10.1103/PhysRevD.78.064005} {\bibfield
  {journal} {\bibinfo  {journal} {\prd}\ }\textbf {\bibinfo {volume} {78}},\
  \bibinfo {eid} {064005} (\bibinfo {year} {2008})},\ \Eprint
  {http://arxiv.org/abs/0804.0332} {arXiv:0804.0332 [gr-qc]} \BibitemShut
  {NoStop}%
\bibitem [{\citenamefont {{Berti}}\ \emph {et~al.}(2006)\citenamefont
  {{Berti}}, \citenamefont {{Cardoso}},\ and\ \citenamefont
  {{Will}}}]{Ber06_LISA}%
  \BibitemOpen
  \bibfield  {author} {\bibinfo {author} {\bibfnamefont {E.}~\bibnamefont
  {{Berti}}}, \bibinfo {author} {\bibfnamefont {V.}~\bibnamefont {{Cardoso}}},
  \ and\ \bibinfo {author} {\bibfnamefont {C.~M.}\ \bibnamefont {{Will}}},\
  }\href {\doibase 10.1103/PhysRevD.73.064030} {\bibfield  {journal} {\bibinfo
  {journal} {\prd}\ }\textbf {\bibinfo {volume} {73}},\ \bibinfo {eid} {064030}
  (\bibinfo {year} {2006})},\ \Eprint {http://arxiv.org/abs/gr-qc/0512160}
  {arXiv:gr-qc/0512160 [gr-qc]} \BibitemShut {NoStop}%
\bibitem [{\citenamefont {{Baibhav}}\ \emph {et~al.}(2020)\citenamefont
  {{Baibhav}}, \citenamefont {{Berti}},\ and\ \citenamefont
  {{Cardoso}}}]{Bai20}%
  \BibitemOpen
  \bibfield  {author} {\bibinfo {author} {\bibfnamefont {V.}~\bibnamefont
  {{Baibhav}}}, \bibinfo {author} {\bibfnamefont {E.}~\bibnamefont {{Berti}}},
  \ and\ \bibinfo {author} {\bibfnamefont {V.}~\bibnamefont {{Cardoso}}},\
  }\href {\doibase 10.1103/PhysRevD.101.084053} {\bibfield  {journal} {\bibinfo
   {journal} {\prd}\ }\textbf {\bibinfo {volume} {101}},\ \bibinfo {eid}
  {084053} (\bibinfo {year} {2020})},\ \Eprint
  {http://arxiv.org/abs/2001.10011} {arXiv:2001.10011 [gr-qc]} \BibitemShut
  {NoStop}%
\bibitem [{\citenamefont {{Pitte}}\ \emph {et~al.}(2023)\citenamefont
  {{Pitte}}, \citenamefont {{Baghi}}, \citenamefont {{Marsat}}, \citenamefont
  {{Besan{\c{c}}on}},\ and\ \citenamefont {{Petiteau}}}]{Pit23_LISA}%
  \BibitemOpen
  \bibfield  {author} {\bibinfo {author} {\bibfnamefont {C.}~\bibnamefont
  {{Pitte}}}, \bibinfo {author} {\bibfnamefont {Q.}~\bibnamefont {{Baghi}}},
  \bibinfo {author} {\bibfnamefont {S.}~\bibnamefont {{Marsat}}}, \bibinfo
  {author} {\bibfnamefont {M.}~\bibnamefont {{Besan{\c{c}}on}}}, \ and\
  \bibinfo {author} {\bibfnamefont {A.}~\bibnamefont {{Petiteau}}},\ }\href
  {\doibase 10.1103/PhysRevD.108.044053} {\bibfield  {journal} {\bibinfo
  {journal} {\prd}\ }\textbf {\bibinfo {volume} {108}},\ \bibinfo {eid}
  {044053} (\bibinfo {year} {2023})},\ \Eprint
  {http://arxiv.org/abs/2304.03142} {arXiv:2304.03142 [gr-qc]} \BibitemShut
  {NoStop}%
\bibitem [{\citenamefont {{Amaro-Seoane}}\ \emph {et~al.}(2023)\citenamefont
  {{Amaro-Seoane}} \emph {et~al.}}]{LISA_astrophysics}%
  \BibitemOpen
  \bibfield  {author} {\bibinfo {author} {\bibfnamefont {P.}~\bibnamefont
  {{Amaro-Seoane}}} \emph {et~al.},\ }\href {\doibase
  10.1007/s41114-022-00041-y} {\bibfield  {journal} {\bibinfo  {journal}
  {Living Reviews in Relativity}\ }\textbf {\bibinfo {volume} {26}},\ \bibinfo
  {eid} {2} (\bibinfo {year} {2023})},\ \Eprint
  {http://arxiv.org/abs/2203.06016} {arXiv:2203.06016 [gr-qc]} \BibitemShut
  {NoStop}%
\bibitem [{\citenamefont {{Gair}}\ \emph {et~al.}(2008)\citenamefont {{Gair}},
  \citenamefont {{Mandel}},\ and\ \citenamefont
  {{Wen}}}]{Gai08_LISA_harmonics}%
  \BibitemOpen
  \bibfield  {author} {\bibinfo {author} {\bibfnamefont {J.~R.}\ \bibnamefont
  {{Gair}}}, \bibinfo {author} {\bibfnamefont {I.}~\bibnamefont {{Mandel}}}, \
  and\ \bibinfo {author} {\bibfnamefont {L.}~\bibnamefont {{Wen}}},\ }\href
  {\doibase 10.1088/0264-9381/25/18/184031} {\bibfield  {journal} {\bibinfo
  {journal} {Classical and Quantum Gravity}\ }\textbf {\bibinfo {volume}
  {25}},\ \bibinfo {eid} {184031} (\bibinfo {year} {2008})},\ \Eprint
  {http://arxiv.org/abs/0804.1084} {arXiv:0804.1084 [gr-qc]} \BibitemShut
  {NoStop}%
\bibitem [{\citenamefont {{Ho-Yeuk Cheung}}\ \emph {et~al.}(2025)\citenamefont
  {{Ho-Yeuk Cheung}}, \citenamefont {{Wadekar}}, \citenamefont {{Mehta}},
  \citenamefont {{Islam}}, \citenamefont {{Roulet}}, \citenamefont {{Berti}},
  \citenamefont {{Venumadhav}}, \citenamefont {{Zackay}},\ and\ \citenamefont
  {{Zaldarriaga}}}]{Che25_IMRI_search}%
  \BibitemOpen
  \bibfield  {author} {\bibinfo {author} {\bibfnamefont {M.}~\bibnamefont
  {{Ho-Yeuk Cheung}}}, \bibinfo {author} {\bibfnamefont {D.}~\bibnamefont
  {{Wadekar}}}, \bibinfo {author} {\bibfnamefont {A.~K.}\ \bibnamefont
  {{Mehta}}}, \bibinfo {author} {\bibfnamefont {T.}~\bibnamefont {{Islam}}},
  \emph {et~al.},\ }\href@noop {} {\bibfield  {journal} {\bibinfo  {journal}
  {arXiv e-prints}\ ,\ \bibinfo {eid} {arXiv:2507.01083}} (\bibinfo {year}
  {2025})},\ \Eprint {http://arxiv.org/abs/2507.01083} {arXiv:2507.01083
  [gr-qc]} \BibitemShut {NoStop}%
\bibitem [{\citenamefont {Abac}\ \emph {et~al.}(2025)\citenamefont {Abac} \emph
  {et~al.}}]{GW231123}%
  \BibitemOpen
  \bibfield  {author} {\bibinfo {author} {\bibfnamefont {A.~G.}\ \bibnamefont
  {Abac}} \emph {et~al.} (\bibinfo {collaboration} {LIGO Scientific, VIRGO,
  KAGRA}),\ }\href@noop {} {\  (\bibinfo {year} {2025})},\ \Eprint
  {http://arxiv.org/abs/2507.08219} {arXiv:2507.08219 [astro-ph.HE]}
  \BibitemShut {NoStop}%
\end{thebibliography}%
\end{document}